\documentclass{article}
\usepackage{amssymb}
\usepackage{amsfonts}
\usepackage{amsmath}

\input{tcilatex}
\begin{document}

\begin{center}
\bigskip

{\large From 2-d Polyakov Action to the 4-d Pseudo-Conformal Field Theory}

{\large \ }\bigskip

C. N. Ragiadakos

email: ragiadak@gmail.com

\bigskip

\textbf{ABSTRACT}
\end{center}

The characteristic property of the 2-dimensional Polyakov action is its
independence on the metric tensor, without being topological. A
renormalizable 4-dimensional action is found satisfying this fundamental
property. The fundamental quantity of this pseudo-conformal field theory
(PCFT) is the lorentzian Cauchy-Riemann (LCR) structure. This action
describes all current phenomenology: 1) The Poincar\'{e} group is
determined. 2) Stable solitonic LCR-tetrads are found, which belong to
representations of the Poincar\'{e} group and they are determined by the
irreducible and reducible algebraic quadratic surfaces of CP3. 3) The static
(irreducible) LCR-structure implies the Kerr-Newman manifold with g=2
gyromagnetic ratio and it is identified with the electron. The stationary
(reducible) LCR-structure is identified with the neutrino. The antiparticles
have conjugate LCR-structures. The Hawking-Penrose singularity theorems are
bypassed in the electron LCR-manifold. 4) The LCR-tetrad defines Einstein's
metric and the U(2) electroweak connection. 5) An effective leptonic
standard model action is derived using the Bogoliubov-Scharf recursive
procedure. 6) The three generations of flavors are implied by the limited
number (for curved spacetime) of permitted algebraic surfaces of CP3. 7) For
every LCR-structure there exists a solitonic distributional gauge field
configuration, identified with the corresponding quark, which explains the
lepton-quark correspondence. It is explicitly computed for the static
LCR-structure. 8) The derivation of a proper geometric SU(3) Cartan
connection opens up the possibility to achieve Einstein's goal to derive all
interactions from the pure geometric LCR-structure.\newpage 

\bigskip

{\LARGE Contents}

\textbf{1. INTRODUCTION}

\textbf{2. LORENTZIAN\ CR-STRUCTURE}

\textbf{3. THE\ COVARIANT\ ACTION\ OF\ PCFT}

\textbf{4. DERIVATION\ OF\ EINSTEIN'S GRAVITY}

\textbf{5. ELECTRON\ AND\ ELECTRODYNAMICS}

\qquad 5.1 Microlocal analysis of the electron

\qquad 5.2 Derivation of quantum electrodynamics

\qquad 5.3 LCR-ray tracing in the electron LCR-manifold

\textbf{6. NEUTRINO\ AND\ STANDARD\ MODEL}

\qquad 6.1 The electroweak-U(2) gauge fields

\qquad 6.2 Derivation of standard model action

\textbf{7. THE\ UP\ AND\ DOWN\ QUARKS}

\qquad 7.1 A quark confining mechanism

\textbf{8. A SU(3) CONNECTION\ FROM\ THE\ LCR-STRUCTURE}

\textbf{9. PERSPECTIVES}

\qquad \newpage

\bigskip

\renewcommand{\theequation}{\arabic{section}.\arabic{equation}}

\section{INTRODUCTION}

\setcounter{equation}{0}

The 2-dimensional Polyakov (string) action has the remarkable property to be
metric independent without being topological (i.e. a pure surface integral).
In fact this particular property is the essential origin of its mathematical
beauty. The higher dimensional conformal field theories (the
Weyl-transformation invariant covariant forms) are not metric independent,
therefore they cannot be considered as the 4-dimensional versions of the
Polyakov action. I found\cite{RAG1990} and studied a 4-dimensional generally
covariant action, which is metric independent without being topological.
This action describes the current phenomenology without needing
supersymmetry, which has not been observed. Besides gravity, it describes%
\cite{RAG2018a} the standard model as an effective field theory, with the
only essential difference the hadronic sector. The strong interactions are
described\cite{RAG2018b} from a 4-dimensional gauge field (gluon) which
explicitly appears with a metric independent non-laplacian lagrangian (with
first order derivatives) .

If we clarify the origin of the metric independence of the Polyakov action 
\begin{equation}
\begin{array}{l}
I_{S}=\frac{1}{2}\int d^{2}\!\xi \ \sqrt{-g}\ g^{\alpha \beta }\ \partial
_{\alpha }X^{\mu }\partial _{\beta }X^{\nu }\eta _{\mu \nu } \\ 
\end{array}
\label{i-1}
\end{equation}%
the invention of the 4-dimensional action is rather simple. Notice that the
metric independence is caused by the general property of the 2-dimensional
metrics to admit a coordinate system (the light-cone coordinates), which
makes them off-diagonal, i.e. 
\begin{equation}
\begin{array}{l}
ds^{2}=2g_{01}d\xi _{+}d\xi _{-} \\ 
\\ 
I_{S}=\int d^{2}\!\xi \ \partial _{-}X^{\mu }\partial _{+}X^{\nu }\eta _{\mu
\nu } \\ 
\end{array}
\label{i-2}
\end{equation}

Apparently the 4-dimensional spacetime metrics cannot generally take an
analogous off-diagonal form. Only metrics, which admit two geodetic and
shear-free null congruences $\ell ^{\mu }\partial _{\mu },\ n^{\mu }\partial
_{\mu }$ can take\cite{FLAHE1974}$^{,}$\cite{FLAHE1976} this form 
\begin{equation}
\begin{array}{l}
ds^{2}=2g_{a\widetilde{\beta }}dz^{\alpha }dz^{\widetilde{\beta }}\quad
,\quad \alpha ,\widetilde{\beta }=0,1 \\ 
\end{array}
\label{i-3}
\end{equation}%
where $z^{b}=(z^{\alpha }(x),z^{\widetilde{\beta }}(x))$ are generally
complex coordinates. In this case we can write down the following metric
independent Yang-Mills-like integral 
\begin{equation}
\begin{array}{l}
I_{G}=\int d^{4}\!z\ \sqrt{-g}g^{\alpha \widetilde{\alpha }}g^{\beta 
\widetilde{\beta }}F_{\!j\alpha \beta }F_{\!j\widetilde{\alpha }\widetilde{%
\beta }}=\int d^{4}\!z\ F_{\!j01}F_{\!j\widetilde{0}\widetilde{1}} \\ 
\\ 
F_{j_{ab}}=\partial _{a}A_{jb}-\partial _{a}A_{jb}-\gamma
\,f_{jik}A_{ia}A_{kb}%
\end{array}
\label{i-4}
\end{equation}%
which depends on the coordinates $(z^{\alpha }(x),z^{\widetilde{\beta }}(x))$%
, and it does not depend on the metric. This integral is apparently complex,
because the structure coordinates are complex. Therefore the real spacetime
action must be either its real or imaginary part. It will be clarified in
section III, where the precise covariant action will be derived. The
restriction on the metrics which admit two geodetic and shear free
congruences, should not physically bother us, because the black-holes have
this property. On the contrary, it is rather encouraging, because it
provides an argument why all the observed spacetimes are Schwarzschild type.
The complete generally covariant action will be described in section III,
after introducing the necessary mathematical notions.

Our first observation is on the fields necessary to achieve metric
independence in two and four dimensions. The Polyakov action needs a scalar
field $X^{\mu }(x)$, which string theory interprets as the embedding
function of the Riemann surfaces in the 26-dimensional ordinary spacetime.
My 4-dimensional metric independent action needs a peculiar gauge field,
which is not of the laplacian type. The implied spherically symmetric
equation and its static "potential" is 
\begin{equation}
\begin{array}{l}
(\frac{\partial ^{2}}{\partial t^{2}}-\frac{\partial ^{2}}{\partial r^{2}}%
)\Phi =[source]\quad ,\quad \Phi =a+br \\ 
\\ 
(\frac{\partial ^{2}}{\partial t^{2}}-\frac{1}{r}\frac{\partial ^{2}}{%
\partial r^{2}}r)\Phi =[source]\quad ,\quad \Phi =\frac{Q}{4\pi r} \\ 
\end{array}
\label{i-5}
\end{equation}%
where in the second line, I write the spherically symmetric laplacian field
equation and its corresponding static "potential". Notice the essential
difference. The present 4-dimensional action gives a confining linear
"potential" in the place of the ($\frac{1}{r}$) potential of the laplacian.
Therefore the present gauge field will be identified with the gluon field.
But the mathematical procedure will be quite sophisticated, because the
symmetries of the present action do not permit the introduction of fermionic
fields, as it happens in ordinary quantum field theory (QFT). That is the
term $[source]$ cannot now be put in the action by hand, because it would
destroy the renormalizability of the action. It must be derived. It is
possible, because of the great advantage of PCFT to admit solitonic
solutions, which are generalized functions (with distributional sources)\cite%
{GELF1}. The function, outside the compact (closed and bounded) source, is
the "potential" and its "source" is the fermionic particle. A typical
example is the defined electromagnetic field with its source been the
electron. Or equivalently the electron in PCFT is a dressing electric field
with its singularity. It is a fermion, because of the well known Carter
observation\cite{CARTER} that the Kerr-Newman spacetime has a fermionic
gyromagnetic ratio. This puzzle, that surprised general relativists\cite%
{NEWM2016}, finds its "raison d'\^{e}tre" in the present theory. In the case
of the gauge field, distributional configurations are found with their
sources identified with the quarks. Therefore the 4-dimensional PCFT does
not need supersymmetry to incorporate fermions. In fact my efforts to
supersymmetrize it have also failed. Supersymmetry may not be compatible
with the metric independence of PCFT.

The Polyakov action does not depend on the 2-dimensional metric, but it does
depend on the more general notion of complex structure (after the usual Wick
rotation). Recall that the string functional integral\cite{POL} is an
integration over the 2-dimentional complex manifolds. My 4-dimensional
action does not depend on the metric tensor, but it does depend on a special
Cauchy-Riemann (CR) structure of the spacetime, the lorentzian CR-structure
(LCR-structure), which is viewed as the existence of two geodetic and
shear-free null congruences in the metric (riemannian) language of general
relativity. The CR-structure was called pseudo-conformal by E. Cartan\cite%
{CARTAN} and Tanaka, who first worked on real submanifolds of complex
manifolds. This is the reason that I call the present kind of 4-dimentional
field theory pseudo-conformal field theory (PCFT), in order to stress its
complete mathematical (but not physical) analogy with the 2-dimensional
Polyakov action.

The present action is based on the lorentzian CR-structure\cite{RAG2013b},
which is determined by two real and one complex independent 1-forms ($\ell
,m;n,\overline{m}$) that satisfy the relations 
\begin{equation}
\begin{array}{l}
d\ell =Z_{1}\wedge \ell +i\Phi _{1}m\wedge \overline{m} \\ 
dn=Z_{2}\wedge n+i\Phi _{2}m\wedge \overline{m} \\ 
dm=Z_{3}\wedge m+\Phi _{3}\ell \wedge n \\ 
\\ 
\ell \wedge m\wedge n\wedge \overline{m}\neq 0%
\end{array}
\label{i-6}
\end{equation}%
where the vector fields $Z_{1\mu }\ ,\ Z_{2\mu }\ $ are real, the vector
field$\ Z_{3\mu }$ is complex, the scalar fields (called relative
invariants) $\Phi _{1}\ ,\ \Phi _{2}$ are real and the scalar field$\ \Phi
_{3}$ is complex. One can easily check that these conditions are equivalent
to the metric independent form of the geodetic and shear-free conditions\cite%
{CHAND} 
\begin{equation}
\begin{array}{l}
(\ell ^{\mu }m^{\nu }-\ell ^{\nu }m^{\mu })(\partial _{\mu }\ell _{\nu
})=0\quad ,\quad (\ell ^{\mu }m^{\nu }-\ell ^{\nu }m^{\mu })(\partial _{\mu
}m_{\nu })=0 \\ 
\\ 
(n^{\mu }\overline{m}^{\nu }-n^{\nu }\overline{m}^{\mu })(\partial _{\mu
}n_{\nu })=0\quad ,\quad (n^{\mu }\overline{m}^{\nu }-n^{\nu }\overline{m}%
^{\mu })(\partial _{\mu }\overline{m}_{\nu })=0 \\ 
\end{array}
\label{i-7}
\end{equation}%
on the tetrad ($\ell ,n,m,\overline{m}$). Notice that these conditions do
not depend on a metric. It is a property of a basis of the tangent (and
cotangent) space of a manifold.

The integrability conditions (\ref{i-6}) of the LCR-structure are invariant
under the transformations%
\begin{equation}
\begin{tabular}{l}
$\ell _{\mu }^{\prime }=\Lambda \ell _{\mu }\quad ,\quad n_{\mu }^{\prime
}=Nn_{\mu }\quad ,\quad m_{\mu }^{\prime }=Mm_{\mu }$ \\ 
\\ 
$Z_{1}^{\prime }=Z_{1}+d\ln \Lambda \quad ,\quad Z_{2}^{\prime }=Z_{2}+d\ln
N\quad ,\quad Z_{3}^{\prime }=Z_{3}+d\ln M$ \\ 
$\Phi _{1}^{\prime }=\frac{\Lambda }{M\overline{M}}\Phi _{1}\quad ,\quad
\Phi _{2}^{\prime }=\frac{N}{M\overline{M}}\Phi _{2}\quad ,\quad \Phi
_{3}^{\prime }=\frac{M}{\Lambda N}\Phi _{2}$%
\end{tabular}
\label{i-7a}
\end{equation}%
which I will call tetrad-Weyl transformations. The fact that the tetrad-Weyl
parameters $\Lambda ,N,M$ must not vanish, implies that the tetrad-Weyl
transformation cannot annihilate the relative invariants. If they do not
vanish, they may be fixed to take a constant numerical value, which is the
reason of the used term "relative invariant". Notice that if all the
relative invariants do not vanish, they may fix the tetrad-Weyl
transformation.

In brief, the fundamental quantity of PCFT is not the metric (like general
relativity) but the LCR-structure (like the Polyakov action). Starting from
a LCR-structure i.e. a LCR-tetrad (\ref{i-6}), we cannot define a unique
symmetric tensor (an Einstein metric) 
\begin{equation}
\begin{array}{l}
g_{\mu \nu }=\ell _{\mu }n_{\nu }+\ell _{\nu }n_{\mu }-m_{\mu }\overline{m}%
_{\nu }-m_{\nu }\overline{m}_{\mu } \\ 
\end{array}
\label{i-7c}
\end{equation}%
Because of the tetrad-Weyl symmetry, we can only define a class of metrics $%
[g_{\mu \nu }]$, with equivalence relation the tetrad-Weyl transformations (%
\ref{i-7a}). Notice that for $\Lambda N=M\overline{M}$, the tetrad-Weyl
transformation becomes the ordinary Weyl transformation. I will explicitly
show how the charge conservation breaks the tetrad-Weyl symmetry down to the
ordinary Weyl symmetry and the energy-momentum conservation breaks it
farther.

The corresponding integrability conditions and transformations for the
2-dimensional LCR-structure are 
\begin{equation}
\begin{array}{l}
d\ell =Z_{1}\wedge \ell \quad ,\quad dn=Z_{2}\wedge n \\ 
\\ 
\ell _{\mu }^{\prime }=\Lambda \ell _{\mu }\quad ,\quad n_{\mu }^{\prime
}=Nn_{\mu }%
\end{array}
\label{i-8}
\end{equation}%
which are satisfied for all the 2-dimensional independent 1-forms ($\ell ,n$%
). But there is an essential difference. In two dimensions the LCR-structure
is always degenerate and the transformation coincides with the ordinary Weyl
transformation, while in four dimensions the relative invariants $\Phi _{j}$
make the LCR-structure non-degenerate and they generate gravity,
electromagnetism and all the leptonic sector.

I want to point out that the Wick rotation in four dimensions "destroys" the
LCR-structure, because simply the Minkowski metric spacetime does not admit
a (real tensor) hermitian structure\cite{FLAHE1974}$^{,}$\cite{FLAHE1976}.
In fact even in two dimensions, we do not need the Wick rotation to show the
dependence of the Polyakov action on the algebraic curves of $CP^{2}$. The
2-dimensional LCR-manifold may be viewed as the product of two real
submanifolds of a complex manifold (identified with an algebraic curve).

In section II, I will describe the fundamental properties of the
LCR-structure, which will permit us to write down the generally covariant
form of the action of PCFT in section III. The holomorphic Frobenius theorem
conducts to two intersections of the lines with the hypersurfaces of $CP^{3}$%
, the 4-dimensional real submanifolds (spacetimes) of the grassmannian
manifold $G_{4,2}$ and finally the Poincar\'{e} group, which will be
identified with the (observed) conserved group in nature. This
identification will permit us to unfold and describe the vacuum, the static
LCR-manifold, which is identified with the electron and the stationary
LCR-structure, which is identified with the neutrino. In section IV Gravity
is derived\cite{RAG1999}. Electromagnetism is derived in section V, and the
same Bogoliubov method is extended to the standard model derivation\cite%
{RAG2018a} in section VI, where the three particle generations are implied
as a restriction of LCR-structure integrability conditions. In section VII,
stable static solitonic solutions of the gauge field equations are found,
which have distributional sources, identified with the quarks\cite{RAG2018b}%
. A quark-antiquark system, described in section VIII, could be the origin
of quark confinement.

The reader will see that the fundamental mathematical framework of PCFT\cite%
{RAG2017} is essentially analogous to that of string theory. We simply pass
from the algebraic curves to the algebraic surfaces. I will stick to this
analogy as long as it is permitted, in order to facilitate string theory
researchers to understand PCFT.

\section{LORENTZIAN CR-STRUCTURE}

\setcounter{equation}{0}

The 4-dimensional pseudo-conformal field theory (PCFT) is based on the
LCR-structure of the spacetime, like the Einstein relativity is based on the
metric structure of the spacetime. Recall that after the success of general
relativity, Einstein tried (and failed) to extend its fundamental structure
from the lorentzian metric first to a 5-dimensional Kaluza-Klein theory and
after to a metric with torsion. PCFT is an attempt in this direction
providing all the leptonic interactions.

The LCR-structure definition (\ref{i-6}) is an integrability condition that
permits the application of Frobenius theorem. But here there is a subtlety
that is essential in four dimensions. The existence of the complex tangent
1-form $m_{\mu }dx^{\mu }$ makes necessary\cite{BAOU} first to complexify
spacetime and after apply the holomorphic Frobenius theorem.

The complexification locally makes the spacetime a real surface of $%
%TCIMACRO{\U{2102} }%
%BeginExpansion
\mathbb{C}
%EndExpansion
^{4}$ and we have to be restricted to real analytic functions. But this
real-analyticity is not necessary to be on the entire spacetime. It must be
valid on a large connected region of spacetime so that the two (imaginary)
sides of spacetime communicate through the analytic continuation. Hence
there may exist isolated local compact regions, which will be the singular
regions of the considered generalized functions\cite{GELF1}. That is, this
complexification permits the consideration of generalized functions and
especially in the picture of the Sato's hyperfunctions\cite{GRAF}$^{,}$\cite%
{MOR}. The found solitonic configurations are generalized functions
(Schwartz distributions). The distributional sources of these generalized
functions will be identified with the leptons and the quarks and the regular
part of the distributional configurations will be identified with the
gravitational, electromagnetic and gluonic "dressings" of the particles.
This will become clear in section VI, where a microlocal analysis of the
electron-configuration will be presented.

The application of the holomorphic Frobenius theorem implies the existence
of four complex functions $(z^{\alpha },\;z^{\widetilde{\alpha }})$,\ $%
\alpha =0,\ 1$ , such that

\begin{equation}
\begin{array}{l}
dz^{\alpha }=f_{\alpha }\ \ell _{\mu }dx^{\mu }+h_{\alpha }\ m_{\mu }dx^{\mu
}\;\;\;\;,\;\;\;dz^{\widetilde{\alpha }}=f_{\widetilde{\alpha }}\ n_{\mu
}dx^{\mu }+h_{\widetilde{\alpha }}\ \overline{m}_{\mu }dx^{\mu } \\ 
\\ 
\ell _{\mu }dx^{\mu }=\ell _{\alpha }dz^{\alpha }\;\;\;\;,\;\;\;m_{\mu
}dx^{\mu }=m_{\alpha }dz^{\alpha } \\ 
n_{\mu }dx^{\mu }=n_{\widetilde{\alpha }}dz^{\widetilde{\alpha }%
}\;\;\;\;,\;\;\;\overline{m}_{\mu }dx^{\mu }=\overline{m}_{\widetilde{\alpha 
}}dz^{\widetilde{\alpha }} \\ 
\end{array}
\label{l-1}
\end{equation}%
This LCR-structure\cite{RAG2013b} is called realizable or embedable and the
complex functions are called LCR-structure coordinates. Notice that the
corresponding result for the 2-dimensional LCR-structure is the existence of
two structure coordinates $(z^{0},\;z^{\widetilde{0}})$, such that

\begin{equation}
\begin{array}{l}
dz^{0}=f_{0}\ \ell _{\mu }dx^{\mu }\;\;\;\;,\;\;\;dz^{\widetilde{\alpha }%
}=f_{\widetilde{0}}\ n_{\mu }dx^{\mu } \\ 
\\ 
\ell =\ell _{\alpha }dz^{\alpha }\;\;\;\;,\;\;\;n=n_{\widetilde{\alpha }}dz^{%
\widetilde{\alpha }} \\ 
\end{array}
\label{l-2}
\end{equation}%
We will assume them generally complex in order to keep the analogy between
the 2-dimensional and the 4-dimentional LCR-structures.

The tangent 1-forms $\ell _{\mu }dx^{\mu }$ and $n_{\mu }dx^{\mu }$ are
real, and the 1-forms $m_{\mu }dx^{\mu }$ and $\overline{m}_{\mu }dx^{\mu }$
are complex conjugate. These (reality) relations imply

\begin{equation}
\begin{array}{l}
dz^{0}\wedge dz^{1}\wedge d\overline{z^{0}}\wedge d\overline{z^{1}}=0 \\ 
dz^{\widetilde{0}}\wedge dz^{\widetilde{1}}\wedge d\overline{z^{0}}\wedge d%
\overline{z^{1}}=0 \\ 
dz^{\widetilde{0}}\wedge dz^{\widetilde{1}}\wedge d\overline{z^{\widetilde{0}%
}}\wedge d\overline{z^{\widetilde{1}}}=0 \\ 
\\ 
dz^{0}\wedge dz^{1}\wedge dz^{\widetilde{0}}\wedge dz^{\widetilde{1}}\neq 0%
\end{array}
\label{l-3}
\end{equation}%
where the last one is implied by the linear independence of the LCR-tetrad.
It is convenient to write the first three conditions as

\begin{equation}
\begin{array}{l}
\rho _{11}(\overline{z^{\alpha }},z^{\alpha })=0\quad ,\quad \rho
_{12}\left( \overline{z^{\alpha }},z^{\widetilde{\alpha }}\right) =0\quad
,\quad \rho _{22}\left( \overline{z^{\widetilde{\alpha }}},z^{\widetilde{%
\alpha }}\right) =0 \\ 
\\ 
\frac{\partial \rho _{ij}}{\partial z^{b}}\neq 0\neq \frac{\partial \rho
_{ij}}{\partial \overline{z^{b}}} \\ 
\end{array}
\label{l-4}
\end{equation}%
where the two functions $\rho _{11}$ , $\rho _{22}$ are real and $\rho _{12}$
is a complex function (i.e. two real functions). Notice the particular
dependence of these functions on the structure coordinates. The two real
conditions determine two ordinary hypersurface type CR-structures, which are
connected through the complex condition. The LCR-structure is essentially a
special totally real CR-structure\cite{BAOU}. But unlike the ordinary
totally real CR-structures, which are invariant under a general holomorphic
transformation $z^{\prime b}=f^{b}(z^{c})$, the LCR-structure is invariant
(and considered to be equivalent) under the special holomorphic
transformations

\begin{equation}
\begin{array}{l}
z^{\prime \beta }=f^{\beta }(z^{\alpha })\quad ,\quad z^{\prime \widetilde{%
\beta }}=f^{\widetilde{\beta }}(z^{\widetilde{\alpha }}) \\ 
\end{array}
\label{l-5}
\end{equation}%
where the transformations of the tilded and untilded structure coordinates
are independent.

In the case of the 2-dimensional LCR-structure (on which the Polyakov action
is based) the corresponding defining functions and the LCR-transformations
are

\begin{equation}
\begin{array}{l}
\rho _{1}(\overline{z^{0}},z^{0})=0\quad ,\quad \rho _{2}(\overline{z^{%
\widetilde{0}}},z^{\widetilde{0}})=0 \\ 
\frac{\partial \rho _{i}}{\partial z^{b}}\neq 0\neq \frac{\partial \rho _{i}%
}{\partial \overline{z^{b}}} \\ 
\\ 
z^{\prime 0}=f^{0}(z^{0})\quad ,\quad z^{\prime \widetilde{0}}=f^{\widetilde{%
0}}(z^{\widetilde{0}})%
\end{array}
\label{l-6}
\end{equation}

I want to point out that I have not yet introduced any riemannian metric.
The CR-structure does not need the metric structure. The Einstein metric
will be defined in section IV, where we will study its limitations and
really amazing consequences. The vector and their dual 1-form tetrads are
related by the following inversion relations

\begin{equation}
\begin{array}{l}
e_{\mu }^{0}dx^{\mu }\equiv \ell _{\mu }dx^{\mu }\;,\;e_{\mu }^{1}dx^{\mu
}\equiv m_{\mu }dx^{\mu }\;,\;e_{\mu }^{\widetilde{0}}dx^{\mu }\equiv n_{\mu
}dx^{\mu }\;,\;e_{\mu }^{\widetilde{1}}dx^{\mu }\equiv \overline{m}_{\mu
}dx^{\mu } \\ 
e_{0}^{\mu }\partial _{\mu }\equiv n^{\mu }\partial _{\mu }\;,\;e_{1}^{\mu
}\partial _{\mu }\equiv -\overline{m}^{\mu }\partial _{\mu }\;,\;e_{%
\widetilde{0}}^{\mu }\partial _{\mu }\equiv \ell ^{\mu }\partial _{\mu
}\;,\;e_{\widetilde{1}}^{\mu }\partial _{\mu }\equiv -m^{\mu }\partial _{\mu
} \\ 
\\ 
e_{a}^{\mu }e_{\mu }^{b}=\delta _{a}^{b}\;\;,\;\;e_{a}^{\mu }e_{\nu
}^{a}=\delta _{\nu }^{\mu } \\ 
\end{array}
\label{l-6a}
\end{equation}%
No lowering and raising index mechanism has been defined yet, because we
have not defined the metric.

In order to clarify the relative essential differences between the
diffeomorphic CR-transformations and the LCR-transformations, I will recall
the historical discovery of 2-dimensional CR-structure by Poincar\'{e}. I
think it is well known that any real submanifold $\rho (x^{\mu })=0$ can
take the one coordinate form $y=0$ after a diffeomorphic transformation
(using the implicit function theorem). A real submanifold (curve) $\rho (%
\overline{z},z)=0$ of the complex plane $%
%TCIMACRO{\U{2102} }%
%BeginExpansion
\mathbb{C}
%EndExpansion
$ can take the real axis form $z-\overline{z}=0$ after a holomorphic
transformation. But Poincar\'{e} showed\cite{BAOU} that this is not possible
for real subsurfaces of $%
%TCIMACRO{\U{2102} }%
%BeginExpansion
\mathbb{C}
%EndExpansion
^{2}$. In higher dimensional complex manifolds the \textbf{holomorphic}
transformations \textbf{cannot} transform a real surface\ to any other real
surface of the same dimension.

In the case of the LCR-structure transformations we have an analogous
restriction. The 2-dimensional LCR-transformations can give the (real
analytic at the neighborhood of a point) defining functions (\ref{l-6}) the
simple (trivial) form

\begin{equation}
\begin{array}{l}
z^{0}-\overline{z^{0}}=0\quad ,\quad z^{\widetilde{0}}-\overline{z^{%
\widetilde{0}}}=0 \\ 
\end{array}
\label{l-7}
\end{equation}%
But in four dimensions there is a restriction. A LCR-transformation can
simplify a real analytic structure (\ref{l-4}) to the non-trivial form 
\begin{equation}
\begin{array}{l}
\func{Im}z^{0}=\phi _{11}(\overline{z^{1}},z^{1},\func{Re}z^{0})\quad ,\quad 
\func{Im}z^{\widetilde{0}}=\phi _{22}(\overline{z^{\widetilde{1}}},z^{%
\widetilde{1}},\func{Re}z^{\widetilde{0}}) \\ 
\\ 
z^{\widetilde{1}}-\overline{z^{1}}=\phi _{12}(\overline{z^{a}},z^{\widetilde{%
0}}) \\ 
\\ 
\phi _{11}(0)=\phi _{22}(0)=\phi _{12}(0)=0\quad ,\quad d\phi _{11}(0)=d\phi
_{22}(0)=d\phi _{12}(0)=0%
\end{array}
\label{l-8}
\end{equation}%
and the corresponding coordinates are called regular LCR-coordinates. The
LCR-transformations cannot completely remove (annihilate) the real analytic
functions $\phi _{ij}$. But a general holomorphic transformation $z^{\prime
b}=f^{b}(z^{c})$ can remove these functions. That is, a general holomorphic
transformation makes a real analytic LCR-structure equivalent to the
degenerate totally real CR-structure\cite{BAOU}, which cannot be generally
done with a LCR-transformation.

The 2-dimensional LCR-structure has two disconnected structure coordinates ($%
z^{0},z^{\widetilde{0}}$), which in string theory are directly related to
the two chiral sectors. In the degenerate 4-dimensional LCR-structure the
tilded and untilded chiral regular coordinates are connected with the
relation $z^{\widetilde{1}}-\overline{z^{1}}=0$, which are the two chiral
representations of the Lorentz group. This indicates the pathway to reveal
the Poincar\'{e} group, which will be identified with the corresponding
observed symmetry group of nature. I postpone this derivation for section
IV, where the Einstein metric is defined and the flat geodetic and
shear-free null congruence conditions are solved through the Kerr theorem
and its homogeneous holomorphic function $K(Z^{m})$.

In order to understand the present work and applied mathematical techniques,
the reader must be aware of the two mathematical approaches to study the
geometry of a complex manifold. In the present case, it is the ambient
complex space of the LCR-manifold. The first is the algebraic approach,
through the possible embedding of the complex manifold into a projective
space (external approach). The second (internal) approach is the sheaf
cohomology. The interplay between these two approaches is regulated (for
compact complex manifolds) by the Kodaira theorem\cite{GRIF}. Below I will
describe the algebraic approach, which provides a clear-cut definition of
the Poincar\'{e} group, which is essential to the understanding of the
implied quantum field theory. After I will describe a "physical" view of the
internal approach, which is better suited for general relativists.

The defining relations (\ref{l-4}) of the quite general class of
LCR-manifolds\cite{RAG2013b} take the following form of real surfaces of the
grassmannian manifold $G_{4,2}$ 
\begin{equation}
\begin{array}{l}
\rho _{11}(\overline{X^{m1}},X^{n1})=0\quad ,\quad \rho _{22}(\overline{%
X^{m2}},X^{n2})=0\quad ,\quad \rho _{12}(\overline{X^{m1}},X^{n2})=0 \\ 
\\ 
K(X^{mj})=0 \\ 
\end{array}
\label{l-9}
\end{equation}%
where all the functions are homogeneous relative to the coordinates $X^{n1}$
and $X^{n2}$\ independently, which must be roots of the homogeneous
holomorphic Kerr polynomial $K(X^{ni})$. The charts of its typical
non-homogeneous (projective) coordinates are determined by the invertible
pairs of rows. If the first two rows constitute an invertible matrix, the
chart is determined by $\det \lambda ^{Aj}\neq 0$ and the corresponding
projective (affine space coordinates) $r_{A^{\prime }A}$ are defined by 
\begin{equation}
\begin{array}{l}
X=%
\begin{pmatrix}
X^{01} & X^{02} \\ 
X^{11} & X^{12} \\ 
X^{21} & X^{22} \\ 
X^{31} & X^{32}%
\end{pmatrix}%
\equiv \left( 
\begin{array}{c}
\lambda ^{Aj} \\ 
-ir_{A^{\prime }A}\lambda ^{Aj}%
\end{array}%
\right) \\ 
\\ 
r_{A^{\prime }A}=\eta _{ab}r^{a}\sigma _{A^{\prime }A}^{b} \\ 
\end{array}
\label{l-10}
\end{equation}%
The matrix $\eta _{ab}$ is the ordinary Minkowski metric and $\sigma
_{A^{\prime }A}^{b}$ are the identity and the three hermitian Pauli matrices%
\begin{equation}
\begin{array}{l}
\sigma _{A^{\prime }B}^{0}=\left( 
\begin{array}{cc}
1 & 0 \\ 
0 & 1%
\end{array}%
\right) \ ,\ \sigma _{A^{\prime }B}^{1}=\left( 
\begin{array}{cc}
0 & 1 \\ 
1 & 0%
\end{array}%
\right) \\ 
\\ 
\sigma _{A^{\prime }B}^{2}=\left( 
\begin{array}{cc}
0 & -i \\ 
i & 0%
\end{array}%
\right) \ ,\ \sigma _{A^{\prime }B}^{3}=\left( 
\begin{array}{cc}
1 & 0 \\ 
0 & -1%
\end{array}%
\right) \\ 
\end{array}
\label{l-11}
\end{equation}%
and the spinor indices are lowered and raised with the antisymmetric matrix%
\begin{equation}
\begin{array}{l}
\lambda ^{A}=\epsilon ^{AB}\lambda _{B}\quad ,\quad \lambda _{C}=\lambda
^{B}\epsilon _{BC} \\ 
\\ 
\lambda ^{A}\xi _{A}=\lambda ^{A}\xi ^{B}\epsilon _{BA}=-\lambda _{B}\xi
^{B}\quad ,\quad \lambda ^{A}\lambda _{A}=0 \\ 
\\ 
\epsilon ^{AB}=\epsilon _{AB}=%
\begin{pmatrix}
0 & 1 \\ 
-1 & 0%
\end{pmatrix}%
\quad ,\quad \epsilon _{A}^{B}=\epsilon ^{BC}\epsilon _{AC}=%
\begin{pmatrix}
1 & 0 \\ 
0 & 1%
\end{pmatrix}
\\ 
\end{array}
\label{l-12}
\end{equation}%
I point out that this notation is not exactly that used in the classical
book of Penrose and Rindler\cite{P-R}.

The grassmannian manifold $G_{4,2}$ is the projective space of the lines of $%
CP^{3}$. The homogeneous coordinates $X^{mi}$ matrix of $G_{4,2}$ are two
points $X^{mi}$ of the hypersurface of $CP^{3}$ determined by the
irreducible or reducible Kerr polynomial $K(Z^{n})$. From the above
LCR-structure conditions (\ref{l-9}) we see that the untilded structure
coordinates $z^{\alpha }$ determine the point $X^{m1}$ and the tilded
structure coordinates $z^{\widetilde{\alpha }}$ determine the point $X^{m2}$%
. I point out that the embedding of the ambient complex manifold of the
LCR-manifold into the grassmannian manifold is the essential step to relate
PCFT with the particle physics. Recall that particles are representations of
the Poincar\'{e} group, and in this review I simply explain my efforts to
find the stable LCR-manifolds which are irreducible representations of the
Poincar\'{e} algebra.

We must be careful with the passage to the "physical" Poincar\'{e} group
from the general projective $SL(4,%
%TCIMACRO{\U{2102} }%
%BeginExpansion
\mathbb{C}
%EndExpansion
)$ symmetry of the $G_{4,2}$ geometry. The general complex Poincar\'{e}
group is an affine subgroup of $SL(4,%
%TCIMACRO{\U{2102} }%
%BeginExpansion
\mathbb{C}
%EndExpansion
)$. It is directly related with an affine chart of $G_{4,2}$. In the present
work our "physical" Poincar\'{e} group will be that imposed by the condition 
$\det \lambda ^{Aj}\neq 0$.

The parameterization\cite{GRIF} of the algebraic manifolds is a very useful
tool to study algebraic surfaces. The Newman generally complex trajectory%
\cite{NEWM1973} is a physically intuitive parameterization, where the Kerr
holomorphic function $K(Z^{m})$ is replaced\cite{RAG2013b} by a trajectory $%
\xi ^{b}(\tau )$ and the following form of the homogeneous coordinates 
\begin{equation}
\begin{array}{l}
X=\left( 
\begin{array}{c}
\lambda ^{Aj} \\ 
-ir_{A^{\prime }A}\lambda ^{Aj}%
\end{array}%
\right) =%
\begin{pmatrix}
\lambda ^{A1} & \lambda ^{A2} \\ 
-i\xi _{A^{\prime }A}(\tau _{1})\lambda ^{A1} & -i\xi _{A^{\prime }A}(\tau
_{2})\lambda ^{A2}%
\end{pmatrix}
\\ 
\\ 
(r_{A^{\prime }A}-\xi _{A^{\prime }A}(\tau ))\lambda ^{A}=0\quad \rightarrow
\quad (r^{a}-\xi ^{a}(\tau ))(r^{b}-\xi ^{b}(\tau ))\eta _{ab}=0 \\ 
\end{array}
\label{l-12a}
\end{equation}%
The last condition assures the existence of a non-vanishing solution of $%
\lambda ^{Ai}$ and permits the computation of $\tau $ as a function of $%
r^{a} $. Notice that this procedure of one trajectory must provide at least
two solutions $\tau _{1}$ and $\tau _{2}$\ which are used to determine the
structure coordinates of the two columns. Apparently we may take two
independent complex trajectories. A general complex linear trajectory
corresponds to the following quadratic polynomial 
\begin{equation}
\begin{array}{l}
\xi ^{a}(\tau )=v^{a}\tau +c^{a}\quad ,\quad v^{a}v^{b}\eta _{ab}=1 \\ 
\\ 
(v_{1^{\prime }0}Z^{0}+v_{1^{\prime }1}Z^{1})(iZ^{2}-c_{0^{\prime
}0}Z^{0}+c_{0^{\prime }1}Z^{1})-(v_{0^{\prime }0}Z^{0}+v_{0^{\prime
}1}Z^{1})(iZ^{3}-c_{1^{\prime }0}Z^{0}+c_{1^{\prime }1}Z^{1})=0 \\ 
\end{array}
\label{l-12b}
\end{equation}%
In fact, the linear trajectory is the rational parameterization of this
precise (Poincar\'{e} invariant) parametrized quadric. In this case we
usually assume ($z^{0}=\tau _{1}\ ,\ z^{\widetilde{0}}=\tau _{2}$). The
Newman complex trajectory is mathematically implied by considering ruled
surfaces of $CP^{3}$. They are surfaces which contain a straight line 
\begin{equation}
\begin{array}{l}
Z^{m}(\tau ,s)=(1-s)Z_{1}^{m}(\tau )+sZ_{2}^{m}(\tau )= \\ 
\qquad =Z_{1}^{m}(\tau )+sT^{m}(\tau ) \\ 
\\ 
T^{m}(\tau ):=Z_{2}^{m}(\tau )-Z_{1}^{m}(\tau )%
\end{array}
\label{l-12c}
\end{equation}%
where $Z_{1}^{m}(\tau )$ is a curve of $CP^{3}$ and $T^{m}(\tau )$ indicates
the direction\ of the generating line which meets $Z_{1}^{m}(\tau )$ (the
generatrix) at $\tau $. This line is a point of $G_{4,2}$, determined by 
\begin{equation}
\begin{array}{l}
\xi (\tau )=:iX_{2}X_{1}^{-1}=:%
\begin{pmatrix}
\xi ^{0}-\xi ^{3} & -(\xi ^{1}-i\xi ^{2}) \\ 
-(\xi ^{1}+i\xi ^{2}) & \xi ^{0}+\xi ^{3}%
\end{pmatrix}
\\ 
\\ 
X_{1}=:%
\begin{pmatrix}
Z_{1}^{0} & Z_{2}^{0} \\ 
Z_{1}^{1} & Z_{2}^{1}%
\end{pmatrix}%
\quad ,\quad X_{2}=:%
\begin{pmatrix}
Z_{1}^{2} & Z_{2}^{2} \\ 
Z_{1}^{3} & Z_{2}^{3}%
\end{pmatrix}%
\end{array}
\label{l-12d}
\end{equation}%
The curve is called non-degenerate\cite{GrHa2} if the following determinant
does not identically vanish 
\begin{equation}
\begin{array}{l}
\det [Z_{1}^{n},Z_{2}^{n},\frac{dZ_{1}^{n}}{d\tau },\frac{dZ_{2}^{n}}{d\tau }%
]=\det 
\begin{pmatrix}
X_{1} & \overset{.}{X_{1}} \\ 
-i\xi X_{1} & -i(\overset{.}{\xi }X_{1}+\xi \overset{.}{X_{1}})%
\end{pmatrix}%
= \\ 
=\det [%
\begin{pmatrix}
1 & 0 \\ 
-i\xi & 1%
\end{pmatrix}%
\begin{pmatrix}
X_{1} & \overset{.}{X_{1}} \\ 
0 & -i\overset{.}{\xi }X_{1}%
\end{pmatrix}%
]=-\det (\overset{.}{\xi })(\det X_{1})^{2}%
\end{array}
\label{l-12e}
\end{equation}%
This happens if and only if $\overset{.}{\xi }^{a}\overset{.}{\xi }^{b}\eta
_{ab}\neq 0$. This condition will differentiate the massive from the
massless partner (neutrino) of a leptonic generation. The complex trajectory
is related to the ordinary classical trajectory of the particle viewed as a
soliton. If they are real, they are identified with the well known
trajectories of the Lienard-Wiechert potential. The degenerate trajectory
occurs for $T^{m}(\tau )=$ $\frac{dZ_{1}^{n}}{d\tau }$, which are called
developable surfaces. Hence a non-degenerate rulled surface \ ($\overset{.}{%
\xi }^{a}\overset{.}{\xi }^{b}\eta _{ab}\neq 0$) corresponds to the massive
particle (say electron) and the developable surface ($\overset{.}{\xi }^{a}%
\overset{.}{\xi }^{b}\eta _{ab}=0$) corresponds to the massless particle
(neutrini) of the leptonic generation (family).

Let us now turn into the internal "physical" approach, noticing that the
LCR-tetrad defines and is defined by the following classes of symmetric and
antisymmetric tensors 
\begin{equation}
\begin{array}{l}
\lbrack g_{\mu \nu }]=\ell _{\mu }n_{\nu }+\ell _{\nu }n_{\mu }-m_{\mu }%
\overline{m}_{\nu }-m_{\nu }\overline{m}_{\mu } \\ 
\lbrack J_{\mu \nu }]=\ell _{\mu }n_{\nu }-\ell _{\nu }n_{\mu }-m_{\mu }%
\overline{m}_{\nu }+m_{\nu }\overline{m}_{\mu } \\ 
\\ 
\det (g_{\mu \nu })\neq 0\neq \det (J_{\mu \nu })%
\end{array}
\label{l-13}
\end{equation}%
The class is defined relative to the regular tetrad-Weyl transformations (%
\ref{i-7a}), which have non-vanishing factors for every coordinate patch,
with the appropriate fitting relations in the intersections of the patches
(sheaf requirement). The equivalent properties are that the metric $g_{\mu
\nu }$ admits 2-geodetic and shear-free congruences ($\ell ^{\mu }$ and $%
n^{\mu }$) and that $J_{\ \nu }^{\mu }$ satisfies\cite{FLAHE1974}$^{,}$\cite%
{FLAHE1976} the equivalent Nijenhuis condition. That is Einstein's gravity
with 2-geodetic and shear-free congruences determines back the LCR-structure.

A typical example of LCR-structure is%
\begin{equation}
\begin{array}{l}
\ell _{\mu }dx^{\mu }=dt-\frac{\rho ^{2}}{\Delta }dr-a\sin ^{2}\theta
d\varphi \\ 
n_{\mu }dx^{\mu }=\frac{\Delta }{2\rho ^{2}}(dt+\frac{\rho ^{2}}{\Delta }%
dr-a\sin ^{2}\theta d\varphi ) \\ 
m_{\mu }dx^{\mu }=\frac{1}{\eta \sqrt{2}}(ia\sin \theta dt-\rho ^{2}d\theta
-i(r^{2}+a^{2})\sin \theta d\varphi ) \\ 
\\ 
\eta \equiv r+ia\cos \theta \quad ,\quad \rho ^{2}\equiv \eta \overline{\eta 
}\quad ,\quad \sqrt{-g}=\rho ^{2}\sin \theta \\ 
\Delta \equiv r^{2}-2Mr+a^{2}+q^{2}%
\end{array}
\label{l-13a}
\end{equation}%
which corresponds to the static trajectory $\xi ^{a}=(\tau ,0,0,ia)$.
Because of the tetrad-Weyl symmetry, the tetrad does not need multiplicative
factors in order to fix an LCR-structure. The precise above form is the
geodetic and shear-free null tetrad of the Kerr-Newman spacetime\cite{CHAND}%
. Its contravarient components are 
\begin{equation}
\begin{array}{l}
\ell ^{\mu }\partial _{\mu }=\frac{1}{\Delta }((r^{2}+a^{2})\partial
_{t}+\Delta \partial _{r}+a\partial _{\varphi }) \\ 
n^{\mu }\partial _{\mu }=\frac{1}{2\rho ^{2}}((r^{2}+a^{2})\partial
_{t}-\Delta \partial _{r}+a\partial _{\varphi }) \\ 
m^{\mu }\partial _{\mu }=\frac{1}{\eta \sqrt{2}}(ia\sin \theta \partial
_{t}+\partial _{\theta }+\frac{i}{\sin \theta }\partial _{\varphi }) \\ 
\end{array}
\label{l-14}
\end{equation}%
and its Newman-Penrose (NP) spin coefficients are%
\begin{equation}
\begin{array}{l}
\varepsilon =0\quad ,\quad \beta =\frac{\cos \theta }{\sin \theta \eta 2%
\sqrt{2}}\quad ,\quad \pi =\frac{ia\sin \theta }{(\overline{\eta })^{2}\sqrt{%
2}} \\ 
\tau =-\frac{ia\sin \theta }{\rho ^{2}\sqrt{2}}\quad ,\quad \rho =-\frac{1}{%
\overline{\eta }}\quad ,\quad \mu =-\frac{\Delta }{2\rho ^{2}\overline{\eta }%
} \\ 
\gamma =-\frac{\Delta }{2\rho ^{2}\overline{\eta }}+\frac{r-M}{2\rho ^{2}}%
\quad ,\quad \alpha =\pi -\overline{\beta }=\frac{ia\sin \theta }{(\overline{%
\eta })^{2}\sqrt{2}}-\frac{\cos \theta }{\sin \theta \overline{\eta }2\sqrt{2%
}}%
\end{array}
\label{l-15}
\end{equation}%
The reader should not confuse the symbol $\rho ^{2}\equiv \eta \overline{%
\eta }$ with the spin-coefficient $\rho $. The tetrad-Weyl gauge fields
1-forms $Z_{j\mu }dx^{\mu }$ and the relative invariants $\Phi _{i}$ are
found using the standard relations\ 
\begin{equation}
\begin{array}{l}
d\ell =[(\varepsilon +\overline{\varepsilon })n-(\alpha +\overline{\beta }-%
\overline{\tau })m-(\overline{\alpha }+\beta -\tau )\overline{m}]\wedge \ell
+(\rho -\overline{\rho })m\wedge \overline{m} \\ 
dn=[-(\gamma +\overline{\gamma })\ell +(\alpha +\overline{\beta }-\pi )m+(%
\overline{\alpha }+\beta -\overline{\pi })\overline{m}]\wedge n+(\mu -%
\overline{\mu })m\wedge \overline{m} \\ 
dm=[(\gamma -\overline{\gamma }+\overline{\mu })\ell +(\varepsilon -%
\overline{\varepsilon }-\rho )n-(\beta -\overline{\alpha })\overline{m}%
]\wedge m-(\tau +\overline{\pi })\ell \wedge n \\ 
\end{array}
\label{l-16}
\end{equation}%
where I have assumed that the tetrad is geodetic and shear-free $\kappa
=\sigma =0=\lambda =\nu $. Notice that this LCR-structure has non-vanishing
relative invariants\ 
\begin{equation}
\begin{array}{l}
\Phi _{1}=\frac{\rho -\overline{\rho }}{i}=-\frac{2a\cos \theta }{\eta 
\overline{\eta }} \\ 
\Phi _{2}=\frac{\mu -\overline{\mu }}{i}=-\frac{\Delta a\cos \theta }{(\eta 
\overline{\eta })^{2}} \\ 
\Phi _{3}=-(\tau +\overline{\pi })=\frac{2ar\cos \theta }{\eta \overline{%
\eta }^{2}} \\ 
\end{array}
\label{l-17}
\end{equation}

The structure coordinates are%
\begin{equation}
\begin{array}{l}
z^{0}=t-f_{0}(r)+ia\cos \theta -ia\quad ,\quad z^{1}=e^{i\varphi
}e^{-iaf_{1}(r)}\tan \frac{\theta }{2} \\ 
z^{\widetilde{0}}=t+f_{0}(r)-ia\cos \theta +ia\quad ,\quad z^{\widetilde{1}%
}=e^{-i\varphi }e^{-iaf_{1}(r)}\tan \frac{\theta }{2} \\ 
\\ 
f_{0}(r)=\tint \frac{r^{2}+a^{2}}{\Delta }dr\quad ,\quad f_{1}(r)=\tint 
\frac{1}{\Delta }dr%
\end{array}
\label{l-18}
\end{equation}%
After straightforward calculations I find the following relations between
the structure coordinates and the LCR-tetrad%
\begin{equation}
\begin{array}{l}
dz^{0}=dt-\frac{r^{2}+a^{2}}{\Delta }dr-ia\sin \theta d\theta =\frac{%
r^{2}+a^{2}}{\eta \overline{\eta }}\ell _{\mu }dx^{\mu }+\frac{ia\sqrt{2}%
\sin \theta }{\overline{\eta }}m_{\mu }dx^{\mu } \\ 
d\ln z^{1}=\frac{-ia}{\Delta }dr+\frac{1}{\sin \theta }d\theta +id\varphi =%
\frac{ia}{\eta \overline{\eta }}\ell _{\mu }dx^{\mu }-\frac{\sqrt{2}}{%
\overline{\eta }\sin \theta }m_{\mu }dx^{\mu } \\ 
\\ 
\ell _{\mu }dx^{\mu }=dz^{0}+ia\sin ^{2}\theta d\ln z^{1} \\ 
m_{\mu }dx^{\mu }=\frac{ia\sin \theta }{\eta \sqrt{2}}dz^{0}-\frac{%
(r^{2}+a^{2})\sin \theta }{\sqrt{2}\eta }d\ln z^{1}%
\end{array}
\label{l-19}
\end{equation}%
and%
\begin{equation}
\begin{array}{l}
dz^{\widetilde{0}}=dt+\frac{r^{2}+a^{2}}{\Delta }dr+ia\sin \theta d\theta =%
\frac{2(r^{2}+a^{2})}{\Delta }n_{\mu }dx^{\mu }-\frac{ia\sqrt{2}\sin \theta 
}{\eta }\overline{m}_{\mu }dx^{\mu } \\ 
d\ln z^{\widetilde{1}}=\frac{-ia}{\Delta }dr+\frac{1}{\sin \theta }d\theta
-id\varphi =-\frac{2ia}{\Delta }n_{\mu }dx^{\mu }-\frac{\sqrt{2}}{\eta \sin
\theta }\overline{m}_{\mu }dx^{\mu } \\ 
\\ 
n_{\mu }dx^{\mu }=\frac{\Delta }{2\rho ^{2}}dz^{\widetilde{0}}-\frac{%
ia\Delta \sin ^{2}\theta }{2\rho ^{2}}d\ln z^{\widetilde{1}} \\ 
\overline{m}_{\mu }dx^{\mu }=-\frac{ia\sin \theta }{\overline{\eta }\sqrt{2}}%
dz^{\widetilde{0}}-\frac{(r^{2}+a^{2})\sin \theta }{\sqrt{2}\overline{\eta }}%
d\ln z^{\widetilde{1}}%
\end{array}
\label{l-20}
\end{equation}

This static LCR-manifold is a stable soliton. In the context of PCFT the
term soliton should not be confused with that in QFT without gravity. The
fact that gravity is contained in PCFT, the energy-momentum and angular
momentum of the configuration is derived from the source integrals of
linearized Einstein general relativity. On the other hand, besides the
topological invariants, the LCR-manifolds have the relative invariants,
which take discrete values and act as stabilizers. We have already found
that all the relative invariants of the present LCR-structure do not vanish,
which is not the case of the neutrino LCR-manifold, as I will show in
section VI.

I want to point out that the gravitation of the particle is a generalized
function (distribution). The singular support of the gravitational and
electromagnetic fields are the "locations" of the electron, and the regular
supports are the gravitational and electromagnetic "dressings" of the
electron.

In order to stress the physical significance of the LCR-structure, let me
mention that it is exactly this common property, that implies the observed
correspondence between the leptonic and hadronic sectors. That is, the up
and down quarks have the same LCR-structures with the neutrino and the
electron, with additional solitonic solutions (with distributional sources)
of the non-abelian gauge field. This will be extensively described in
section VII.

\section{THE\ COVARIANT ACTION\ OF\ PCFT}

\setcounter{equation}{0}

The integral (\ref{i-4}) is complex and not generally covariant. It is
written in the LCR-structure (chiral) coordinates (where the metric
independence appears) in order to clarify how the metric independence of the
Polyakov action triggered the search, discovery and study of the dynamical
content of the 4-dimentional PCFT.

The fact that the structure coordinates are generally complex implies that
the original metric independent form (\ref{i-4}) is complex, while the final
action must be real. In order to make things clear, I will start from the
LCR compatible gauge connection and its curvature 
\begin{equation}
\begin{array}{l}
(D_{\alpha })_{ij}=\partial _{\alpha }\delta _{ij}-\gamma f_{ikj}A_{k\alpha
}\quad ,\quad (D_{\widetilde{\beta }})_{ij}=\partial _{\widetilde{\beta }%
}\delta _{ij}-\gamma f_{ikj}A_{k\widetilde{\beta }} \\ 
F_{i\alpha \beta }=\partial _{\alpha }A_{i\beta }-\partial _{\beta
}A_{i\alpha }-\gamma f_{ikj}A_{j\alpha }A_{k\beta }\quad ,\quad F_{i%
\widetilde{\alpha }\widetilde{\beta }}=\partial _{\widetilde{\alpha }}A_{i%
\widetilde{\beta }}-\partial _{\widetilde{\beta }}A_{i\widetilde{\alpha }%
}-\gamma f_{ikj}A_{j\widetilde{\alpha }}A_{k\widetilde{\beta }} \\ 
\end{array}
\label{p-0}
\end{equation}%
in structure coordinates. The gauge invariant and metric independent 4-form
is 
\begin{equation}
\begin{array}{l}
F\wedge \widetilde{F}=(\frac{1}{2}F_{i\alpha \beta }dz^{\alpha }\wedge
dz^{\beta })\wedge (\frac{1}{2}F_{i\widetilde{\alpha }\widetilde{\beta }}dz^{%
\widetilde{\alpha }}\wedge dz^{\widetilde{\beta }})=F_{i01}F_{i\widetilde{0}%
\widetilde{1}}dz^{0}\wedge dz^{1}\wedge dz^{\widetilde{0}}\wedge dz^{%
\widetilde{1}} \\ 
\end{array}
\label{p-0a}
\end{equation}%
Using the identity 
\begin{equation}
\begin{array}{l}
\delta _{\nu }^{\mu }=\ell ^{\mu }n_{\nu }+n^{\mu }\ell _{\nu }-m^{\mu }%
\widetilde{m}_{\nu }-\widetilde{m}^{\mu }m_{\nu } \\ 
\\ 
\delta _{\beta }^{\alpha }=n^{\alpha }\ell _{\beta }-\widetilde{m}^{\alpha
}m_{\beta }\quad ,\quad \delta _{\widetilde{\beta }}^{\widetilde{\alpha }%
}=\ell ^{\widetilde{\alpha }}n_{\widetilde{\beta }}-m^{\widetilde{\alpha }}%
\widetilde{m}_{\widetilde{}} \\ 
\end{array}
\label{p-0b}
\end{equation}%
in structure coordinates, the complexified 4-form becomes 
\begin{equation}
\begin{array}{l}
F\wedge \widetilde{F}=\ell \wedge m\wedge n\wedge \widetilde{m}(\ell ^{\mu
}m^{\nu }F_{i\mu \nu })(n^{\rho }\widetilde{m}^{\sigma }F_{i\rho \sigma })
\\ 
\end{array}
\label{p-0c}
\end{equation}

When we return back to the real spacetime, it becomes the complex 4-form 
\begin{equation}
\begin{array}{l}
(F\wedge \widetilde{F})|_{S}=\ell \wedge m\wedge n\wedge \overline{m}(\ell
^{\mu }m^{\nu }F_{i\mu \nu })(n^{\rho }\overline{m}^{\sigma }F_{i\rho \sigma
}) \\ 
\ell \wedge m\wedge n\wedge \overline{m}=d^{4}\!x\sqrt{-g}i \\ 
\\ 
g=\det (g_{\mu \nu })=\det (\eta _{ab})[\det (e_{\mu }^{a})]^{2}=[\det
(e_{\mu }^{a})]^{2} \\ 
\eta _{ab}=%
\begin{pmatrix}
0 & 0 & 1 & 0 \\ 
0 & 0 & 0 & -1 \\ 
1 & 0 & 0 & 0 \\ 
0 & -1 & 0 & 0%
\end{pmatrix}%
\end{array}
\label{p-0d}
\end{equation}%
Hence we may assume as gauge field action either its real or its imaginary
part 
\begin{equation}
\begin{array}{l}
I_{R}=\tint d^{4}\!x\sqrt{-g}i\{(\ell ^{\mu }m^{\nu }F_{i\mu \nu })(n^{\rho }%
\overline{m}^{\sigma }F_{i\rho \sigma })-(\ell ^{\mu }\overline{m}^{\nu
}F_{i\mu \nu })(n^{\rho }m^{\sigma }F_{i\rho \sigma })\} \\ 
\\ 
I_{I}=\tint d^{4}\!x\sqrt{-g}\{(\ell ^{\mu }m^{\nu }F_{i\mu \nu })(n^{\rho }%
\overline{m}^{\sigma }F_{i\rho \sigma })+(\ell ^{\mu }\overline{m}^{\nu
}F_{i\mu \nu })(n^{\rho }m^{\sigma }F_{i\rho \sigma })\} \\ 
\\ 
F_{j\mu \nu }=\partial _{\mu }A_{j\nu }-\partial _{\nu }A_{j\mu }-\gamma
\,f_{jik}A_{i\mu }A_{k\nu }%
\end{array}
\label{p-1}
\end{equation}%
Both actions are apparently invariant under the tetrad-Weyl transformation.
Notice that only the null self-dual 2-forms appear in the actions. The
non-null self-dual component does not appear in the action, because simply
it is not multiplicatively transformed relative to the tetrad-Weyl
transformation.

In fact these two actions are strongly related. The appearing gauge tensors $%
F_{i\mu \nu }$ are each other duals, because $\ell ^{\lbrack \mu }m^{\nu ]}$
and $n^{[\rho }\overline{m}^{\sigma ]}$ are self-duals (relative to their
corresponding metric). One of these two actions will be the starting point
for the emergence of chromodynamics in the context of PCFT. In the hadronic
sector, we will see how the $I_{R}$\ action implies field equations, which
admit distributional solitons, which could be identified with the quarks\cite%
{RAG2018b}.

We saw that the existence of a globally defined LCR-structure is the new
(fundamental) mathematical notion, which corresponds to the metric structure
of general relativity. In two dimensions all the smooth manifolds are
LCR-manifolds, therefore in Polyakov functional integral we simply integrate
over all 2-dimensional manifolds. But in four dimensions we have to consider
only the LCR-manifolds. The simple way to impose this restriction is to use
the Lagrange multiplier technique to add the following action term with the
integrability conditions (\ref{i-7}) on the tetrad 
\begin{equation}
\begin{array}{l}
I_{C}=\int d^{4}\!x\ \sqrt{-g}\{\phi _{0}(\ell ^{\mu }m^{\nu }-\ell ^{\nu
}m^{\mu })(\partial _{\mu }\ell _{\nu })+ \\ 
\qquad +\phi _{1}(\ell ^{\mu }m^{\nu }-\ell ^{\nu }m^{\mu })(\partial _{\mu
}m_{\nu })+\phi _{\widetilde{0}}(n^{\mu }\overline{m}^{\nu }-n^{\nu }%
\overline{m}^{\mu })(\partial _{\mu }n_{\nu })+ \\ 
\qquad +\phi _{\widetilde{1}}(n^{\mu }\overline{m}^{\nu }-n^{\nu }\overline{m%
}^{\mu })(\partial _{\mu }\overline{m}_{\nu })+c.conj.\} \\ 
\end{array}
\label{p-2}
\end{equation}%
These Lagrange multipliers make the complete action $I=I_{R}+I_{C}$
self-consistent and the usual quantization techniques may be applied\cite%
{RAG1992}. The action is formally renormalizable\cite{RAG2008a}, because it
is dimensionless and metric independent. Recall that even the (ordinary Weyl
symmetric) conformal action is renormalizable, with the problem being that
it contains non-removable negative-norm states, because of its higher order
derivatives. The path-integral quantization of PCFT is also formulated\cite%
{RAG2017} as functional summation of open and closed 4-dimensional
LCR-manifolds in complete analogy to the summation of 2-dimensional surfaces
in string theory\cite{POL}. These transition amplitudes of a quantum theory
of LCR-manifolds provide (in principle) the self-consistent algorithms for
the computation of the physical quantities.

The LCR-manifolds are defined with the existence of a tetrad ($\ell ,m;n,%
\overline{m}$), which satisfies the integrability conditions. But if they
are realizable\cite{BAOU}, i.e. they admit structure coordinates ($z^{\alpha
}(x);z^{\widetilde{\alpha }}(x)$), which satisfy the conditions (\ref{l-3}),
they may be considered as real submanifolds of complex manifolds. In this
case the structure coordinates may replace the tetrad as dynamical
variables. Then the LCR-transformation may be viewed as a proper vector
bundle on a LCR-manifold, which will permit us better understand the gauge
field solitonic solutions, which will be identified with the quarks.

The ambient complex manifold of the LCR-manifold (implied by the holomorphic
Frobenius theorem) has two commuting complex structures. The trivial one
defined by the complexification of the real spacetime coordinates and the
second one defined by the structure coordinates ($z^{\alpha }(x);z^{%
\widetilde{\alpha }}(x)$). The holomorphic (relative to the trivial complex
structure) transformation between these two complex structures is ($%
z^{\alpha }(r^{b});z^{\widetilde{\alpha }}(r^{b})$). This permit us to
separate the total ($d$) and partial ($\partial ,\overline{\partial }$)
exterior derivatives into the LCR-exterior derivative ($\partial ^{\prime
},\partial ^{\prime \prime }$) as follows

\begin{equation}
\begin{array}{l}
d=\partial +\overline{\partial }=(\partial ^{\prime }+\partial ^{\prime
\prime })+(\overline{\partial ^{\prime }}+\overline{\partial ^{\prime \prime
}}) \\ 
\\ 
\partial ^{\prime }f=\frac{\partial f}{\partial z^{\alpha }}dz^{\alpha
}\quad ,\quad \partial ^{\prime \prime }f=\frac{\partial f}{\partial z^{%
\widetilde{\alpha }}}dz^{\widetilde{\alpha }} \\ 
\\ 
A_{b}dr^{b}=A_{\alpha }^{\prime }dz^{\alpha }+A_{\widetilde{\alpha }%
}^{\prime \prime }dz^{\widetilde{\alpha }} \\ 
\end{array}
\label{p-3}
\end{equation}%
In the last line I separate the 1-forms into marked 1-LCR-forms. In order to
familiarize the reader with this new formalism we make the transcription $%
A\rightarrow A^{\prime }+A^{\prime \prime }$ in details 
\begin{equation}
\begin{array}{l}
A_{\mu }dr^{\mu }=A_{\mu }\delta _{\nu }^{\mu }dr^{\nu }=A_{\mu }(\ell ^{\mu
}n_{\nu }+n^{\mu }\ell _{\nu }-\overline{m}^{\mu }m_{\nu }-m^{\mu }\overline{%
m}_{\nu })dr^{\nu }= \\ 
\\ 
=[(n^{\mu }A_{\mu })\ell _{\alpha }-(\overline{m}^{\mu }A_{\mu })m_{\alpha
}]dz^{\alpha }+[(\ell ^{\mu }A_{\mu })n_{\widetilde{\alpha }}-(m^{\mu
}A_{\mu })\overline{m}_{\widetilde{\alpha }}]dz^{\widetilde{\alpha }}= \\ 
\\ 
=A_{\alpha }^{\prime }dz^{\alpha }+A_{\widetilde{\alpha }}^{\prime \prime
}dz^{\widetilde{\alpha }} \\ 
\end{array}
\label{p-4}
\end{equation}%
The reader should be careful with the "complex bar" on $m$. After the
complexification of $x$, I had to replace it with a tilde, but I hope it
will be understood from the general content. Then the connections of the
LCR-bundle take the form 
\begin{equation}
\begin{array}{l}
D^{\prime }=\partial ^{\prime }+A^{\prime }\quad ,\quad D^{\prime \prime
}=\partial ^{\prime \prime }+A^{\prime \prime } \\ 
\end{array}
\label{p-5}
\end{equation}%
where the connection belongs to the Lie algebra of the gauge group.

If the ambient complex manifold is considered as a submanifold of the
grassmannian space $G_{4,2}$, the connection is essentially identified with
the connection on the hypersurface of $CP^{3}$ determined by the Kerr
polynomial. The connections $A^{\prime }$ and $A^{\prime \prime }$
correspond to the two branches of the hypersurface, which are necessary to
define the LCR-structure. That is, to the left and right columns of the
homogeneous coordinates of $G_{4,2}$. This point of view and the chirality
of gauge field solitonic solutions may explain why the pions are
pseudoscalars. This will be explained in section VII.

Using the LCR-connection, the action takes the following compact form 
\begin{equation}
\begin{array}{l}
I_{G}=\int_{M}(\partial ^{\prime }A_{j}^{\prime }+\gamma
f_{jlk}A_{l}^{\prime }\wedge A_{k}^{\prime })\wedge (\partial ^{\prime
\prime }A_{j}^{\prime \prime }+\gamma f_{jim}A_{i}^{\prime \prime }\wedge
A_{m}^{\prime \prime })+c.c. \\ 
\\ 
I_{C}=\int_{M}\{\phi _{0}dz^{0}\wedge dz^{1}\wedge \overline{dz^{0}}\wedge 
\overline{dz^{1}}+\phi _{\widetilde{0}}dz^{\widetilde{0}}\wedge dz^{%
\widetilde{1}}\wedge \overline{dz^{\widetilde{0}}}\wedge \overline{dz^{%
\widetilde{1}}}+ \\ 
\qquad +\phi dz^{\widetilde{0}}\wedge dz^{\widetilde{1}}\wedge \overline{%
dz^{0}}\wedge \overline{dz^{1}}+\overline{\phi }\overline{dz^{\widetilde{0}}}%
\wedge \overline{dz^{\widetilde{1}}}\wedge dz^{0}\wedge dz^{1}\} \\ 
\end{array}
\label{p-6}
\end{equation}%
The indication of the LCR-manifold $M$ in the integral sign is not necessary
here, because the Lagrange multipliers assure that $M$ admits a
LCR-structure. I want only to stress the natural emergence of integral
geometry, which will be very helpful to define the LCR-structure measure for
the functional integration.

Using the relations (\ref{l-1}), the action $I_{G}$ takes the better
manageable form 
\begin{equation}
\begin{array}{l}
I_{G}=\int d^{4}x\left[ \det (\partial _{\lambda }z^{a})\ \left\{ (\partial
_{0}x^{\mu })(\partial _{1}x^{\nu })F_{j\mu \nu }\right\} \{(\partial _{%
\widetilde{0}}x^{\rho })(\partial _{\widetilde{1}}x^{\sigma })F_{j\rho
\sigma }\}+c.\ c.\right] \\ 
\\ 
I_{C}=\int d^{4}x\ \epsilon ^{\mu \nu \rho \sigma }[\phi _{0}(\partial _{\mu
}z^{0})(\partial _{\nu }z^{1})(\partial _{\rho }\overline{z^{0}})(\partial
_{\sigma }\overline{z^{1}})+\phi _{\widetilde{0}}(\partial _{\mu }z^{%
\widetilde{0}})(\partial _{\nu }z^{\widetilde{1}})(\partial _{\rho }%
\overline{z^{\widetilde{0}}})(\partial _{\sigma }\overline{z^{\widetilde{1}}}%
)+ \\ 
\qquad \qquad +\phi (\partial _{\mu }\overline{z^{0}})(\partial _{\nu }%
\overline{z^{1}})(\partial _{\rho }z^{\widetilde{0}})(\partial _{\sigma }z^{%
\widetilde{1}})+\overline{\phi }(\partial _{\mu }z^{0})(\partial _{\nu
}z^{1})(\partial _{\rho }\overline{z^{\widetilde{0}}})(\partial _{\sigma }%
\overline{z^{\widetilde{1}}})] \\ 
\end{array}
\label{p-7}
\end{equation}%
where the $4\times 4$ matrix ($\partial _{b}x^{\mu }$) is the inverse of ($%
\partial _{\mu }z^{b}$). This form of the action permits the direct use of
LCR-transformations to define conserved currents applying Noether's theorem.
Energy-momentum and angular momentum are defined as charges of such currents.

The action is invariant under the following two infinitesimal
pseudo-conformal (LCR-structure preserving) transformations%
\begin{equation}
\begin{array}{l}
\delta z^{\beta }\simeq \varepsilon \psi ^{\beta }(z^{\gamma })\quad ,\quad
\delta z^{\widetilde{\beta }}\simeq \widetilde{\varepsilon }\psi ^{%
\widetilde{\beta }}(z^{\widetilde{\gamma }}) \\ 
\\ 
\delta \phi _{0}=-\phi _{0}[(\partial _{\alpha }\psi ^{\alpha })\varepsilon
+(\overline{\partial _{\alpha }\psi ^{\alpha }})\overline{\varepsilon }] \\ 
\delta \phi _{\widetilde{0}}=-\phi _{\widetilde{0}}[(\partial _{\widetilde{%
\alpha }}\psi ^{\widetilde{\alpha }})\widetilde{\varepsilon }+(\overline{%
\partial _{\widetilde{\alpha }}\psi ^{\widetilde{\alpha }}})\overline{%
\widetilde{\varepsilon }}] \\ 
\delta \phi =-\phi \lbrack (\partial _{\alpha }\psi ^{\alpha })\varepsilon +(%
\overline{\partial _{\widetilde{\alpha }}\psi ^{\widetilde{\alpha }}})%
\overline{\widetilde{\varepsilon }}]%
\end{array}
\label{p-8}
\end{equation}%
Notice that the transformations of the "left" and "right" structure
coordinates are independent, like the conformal transformations in the
ordinary 2-dimensional conformal field theory (the Polyakov action).

Using such a general transformation we derive the conservation of the
following "left" and "right" LCR-currents%
\begin{equation}
\begin{array}{l}
J^{\lambda }\equiv -\det (\partial _{\tau }z^{a})\ F_{j01}\psi ^{\gamma
}F_{j\gamma \widetilde{\alpha }}\epsilon ^{\widetilde{\alpha }\widetilde{%
\beta }}(\partial _{\widetilde{\beta }}x^{\lambda })- \\ 
\quad -\epsilon _{\alpha \beta }\psi ^{\alpha }\epsilon ^{\lambda \nu \rho
\sigma }(\partial _{\nu }z^{\beta })[\phi _{0}(\partial _{\rho }\overline{%
z^{0}})(\partial _{\sigma }\overline{z^{1}})+\overline{\phi }(\partial
_{\rho }\overline{z^{\widetilde{0}}})(\partial _{\sigma }\overline{z^{%
\widetilde{1}}})] \\ 
\\ 
\widetilde{J}^{\lambda }\equiv -\det (\partial _{\tau }z^{a})\ \psi ^{%
\widetilde{\gamma }}\epsilon ^{\alpha \beta }F_{j\alpha \widetilde{\gamma }%
}(\partial _{\beta }x^{\lambda })F_{j\widetilde{0}\widetilde{1}}- \\ 
\quad -\epsilon _{\widetilde{\alpha }\widetilde{\beta }}\psi ^{\widetilde{%
\alpha }}\epsilon ^{\lambda \nu \rho \sigma }(\partial _{\nu }z^{\widetilde{%
\beta }})[\phi _{\widetilde{0}}(\partial _{\rho }\overline{z^{\widetilde{0}}}%
)(\partial _{\sigma }\overline{z^{\widetilde{1}}})+\phi (\partial _{\rho }%
\overline{z^{0}})(\partial _{\sigma }\overline{z^{1}})] \\ 
\end{array}
\label{p-9}
\end{equation}%
An appropriate definition of the structure coordinates and their relation to
the Poincar\'{e} group permit us to find the energy-momentum and angular
momentum conserving currents. Notice that the explicit contribution of the
gluon field indicate that we could in principle find a way to calculate the
mass differences between leptons and hadrons.

The canonical and BRST quantization\cite{RAG1992} of the PCFT action is
straightforward and I will not repeat it here. The path-integral
quantization is analogous to that of the Polyakov action, where the measures
are geometric. Here we sum over the 4-dimensional LCR-structures instead of
the 2-dimensional complex structures, because the Wick rotation destroys the
LCR-structure. On the other hand, as we will see below, here the elementary
particles are solitonic (distributional) configurations. The computation of
path-integrals for soliton-soliton scattering processes looks quite
formidable. Therefore I will use more intuitive solitonic technics to
describe the experimental consequences of PCFT.

\section{DERIVATION\ OF\ EINSTEIN'S\ GRAVITY}

\setcounter{equation}{0}

In this section I will properly define gravity from the LCR-structure
defining conditions (\ref{l-9}). As I mentioned in the introduction, because
of its tetrad-Weyl symmetry a LCR-structure does not uniquely define a
tetrad ($\ell ,m;n,\overline{m}$). Hence 
\begin{equation}
\begin{array}{l}
\lbrack g_{\mu \nu }]=\ell _{\mu }n_{\nu }+\ell _{\nu }n_{\mu }-m_{\mu }%
\overline{m}_{\nu }-m_{\nu }\overline{m}_{\mu } \\ 
\end{array}
\label{g-1}
\end{equation}%
defines a class of symmetric tensors. In this form the LCR-tetrad is the
null tetrad of the Newman-Penrose (NP) formalism\cite{CHAND}, for all the
metrics of the class. Recall that the NP formalism is essentially the Cartan
formalism adapted to the null tetrad. The LCR-conditions (\ref{i-7}) are the
geodetic and shear-free conditions of the null tetrad, which in the NP
formalism coincide with the annihilation of the spin coefficients $\kappa
=\sigma =\lambda =\nu =0$. This imposes the restriction to the Einstein
metric to admit a geodetic and shear-free null tetrad (two geodetic and
shear-free congruences). This restriction is experimentally in favor to the
PCFT, because all the observed spacetimes have this property.

In the context of riemannian geometry, where the metric is the fundamental
structure and the tetrad is derived, we have the local $SO(1,3)$ symmetry of
the tetrad. But in PCFT, where the LCR-structure is the dynamical variable,
there is no local $SO(1,3)$ symmetry. Instead we have the tetrad-Weyl
symmetry (\ref{i-7a}).

Let us now consider the class of metrics $[\eta _{\mu \nu }]$,\ which are
compatible with the Minkowski spacetime. The Penrose form\cite{P-R} of the
Kerr solution for two geodetic and shear-free flat congruences is 
\begin{equation}
\begin{array}{l}
X^{mi}E_{mn}X^{nj}=0\quad ,\quad K(X^{mi})=0 \\ 
E_{mn}=%
\begin{pmatrix}
0 & 0 & 1 & 0 \\ 
0 & 0 & 0 & 1 \\ 
1 & 0 & 0 & 0 \\ 
0 & 1 & 0 & 0%
\end{pmatrix}
\\ 
\end{array}
\label{g-2}
\end{equation}%
where $X^{mi}$ are the homogeneous coordinates of $G_{4,2}$ already defined
in (\ref{l-10}). The first relation implies that the projective coordinates $%
r_{A^{\prime }A}$ are hermitian, which means that the Shilov boundary of the 
$SU(2,2)$ symmetric classical domain is the "real axis" of $%
%TCIMACRO{\U{2102} }%
%BeginExpansion
\mathbb{C}
%EndExpansion
^{4}$, identified with the Minkowski spacetime. Besides, these are exactly
the four conditions (\ref{l-9}), which determine the LCR-structure. The
relation $K(X^{mi})=0$ is the Kerr holomorphic function, which determines
the hypersurface of $CP^{3}$. Notice that for the Minkowski spacetime, the
conditions required for two null congruences to be geodetic and shear-free
are only the (corresponding two or one common) Kerr holomorphic functions.
In the context of PCFT, the important point is that the flat spacetime is
defined by the algebraic conditions of the LCR-structure without direct
reference to the metric.

I want to point out that PCFT is intended to describe the elementary
particles and not the macroscopic bodies, which must be viewed as made up of
elementary particles. The macroscopic spherically symmetric metrics are
simple approximations of macroscopic bodies. In the context of quantum field
theory the crucial property is the Poincar\'{e} Lie algebra. We saw that in
the context of LCR-structures, the identification of a Poincar\'{e} Lie
algebra is achieved through a local embedding (immersion) of the
LCR-structure ambient complex manifold into an affine subspace of $G_{4,2}$,
viewed as a pair of points of $CP^{3}$. The LCR-manifolds, which admit a $%
G_{4,2}$ immersion, will be called "particles", and the rest "unparticles".
Notice that the class of metrics [$g_{\mu \nu }$] can be defined for
unparticles, but a Poincar\'{e} algebra cannot be properly defined.

Let us also clarify the real analyticity problems implied by the projection
of ambient complex manifold down to its LCR-manifold. This is understood as
a limit of the complex structure coordinates in the ambient complex manifold
down to the LCR-subsurface. In higher dimensional surfaces this can be done
in an infinite number of cones with edge the limit point of the LCR-surface.
In a real analytic LCR-structure, all these limits coincide. But it is not
generally necessary. If it does not happen, the structure coordinates $%
z^{b}(x)$ become generalized functions. Notice that this point of view is
essentially the Sato's approach to generalized functions (hyperfunctions)%
\cite{MOR}.

Let us now consider the quadratic surfaces of $CP^{3}$ implied by the linear
trajectory (\ref{l-12b}) with all the constants real. After a Poincar\'{e}
transformation, $v^{a}=(1,0,0,0)$ and $c^{a}=0$. The LCR-structure is 
\begin{equation}
\begin{array}{l}
z^{0}=t-r\quad ,\quad z^{1}=e^{i\varphi }\tan \frac{\theta }{2}=\frac{x+iy}{%
r+z} \\ 
z^{\widetilde{0}}=t+r\quad ,\quad z^{\widetilde{1}}=e^{-i\varphi }\tan \frac{%
\theta }{2}=\overline{z^{1}} \\ 
\\ 
\ell =dz^{0}=\frac{1}{r}(rdt-\overrightarrow{r}d\overrightarrow{r})\quad
,\quad n=dz^{\widetilde{0}}=\frac{1}{r}(rdt+\overrightarrow{r}d%
\overrightarrow{r}) \\ 
m=dz^{1}=\frac{1}{r(r+z)^{2}}\{[(r^{2}+rz+x^{2})+ixy]dx+ \\ 
\qquad +[xy+i(r^{2}+rz+y^{2})]dy+[x(r+z)+iy(r+z)]dz\} \\ 
\ell \wedge n\wedge m\wedge \overline{m}=\frac{-4i}{(r+z)^{2}}dt\wedge
dx\wedge dy\wedge dz \\ 
\end{array}
\label{g-3}
\end{equation}%
which is degenerate ($\equiv $ with vanishing relative invariants). It has
singularities at $r=0=r+z$ and possibly at the projective infinity, where
the affine coordinates are not valid. It is known that the grassmannian
manifold $G_{4,2}$ admits an affine space where the infinite LCR-submanifold
(\ref{g-2}) is compactified. It is apparently interesting to see how our
infinite physical "flat spacetime" globally appears in the precise
"unphysical" affine chart, where the Poincar\'{e} transformation is no
longer quasi-linear. For that we have to find the Cayley transformation
between the corresponding projective coordinates. Consider the transformation%
\begin{equation}
\begin{array}{l}
Y=\left( 
\begin{array}{c}
\mu \\ 
\widehat{z}\mu%
\end{array}%
\right) =B\left( 
\begin{array}{c}
\lambda \\ 
-i\widehat{r}\lambda%
\end{array}%
\right) \quad ,\quad B=\frac{1}{\sqrt{2}}\left( 
\begin{array}{cc}
I & I \\ 
I & -I%
\end{array}%
\right) \\ 
\\ 
E^{\prime }=BEB^{\dagger }=\left( 
\begin{array}{cc}
I & 0 \\ 
0 & -I%
\end{array}%
\right)%
\end{array}
\label{g-3a}
\end{equation}%
which gives the Cayley transformation between the bounded $E^{\prime }$ and
the unbounded realization $E$ of the domain 
\begin{equation}
\begin{array}{l}
\widehat{r}=i(I-\widehat{z})(I+\widehat{z})^{-1}=i(I+\widehat{z})^{-1}(I-%
\widehat{z})\quad ,\quad \frac{\widehat{r}-\overline{\widehat{r}}}{2i}=0 \\ 
\\ 
\widehat{z}=(iI-\widehat{r})(iI+\widehat{r})^{-1}=(iI+\widehat{r})^{-1}(iI-%
\widehat{r})\quad ,\quad I-\widehat{z}^{\dagger }\widehat{z}=0%
\end{array}
\label{g-3b}
\end{equation}%
They are boundaries of the bounded and unbounded realizations of the $%
SU(2,2) $ classical domain. It transforms the unbounded real surface (\ref%
{g-2}) $%
%TCIMACRO{\U{211d} }%
%BeginExpansion
\mathbb{R}
%EndExpansion
^{4}$ to the bounded $U(2)$ boundary of the classical domain. It is a $%
1\Leftrightarrow 2$ mapping. Using the parameterization 
\begin{equation}
\begin{array}{l}
U=e^{i\tau }\left( 
\begin{array}{cc}
\cos \rho +i\sin \rho \cos \sigma & -i\sin \rho \sin \sigma \ e^{-i\chi } \\ 
-i\sin \rho \sin \sigma \ e^{i\chi } & \cos \rho -i\sin \rho \cos \sigma%
\end{array}%
\right) \\ 
\\ 
\tau \in (-\pi ,\pi )\quad ,\quad \rho \in \lbrack 0,2\pi )\quad ,\quad
\sigma \in \lbrack 0,\pi )\quad ,\quad \chi \in (0,2\pi )%
\end{array}
\label{g-3c}
\end{equation}%
we see that the one $%
%TCIMACRO{\U{211d} }%
%BeginExpansion
\mathbb{R}
%EndExpansion
^{4}$ patch is \ 
\begin{equation}
\begin{array}{l}
x_{+}^{0}=\frac{\sin \tau }{\cos \tau \ +\ \cos \rho } \\ 
x_{+}^{1}+ix_{+}^{2}=\frac{\sin \rho }{\cos \tau \ +\cos \rho }\sin \sigma \
e^{i\chi } \\ 
x_{+}^{3}=\frac{\sin \rho }{\cos \tau \ +\ \cos \rho }\cos \sigma \\ 
\tau \in (-\pi ,\pi )\ ,\ \rho \in \lbrack 0,\pi )\ ,\ \sigma \in \lbrack
0,\pi )\ ,\ \chi \in (0,2\pi ) \\ 
\\ 
s:=\frac{\sin \rho }{\cos \tau \ +\cos \rho }>0\quad \leftrightarrow \quad
\cos \tau \ +\ \cos \rho >0%
\end{array}
\label{g-3d}
\end{equation}
and the second $%
%TCIMACRO{\U{211d} }%
%BeginExpansion
\mathbb{R}
%EndExpansion
^{4}$ patch is \ 
\begin{equation}
\begin{array}{l}
x_{-}^{0}=\frac{\sin \tau }{\cos \tau \ +\ \cos \rho } \\ 
x_{-}^{1}+ix_{-}^{2}=-\frac{\sin \rho }{\cos \tau \ +\cos \rho }\sin \sigma
\ e^{i\chi } \\ 
x_{-}^{3}=-\frac{\sin \rho }{\cos \tau \ +\ \cos \rho }\cos \sigma \\ 
\tau \in (-\pi ,\pi )\ ,\ \rho \in \lbrack 0,\pi )\ ,\ \sigma \in \lbrack
0,\pi )\ ,\ \chi \in (0,2\pi ) \\ 
\\ 
s:=\frac{\sin \rho }{\cos \tau \ +\cos \rho }<0\quad \leftrightarrow \quad
\cos \tau \ +\ \cos \rho <0%
\end{array}
\label{g-3e}
\end{equation}
in order to cover the rest of $U(2)$ universe. Aparently these two patches
do not overlap, therefore they cannot form an atlas.

Let us now look for a quadratic polynomial, which is invariant under the
left and right massless Poincar\'{e} transformations [$E=\pm p^{3}$]. From
the detailed analysis (presented in the neutrino section), the unique (in
the chosen Poincar\'{e} group) degenerate quadratic surface of $CP^{3}$ is 
\begin{equation}
\begin{array}{l}
K(Z^{m})=Z^{0}Z^{1}=0 \\ 
\end{array}
\label{g-3f}
\end{equation}%
The structure coordinates are 
\begin{equation}
\begin{array}{l}
z^{\prime 0}=t-z\quad ,\quad z^{\prime 1}=x+iy \\ 
z^{\prime \widetilde{0}}=t+z\quad ,\quad z^{\widetilde{1}}=x-iy=\overline{%
z^{1}} \\ 
\end{array}
\label{g-4}
\end{equation}%
which are regular in $%
%TCIMACRO{\U{211d} }%
%BeginExpansion
\mathbb{R}
%EndExpansion
^{4}$. In compacted coordinates they have the form 
\begin{equation}
\begin{array}{l}
z^{\prime 0}=\frac{\sin \tau -\sin \rho \cos \theta }{\cos \tau +\cos \rho }%
\quad ,\quad z^{\prime 1}=\frac{\sin \rho }{\cos \tau +\cos \rho }\sin
\theta \ e^{i\varphi } \\ 
z^{\prime \widetilde{0}}=\frac{\sin \tau +\sin \rho \cos \theta }{\cos \tau
+\cos \rho }\quad ,\quad z^{\widetilde{1}}=\overline{z^{1}} \\ 
\end{array}
\label{g-4a}
\end{equation}%
with an apparent singularity at $\pm scri$ ($\cos \tau +\cos \rho =0$). It
seems to be a quite reasonable flat vacuum. Notice that the following
relation 
\begin{equation}
\begin{array}{l}
dz^{0}=\frac{r+z}{2r}dz^{\prime 0}-\frac{z^{\prime \widetilde{1}}}{2r}%
dz^{\prime 1}+\frac{r-z}{2r}dz^{\prime \widetilde{0}}-\frac{z^{\prime 1}}{2r}%
dz^{\prime \widetilde{1}} \\ 
\end{array}
\label{g-5}
\end{equation}%
indicates that the LCR-structures (\ref{g-3}) and (\ref{g-3f}) are not
equivalent, despite the fact that both are degenerate.

The algebraic definition of the "flat" class of metrics $[\eta _{\mu \nu }]$
indicates the algebraic definition of gravity. We simply replace the
algebraic LCR-structure conditions (\ref{g-2}) with 
\begin{equation}
\begin{array}{l}
X^{\dagger }EX=%
\begin{pmatrix}
G_{11}(\overline{X^{m1}},X^{m1}) & G_{12}(\overline{X^{m1}},X^{m2}) \\ 
\overline{G_{12}} & G_{22}(\overline{X^{m2}},X^{m2})%
\end{pmatrix}
\\ 
K(X^{mi})=0 \\ 
\end{array}
\label{g-8}
\end{equation}%
where $G_{ij}=G_{ij}(\overline{X^{mi}},X^{mj})$ are homogeneous functions
with this precise dependence on the two points of the algebraic variety
determined by the Kerr polynomial. The non-vanishing of these terms implies
that the complex component of $r^{b}=x^{b}+iy^{b}$ does not vanish and
gravity emerges. In order to compute this gravity $y^{b}(x^{a})$ generated
by LCR-structure, it is convenient to use the projective coordinates of the
grassmannian space $G_{4,2}$ in a precise coordinate patch (affine variety)
with the following spinorial form 
\begin{equation}
\begin{array}{l}
X^{mj}=%
\begin{pmatrix}
\lambda ^{Aj} \\ 
-ir_{A^{\prime }B}\lambda ^{Bj}%
\end{pmatrix}
\\ 
\end{array}
\label{g-9}
\end{equation}%
of the rank-2 matrix $X^{mj}$, and define the tetrad 
\begin{equation}
\begin{array}{l}
L^{a}=\frac{1}{\sqrt{2}}\overline{\lambda }^{A^{\prime }1}\lambda
^{B1}\sigma _{A^{\prime }B}^{a}\quad ,\quad N^{a}=\frac{1}{\sqrt{2}}%
\overline{\lambda }^{A^{\prime }2}\lambda ^{B2}\sigma _{A^{\prime
}B}^{a}\quad ,\quad M^{a}=\frac{1}{\sqrt{2}}\overline{\lambda }^{A^{\prime
}2}\lambda ^{B1}\sigma _{A^{\prime }B}^{a} \\ 
\\ 
\epsilon _{AB}\lambda ^{A1}\lambda ^{B2}=1 \\ 
\end{array}
\label{g-10}
\end{equation}%
which is null relative to the Minkowski metric $\eta _{ab}$. Then the above
relations (\ref{g-8}) take the form 
\begin{equation}
\begin{array}{l}
2\sqrt{2}y^{a}L_{a}=G_{11}(\overline{Y^{m1}},Y^{n1}) \\ 
\\ 
2\sqrt{2}y^{a}\overline{M}_{a}=G_{12}(\overline{Y^{m1}},Y^{n2}) \\ 
\\ 
2\sqrt{2}y^{a}N_{a}=G_{22}(\overline{Y^{m2}},Y^{n2})%
\end{array}
\label{g-11}
\end{equation}%
Recall that $y^{a}$\ is the imaginary part of the projective coordinates $%
r^{a}=x^{a}+iy^{a}$\ defined by the relation $r_{A^{\prime }B}=r^{a}\sigma
_{aA^{\prime }B}$\ and $\sigma _{A^{\prime }B}^{a}$ being the identity and
the three Pauli matrices (\ref{l-10}-\ref{l-12}). The normalization of the
spinors is permitted, because of the homogeneity of the functions. These
conditions are formally "solved" by \ 
\begin{equation}
\begin{array}{l}
y^{a}=\frac{1}{2\sqrt{2}}[G_{22}N^{a}+G_{11}L^{a}-G_{12}M^{a}-\overline{%
G_{12}}\overline{M}^{a}] \\ 
\end{array}
\label{g-12}
\end{equation}%
which combined with the computation of $\lambda ^{Ai}$\ as functions of $%
r^{a}$, using the Kerr conditions $K_{i}(X^{mi})=0$, permit us to
perturbatively compute $y^{a}$ as functions of the real part of $r^{a}$.
From the physical point of view, we may say that this procedure gives the
gravitational "dressing" of the soliton (particle) in the form $y^{a}=$ $%
y^{a}(x)$ of the (totally real) lorentzian CR-submanifold expressed in the
projective coordinates of $G_{4,2}$. The explicit form of $y^{a}(x)$\ is
implied by the precise dependence of $G_{ij}(\overline{X^{mi}},X^{mj})$,
considered real analytic, and their expansion into a series relative to $%
y^{a}$. This is just a simple application of the implicit function theorem.

The definition of the Einstein metric permit us to define energy-momentum
and angular momentum as conserved quantities in the linearized Einstein
gravity approximation\cite{MTW}. We find the following linearized gravity
relations in the limit \ 
\begin{equation}
\begin{array}{l}
g_{\mu \nu }=\eta _{\mu \nu }+kh_{\mu \nu }+O(k^{2}) \\ 
\\ 
\widehat{R}_{\nu \rho \sigma \tau }=\underset{k\rightarrow 0}{\lim }%
(k^{-1}R_{\nu \rho \sigma \tau })=2\partial _{\lbrack \nu }\partial
_{|[\sigma }h_{\tau ]|\rho ]} \\ 
\end{array}
\label{g-14}
\end{equation}%
for the curvature tensor. The second Bianchi identities imply the
conservation condition of the Einstein tensor \ 
\begin{equation}
\begin{array}{l}
\partial _{\mu }\widehat{E}_{\ \nu }^{\mu }=\partial _{\mu }[\widehat{R}_{\
\nu }^{\mu }-\frac{1}{2}\delta _{\nu }^{\mu }\widehat{R}]=0 \\ 
\\ 
\widehat{E}_{\rho \tau }\equiv \widehat{R}_{\rho \tau }-\frac{1}{2}\eta
_{\rho \tau }\widehat{R}=\frac{1}{2}[\partial ^{2}h_{\rho \tau }+\partial
_{\rho }\partial _{\tau }h_{\nu }^{\nu }-\partial _{\rho }(\partial _{\nu
}h_{\tau }^{\nu })--\partial _{\tau }(\partial _{\nu }h_{\rho }^{\nu })]%
\end{array}
\label{g-16}
\end{equation}%
This means that the linearized Einstein tensor defines the energy-momentum
density as a preserved tensor distribution.

The standard model does not explain the existence of only three generations
of leptons and quarks. In the context of PCFT context the three generations
of flavors is imposed by gravity, despite the fact that the standard model
does not contain gravity. It is well known in general relativity\cite{CHAND}%
, that in a geodetic and shear-free null tetrad the first $\Psi _{0}$ and
last $\Psi _{4}$ components of the Weyl tensor in the Newman-Penrose
formalism vanish, i.e. \ 
\begin{equation}
\begin{array}{l}
\Psi _{0}=\Psi _{ABCD}o^{A}o^{B}o^{C}o^{D}=0\quad ,\quad \Psi _{4}=\Psi
_{ABCD}\imath ^{A}\imath ^{B}\imath ^{C}\imath ^{D}=0 \\ 
\end{array}
\label{g-17}
\end{equation}%
where $\Psi _{ABCD}$ is the conformal tensor in spinorial coordinates and $%
o^{A},\imath ^{A}$ is the geodetic and shear-free spinor dyad. In the zero
gravity approximation we have $o^{A}=\lambda ^{A1}$\ and $\imath
^{A}=\lambda ^{A2}$, the two spinors which appear in the homogeneous
coordinates (\ref{l-10}) of a flat LCR-structure. Hence in the linearized
Einstein gravity approximation we have the relations \ 
\begin{equation}
\begin{array}{l}
\Psi _{ABCD}o^{A}o^{B}o^{C}o^{D}\simeq k\widehat{\Psi }_{ABCD}\lambda
^{A1}\lambda ^{B1}\lambda ^{C1}\lambda ^{D1}+O(k^{2})=0 \\ 
\Psi _{ABCD}\imath ^{A}\imath ^{B}\imath ^{C}\imath ^{D}\simeq k\widehat{%
\Psi }_{ABCD}\lambda ^{A2}\lambda ^{B2}\lambda ^{C2}\lambda ^{D2}+O(k^{2})=0%
\end{array}
\label{g-18}
\end{equation}%
That is, at every point of spacetime a gravitating (with non vanishing
conformal tensor) LCR-manifold is implied by at least a quadratic
hypersurface of $CP^{3}$ (already known) and at most to a quartic branched
hypersurface of $CP^{3}$. This restriction imposes the existence of three
generations of solitonic LCR-manifolds, which are identified with leptons.
We will see below that they are the Petrov type D (the generation of the
electron), the Petrov type II (the muon generation) and the Petrov type I
(the tau generation). The Petrov type III spacetimes (LCR-manifolds) may not
be realizable as elementary particles.

Generalizing the above simple examples of scalar LCR-structures we may
consider the general real Newman trajectories $\xi ^{b}(\tau )$, which may
generally viewed as "interacting" LCR-structures. The $CP^{3}$ embedding (%
\ref{l-12a}) imply 
\begin{equation}
\begin{array}{l}
(r_{A^{\prime }A}-\xi _{A^{\prime }A}(\tau ))\lambda ^{A}=0 \\ 
\end{array}
\label{g-18e}
\end{equation}%
which completely determines the holomorphic solutions $\tau (r^{a})$ and
projectively $\lambda ^{A}(r^{a})$. The term "real" here means that all the
parameters of the function $\xi ^{b}(\cdot )$ are real or the mathematically
correct condition that the coefficients of a $r^{a}$-local Taylor expansion
are real numbers, but the solutions $\tau (r^{a})$ , $\lambda ^{A}(r^{a})$
may be complex. The relations (projections) $r^{a}(x^{b})$ down to the real
plane $%
%TCIMACRO{\U{211d} }%
%BeginExpansion
\mathbb{R}
%EndExpansion
^{4}$ is implied by (\ref{g-12}). The number of solutions $\tau (r^{a})$ , $%
\lambda ^{A}(r^{a})$ depends on the spacetime points and essentially
characterize the multiplicity of the generally forth degree polynomial of
the conformal tensor $\Psi _{ABCD}$. Hence we may have the following
possibilities: 1) At the points $r^{a}(x^{b})$ where we have one solution $%
\tau (r^{a}(x))$ and $\lambda ^{A}(r^{a}(x))$, the corresponding
LCR-manifold needs a second trajectory to be defined. Using as second
trajectory that of the vacuum, then we interpret it as an object with only
one chirality. 2) At the points with two solutions $\tau _{i}(r^{a})$ , $%
\lambda ^{Ai}(r^{a})$ with $i=1,2$, is the most common form of the
LCR-manifold, because it provides the classical notion of a generally
interacting particle. This is completely clarified in the newtonian
approximation. I first normalize the parameter $\tau $ as $\xi ^{0}=c\tau $,
where the velocity of light $c$ is explicitly written. Then, the condition
for the existence of non-vanishing solutions $\lambda ^{A}(r^{a})$ from (\ref%
{g-18e}), takes the form 
\begin{equation}
\begin{array}{l}
(cr^{0}-c\tau )^{2}-(r^{j}-\xi ^{j}(\tau ))^{2}=0 \\ 
z^{0}=\tau _{1}=r^{0}-\frac{1}{c}\sqrt{(r^{j}-\xi ^{j}(\tau _{1}))^{2}}%
\simeq t-\frac{1}{c}\sqrt{(r^{j}-\xi ^{j}(t))^{2}}+O(\frac{1}{c^{2}}) \\ 
z^{\widetilde{0}}=\tau _{2}=r^{0}+\frac{1}{c}\sqrt{(r^{j}-\xi ^{j}(\tau
_{2}))^{2}}\simeq t+\frac{1}{c}\sqrt{(r^{j}-\xi ^{j}(t))^{2}}+O(\frac{1}{%
c^{2}}) \\ 
\end{array}
\label{g-18f}
\end{equation}%
We see that the structure coordinates $z^{0},z^{\widetilde{0}}$ are retarded
and advanced "wave" functions indicating that physical causality comes from
the LCR-structure. Besides, these "waves" come from the same point $%
r^{j}-\xi ^{j}(t)=0$, interpreted as the trajectory of the LCR-manifold
(particle). 3) At the points with three solutions $\tau _{i}\ ,\ i=1,2,3$,
we may generally have the three LCR-structures $(\tau _{1},\tau _{2})$ , $%
(\tau _{1},\tau _{3})$ , $(\tau _{2},\tau _{3})$ and their complex
conjugates. Recall that integral curves bifurcate. Hence we may have a
LCR-structure (two roots $(\tau _{1}^{\prime },\tau _{2}^{\prime })$, one
particle) in which and at a given point the integral curve $\ell ^{\mu
}\partial _{\mu }$ bifurcates $\tau _{1}^{\prime }\rightarrow (\tau
_{1},\tau _{2})$. Then the natural implied LCR-structure (particle)
evolution is 
\begin{equation}
\begin{array}{l}
(\tau _{1}^{\prime },\tau _{2}^{\prime })\rightarrow (\tau _{1},\tau
_{2}^{\prime })+(\tau _{2},\tau _{2}^{\prime }) \\ 
\end{array}
\label{g-18g}
\end{equation}%
which is a typical particle disintegration. 4) Using the linearized gravity
approximation (\ref{g-18}), I have already showed that the LCR-manifold with
four solutions is the largest Einstein spacetime (with gravity). The implied
bifurcations are more complicated. The natural particle picture is when two
decoupled stable particles, one LCR-manifold with two decoupled pairs $(\tau
_{1},\tau _{2})$ , $(\tau _{3},\tau _{4})$ of solutions through the
formation of LCR-structures with up to four $\lambda ^{A}(r^{a})$ solutions%
\cite{P-R}, which finally disintegrate into stable particles, considered
embedded into a flat spacetime (which is compatible with any number of
geodetic and shear free congruences).

\section{ELECTRON\ AND\ ELECTRODYNAMICS}

\setcounter{equation}{0}

Like the 2-dimensional Polyakov action (and its supersymmetric evolution),
the present 4-dimensional PCFT does not explicitly contain the observed
particles as independent fields. Therefore they have to be found as stable
configurations. In string theory the guiding clue was the Poincar\'{e} group
of the 26-dimensional Minkowski space, emerging after the identification of
the $X^{\mu }(x)$ field with the embedding function of the string in to the
26-dimensional Minkowski space. It is well known that string theory tried to
identify the observed elementary particles with the lowest string modes. In
the context of PCFT the gluon field is identified with the gauge field (LCR
vector bundle), which explicitly appears in the action, and the observed
elementary particles are identified with precise (distributional) solitons.

If we identify the 4-dimentional flat spacetime with the boundary (\ref{g-2}%
) of the $SU(2,2)$ classical domain, the linear subgroup\cite{PIAT} of $%
SU(2,2)$, which fixes the projective "infinity" (the scri in the Penrose
terminology), becomes the physical Poincar\'{e}$\times $Dilation group. The
particles will emerge as stable solitonic (configurations) generalized
functions (Schwartz distributions or Sato's hyperfunctions) viewed as
potentials of their distributional sources identified with the fermionic
flavors (leptons and quarks). The stable particles (electron, neutrino, and
up and down quarks) admit the automorphisms of time translation and z-axis
rotation, which make them eigenstates of the corresponding generators of a
Poincar\'{e} representation. The unstable (decaying) elementary particles
admit only the z-rotation automorphism. That is we only consider that they
can have only exact spin.

The Poincar\'{e}$\times $Dilation transformation in the Siegel (chiral)
realization is%
\begin{equation}
\begin{array}{l}
\begin{pmatrix}
\lambda ^{\prime } \\ 
-ir^{\prime }\lambda ^{\prime }%
\end{pmatrix}%
=%
\begin{pmatrix}
B & 0 \\ 
-iTB & (B^{\dag })^{-1}%
\end{pmatrix}%
\left( 
\begin{array}{c}
\lambda \\ 
-ir\lambda%
\end{array}%
\right) \\ 
\\ 
r^{\prime }=(B^{-1})^{\dag }r(B^{-1})+T\quad ,\quad T^{\dagger }=T \\ 
\end{array}
\label{e-1}
\end{equation}%
where for $\det B=1$ is the Poincar\'{e} transformation. This is an
automorphism of the degenerate LCR-structure (\ref{g-3}). The proof\cite%
{RAG2017} uses the Newman replacement of the quadratic Kerr polynomial (\ref%
{g-5a}) with a trajectory $\xi ^{a}=(\tau ,0,0,0)$. Under a Poincar\'{e}$%
\times $Dilation transformation this trajectory becomes a real linear
trajectory $\xi ^{a}=v^{a}\tau +c^{a}$ (\ref{l-12b}) with $v^{a}v^{b}\eta
_{ab}=1$. The two structure coordinates $z^{0}=\tau _{1}$ and $z^{\widetilde{%
0}}=\tau _{2}$\ are determined from\ 
\begin{equation}
\begin{array}{l}
(x^{a}-\xi ^{a}(\tau _{j}))^{2}=0 \\ 
\end{array}
\label{e-2}
\end{equation}%
for each column of the homogeneous coordinates $X^{mi}$, and found to be
invariant, because the above defining form is invariant. The spinors $%
\lambda ^{Aj}$\ are the dyad, which determine the null vectors $\Delta
_{(j)}^{a}=(x^{a}-\xi ^{(j)a}(\tau _{j}))$, that is\cite{P-R}\ 
\begin{equation}
\begin{array}{l}
(x^{a}-\xi ^{a}(\tau _{j}))=\lambda ^{j\dag }\sigma ^{a}\lambda ^{j} \\ 
\end{array}
\label{e-3}
\end{equation}%
with the same trivial trajectory for both $j=1,2$ before and after the
Poincar\'{e} transformation. The (real) translation does not affect the
spinors. The Lorentz transformation changes the trajectory, but its form
remains the same for $j=1,2$. We precisely find 
\begin{equation}
\begin{array}{l}
\begin{pmatrix}
\Delta _{(j)}^{0} \\ 
\Delta _{(j)}^{1} \\ 
\Delta _{(j)}^{2} \\ 
\Delta _{(j)}^{3}%
\end{pmatrix}%
=%
\begin{pmatrix}
\Delta \\ 
\Delta _{(j)}\sin \theta \ \cos \varphi \\ 
\Delta _{(j)}\sin \theta \ \sin \varphi \\ 
\Delta _{(j)}\cos \theta%
\end{pmatrix}%
=%
\begin{pmatrix}
\overline{\lambda ^{0j}}\lambda ^{0j}+\overline{\lambda ^{1j}}\lambda ^{1j}
\\ 
\overline{\lambda ^{1j}}\lambda ^{0j}+\overline{\lambda ^{0j}}\lambda ^{1j}
\\ 
i\overline{\lambda ^{1j}}\lambda ^{0j}-i\overline{\lambda ^{0j}}\lambda ^{1j}
\\ 
\overline{\lambda ^{0j}}\lambda ^{0j}-\overline{\lambda ^{1j}}\lambda ^{1j}%
\end{pmatrix}
\\ 
\\ 
\Delta _{(1)}=\Delta \quad ,\quad \Delta _{(2)}=-\Delta \\ 
\end{array}
\label{e-4}
\end{equation}%
Before the Lorentz transformation with zero velocity the LCR-structure
condition $\overline{X^{m1}}E_{mn}X^{n2}=0$ implies $z^{\widetilde{1}}=%
\overline{z^{1}}$. After the transformation $\Delta ,\theta ,\varphi $
change, but the relation of the structure coordinates remains the same, $%
z^{\prime \widetilde{1}}=\overline{z^{\prime 1}}$, i.e. 
\begin{equation}
\begin{array}{l}
z^{1}=\frac{\lambda ^{11}}{\lambda ^{01}}=\frac{\Delta }{2}\sin \theta
e^{i\varphi }\quad ,\quad z^{\widetilde{1}}=-\frac{\lambda ^{02}}{\lambda
^{12}}=\frac{\Delta }{2}\sin \theta e^{-i\varphi } \\ 
\end{array}
\label{e-5}
\end{equation}%
From their definition the two spinors have opposite chiralities. That is,
even the vacuum configurations "see" the two chiralities, on which the
standard model is built up.

We will now look for solitonic LCR-structures. The knowledge of the
"physical" Poincar\'{e} group permit us to look\cite{RAG1999} for static and
axially symmetric LCR-manifolds. That is massive LCR-structures, which admit
time translation and z-rotations as automorphisms. These stable solitons are
states of the Hilbert space and hence eigenstates of the translation and the
z-rotation generators.

For a LCR-manifold embeddable in $G_{4,2}$, I consider the following
structure coordinates and LCR-conditions 
\begin{equation}
\begin{array}{l}
z^{0}\equiv i\frac{X^{21}}{X^{01}}\quad ,\quad z^{1}\equiv \frac{X^{11}}{%
X^{01}}\quad ,\quad z^{\widetilde{0}}\equiv i\frac{X^{32}}{X^{12}}\quad
,\quad z^{\widetilde{1}}\equiv -\frac{X^{02}}{X^{12}} \\ 
\\ 
\frac{z^{0}-\overline{z^{0}}}{2i}-U(\frac{z^{0}+\overline{z^{0}}}{2},z^{1},%
\overline{z^{1}})=0\quad ,\quad z^{\widetilde{1}}-Z(z^{\widetilde{0}},%
\overline{z^{0}},\overline{z^{1}})=0\quad ,\quad \frac{z^{\widetilde{0}}-%
\overline{z^{\widetilde{0}}}}{2i}-V(\frac{z^{\widetilde{0}}-\overline{z^{%
\widetilde{0}}}}{2},z^{\widetilde{1}},\overline{z^{\widetilde{1}}})=0 \\ 
\end{array}
\label{e-6}
\end{equation}%
Then the infinitesimal time-translation and z-rotation are 
\begin{equation}
\begin{array}{l}
\delta X^{0i}=0\quad ,\quad \delta X^{1i}=0\quad ,\quad \delta
X^{2i}=-i\varepsilon ^{0}X^{0i}\quad ,\quad \delta X^{3i}=-i\epsilon
^{0}X^{1i} \\ 
\delta z^{0}=\varepsilon ^{0}\quad ,\quad \delta z^{1}=0\quad ,\quad \delta
z^{\widetilde{0}}=\varepsilon ^{0}\quad ,\quad \delta z^{\widetilde{1}}=0 \\ 
\\ 
\delta X^{0i}=-i\frac{\varepsilon ^{12}}{2}X^{0i}\quad ,\quad \delta X^{1i}=i%
\frac{\varepsilon ^{12}}{2}X^{1i}\quad ,\quad \delta X^{2i}=-i\frac{%
\varepsilon ^{12}}{2}X^{2i}\quad ,\quad \delta X^{3i}=i\frac{\varepsilon
^{12}}{2}X^{3i} \\ 
\delta z^{0}=0\quad ,\quad \delta z^{1}=i\varepsilon ^{12}z^{1}\quad ,\quad
\delta z^{\widetilde{0}}=0\quad ,\quad \delta z^{\widetilde{1}%
}=-i\varepsilon ^{12}z^{\widetilde{1}} \\ 
\end{array}
\label{e-7}
\end{equation}%
and the LCR-structure conditions become%
\begin{equation}
\begin{array}{l}
\frac{z^{0}-\overline{z^{0}}}{2i}-U(z^{1}\overline{z^{1}})=0\quad ,\quad z^{%
\widetilde{1}}-\overline{z^{1}}W(z^{\widetilde{0}}-\overline{z^{0}})=0\quad
,\quad \frac{z^{\widetilde{0}}-\overline{z^{\widetilde{0}}}}{2i}-V(z^{%
\widetilde{1}}\overline{z^{\widetilde{1}}})=0 \\ 
\end{array}
\label{e-8}
\end{equation}%
I also found\cite{RAG2008b} that only the quadratic Kerr polynomial%
\begin{equation}
\begin{array}{l}
K(X^{m})=X^{1}X^{2}-X^{0}X^{3}+2aX^{0}X^{1}=0 \\ 
\end{array}
\label{e-9}
\end{equation}%
admits these automorphisms among all the polynomials of maximal degree four.
Notice that if we try to impose the dilation as an additional automorphism,
we find $a=0$, which is the "spherical" degenerate LCR-structure (\ref{g-5a}%
). The quite general LCR-tetrad (\ref{l-13}) (with $\Delta (r)$ arbitrary)
satisfies these conditions, the additional condition of asymptotic flatness
at null infinity%
\begin{equation}
\begin{array}{l}
X^{m1}E_{mn}X^{n1}=0=X^{m2}E_{mn}X^{n2} \\ 
\\ 
\frac{z^{0}-\overline{z^{0}}}{2i}+2a\frac{z^{1}\overline{z^{1}}}{1+z^{1}%
\overline{z^{1}}}=0\quad ,\quad z^{\widetilde{1}}-\overline{z^{1}}W(z^{%
\widetilde{0}}-\overline{z^{0}})=0\quad ,\quad \frac{z^{\widetilde{0}}-%
\overline{z^{\widetilde{0}}}}{2i}-2a\frac{z^{\widetilde{1}}\overline{z^{%
\widetilde{1}}}}{1+z^{\widetilde{1}}\overline{z^{\widetilde{1}}}}=0%
\end{array}
\label{e-10}
\end{equation}%
and a symmetry between the left and right chiral columns $z^{1}\overline{%
z^{1}}=z^{\widetilde{1}}\overline{z^{\widetilde{1}}}$. Its embedding in $%
G_{4,2}$ is%
\begin{equation}
\begin{array}{l}
X^{mi}=%
\begin{pmatrix}
1 & -z^{\widetilde{1}} \\ 
z^{1} & 1 \\ 
-iz^{0} & iz^{\widetilde{1}}(z^{\widetilde{0}}-2ia) \\ 
-iz^{1}(z^{0}+2ia) & -iz^{\widetilde{0}}%
\end{pmatrix}
\\ 
z^{0}=t-f_{0}(r)-2ia\sin ^{2}\frac{\theta }{2}\quad ,\quad z^{1}=e^{i\varphi
}e^{-iaf_{1}(r)}\tan \frac{\theta }{2} \\ 
z^{\widetilde{0}}=t+f_{0}(r)+2ia\sin ^{2}\frac{\theta }{2}\quad ,\quad z^{%
\widetilde{1}}=e^{-i\varphi }e^{-iaf_{1}(r)}\tan \frac{\theta }{2} \\ 
f_{0}(r)=\tint \frac{r^{2}+a^{2}}{\Delta }dr\quad ,\quad f_{1}(r)=\tint 
\frac{1}{\Delta }dr%
\end{array}
\label{e-10a}
\end{equation}%
The gravitational dressing of the electron can be easily computed. It is
stable relative to the vacuum, because it has non-vanishing all its
relative-invariants $\Phi _{j}$.

A different way to find a static and axially symmetric LCR-structure is
first to solve the problem for flat compatible LCR-structures which satisfy
the Kerr polynomial (\ref{e-9}). After we apply the well known Kerr-Schild
ansatz to find the corresponding curved LCR-structure. The final result\cite%
{RAG1991} is the same LCR-manifold (\ref{l-13}).

The general quadratic form, which is invariant (but not automorphic) under a
Poincar\'{e} transformation is%
\begin{equation}
\begin{array}{l}
A_{mn}Z^{m}Z^{n}=0 \\ 
\\ 
A_{mn}=%
\begin{pmatrix}
\omega & P \\ 
P^{\top } & 0%
\end{pmatrix}%
\quad ,\quad P=%
\begin{pmatrix}
-(p^{1}-ip^{2}) & -p^{0}+p^{3} \\ 
p^{0}+p^{3} & (p^{1}+ip^{2})%
\end{pmatrix}%
=-p\epsilon \\ 
p=%
\begin{pmatrix}
p^{0}-p^{3} & -(p^{1}-ip^{2}) \\ 
-(p^{1}+ip^{2}) & p^{0}+p^{3}%
\end{pmatrix}%
\quad ,\quad \epsilon =%
\begin{pmatrix}
0 & 1 \\ 
-1 & 0%
\end{pmatrix}%
\quad ,\quad \det p\neq 0%
\end{array}
\label{e-10b}
\end{equation}%
The variables $p^{\mu }$ are the momentum (boost) parameters and $\omega $
is the spin. If we first make a boost transformation, we can annihilate the
momenta. After we make a general complex translation%
\begin{equation}
\begin{array}{l}
\begin{pmatrix}
X_{1}^{\prime } \\ 
X_{2}^{\prime }%
\end{pmatrix}%
=%
\begin{pmatrix}
I & 0 \\ 
C & I%
\end{pmatrix}%
\left( 
\begin{array}{c}
X_{1} \\ 
X_{2}%
\end{array}%
\right) \\ 
\\ 
r^{\prime }=r+iC\quad ,\quad C^{\dagger }\neq \pm C%
\end{array}
\label{e-11}
\end{equation}%
Then the spin matrix transforms as follows%
\begin{equation}
\begin{array}{l}
\omega ^{\prime }=\omega +2m%
\begin{pmatrix}
-(C^{1}-iC^{2}) & C^{3} \\ 
C^{3} & (C^{1}+iC^{2})%
\end{pmatrix}
\\ 
\end{array}
\label{e-12}
\end{equation}%
We see that a real translation ($C=-iT)$ cannot remove the spin matrix. But
a complex translation can do it. This means that the spin can be considered
as a complex space translation in $G_{4,2}$. That is, the spin can be
considered as an imaginary space translation in $G_{4,2}$, which explains
why the Newman "magic" complex translation\cite{NEWM2016} of the
Schwartzschild metric implies the Kerr metric. Besides notice that a complex
time translation does not affect the quadric.

\subsection{Microlocal analysis of the electron}

We saw that the LCR-tetrad defines the class of symmetric tensors [$g_{\mu
\nu }$], which appear as the gravitation field. Besides the LCR-tetrad
defines the class of antisymmetric tensors 
\begin{equation}
\begin{array}{l}
\lbrack J_{\mu \nu }]=\ell _{\mu }n_{\nu }-\ell _{\nu }n_{\mu }-m_{\mu }%
\overline{m}_{\nu }+m_{\nu }\overline{m}_{\mu } \\ 
\end{array}
\label{e-12a}
\end{equation}%
Flaherty observed\cite{FLAHE1974} that the metric $g_{\mu \nu }$ defines an
integrable pseudo-complex structure (pseudo, because it is not a real tensor)

\begin{equation}
\begin{array}{l}
J_{\ \nu }^{\mu }=\ell ^{\mu }n_{\nu }-n^{\mu }\ell _{\nu }-m^{\mu }%
\overline{m}_{\nu }+\overline{m}^{\mu }m_{\nu } \\ 
\end{array}
\label{e-12b}
\end{equation}%
Its Nijenhuis integrability conditions coincide with the LCR-structure
conditions. Notice that this tensor is invariant under the tetrad-Weyl
transformation and that the LCR-tetrad are eigenvectors of this tensor. That
is $g_{\mu \nu }$ and $J_{\mu \nu }$\ determine the LCR-structure.

In the special case of the stable LCR-sructure (\ref{l-13}), the self-dual
2-form admits a multiplicative function, which makes it "closed" up to a
singular source.

\begin{equation}
\begin{array}{l}
G^{+}=\frac{2C}{(r+ia\cos \theta )^{2}}(\ell \wedge n-m\wedge \overline{m}%
)=G-i\ \ast G \\ 
\end{array}
\label{e-13}
\end{equation}%
where $C$ is an arbitrary complex constant. That is, it defines an
electromagnetic field $G$ determined by the self-dual 2-form $G^{+}$. It is
closed outside a distributional singularity concentrated at the
ring-singularity of the LCR-manifold, which provides a generally complex
(electric plus magnetic) charge. Hence for an arbitrary complex constant $C$%
, this complex 2-form defines a real 2-form $G$ such that%
\begin{equation}
\begin{array}{l}
dG=-\ast j_{m}\quad ,\quad d\ast G=-\ast j_{e} \\ 
\end{array}
\label{e-14}
\end{equation}%
where $j_{e}$ and $j_{m}$ are the "electric" and "magnetic" currents. These
are apparently analogous to the symmetric Maxwell equations (with both
electric and magnetic monopoles), which were used by Dirac to prove the
quantization\cite{FELS} of the electric charge. It implies that the general
electric charge is quantized\cite{RAG1999}. But the apparent symmetry under
the duality rotation absorbs the magnetic charge (or electric charge)
leaving detectable only one kind of monopoles, as observed in nature%
\begin{equation}
\begin{array}{l}
dG=0\quad ,\quad d\ast G=-\ast j_{e} \\ 
\end{array}
\label{e-15}
\end{equation}%
That is, here we have a "self-quantization" of the electric charge. But once
fixed, the conserved electric charge reduces the general tetrad-Weyl
symmetry (\ref{i-7a}) down to the ordinary Weyl symmetry of the
electromagnetic field. The precise tetrad-Weyl factors used in (\ref{l-14})
give a metric, which coincides with the linearized gravity approximation,
and hence define the Poincar\'{e} conserved quantities. This fact fixes the
remaining ordinary Weyl transformation. That is the precise tetrad-Weyl
factors, which provide the conserved charge, momentum and angular momentum
of the electron, fix (break) the tetrad-Weyl symmetry.

Now it is trivial to show that the positron is the conjugate LCR-structure ($%
\overline{z^{\alpha }},\overline{z^{\widetilde{\beta }}}$), which
corresponds to the tetrad ($\ell ,\overline{m};n,m$). From the definition of
the electromagnetic form (\ref{e-13}) we easily see that its electric charge
has opposite sign from that of the electron LCR-manifold. Hence we have to
identify the conjugate LCR-structure with the antiparticle as long as these
two conjugate structures are not equivalent.

In order to avoid any confusion, I want to point out that the derivation of
the electromagnetic equations (\ref{e-15}) must be interpreted that the
static solitonic LCR-manifold (\ref{l-13}) admits a distributional potential
implied by the closed self-dual 2-form (\ref{e-13}). Other solitonic
LCR-manifolds, having this precise 2-form closed, will be considered to have
an electromagnetic charge. No more generalizations are permitted. The other
important point is to realize the meaning of the ring-singularity, which
essentially determines the electron. In the context of the Einstein gravity
(based on riemannian geometry), the ring-singularity is an essential
singularity. That is, it cannot be removed by a real coordinate
transformation, in contrast to the (soft) horizon singularities, which are
coordinate singularities. In PCFT the ring-singularity comes from the branch
curve of the regular quadratic hypersurface of $CP^{3}$, which is a
coordinate singularity. It is implied by the projection of the two sheets
(branches) of the surface into a $CP^{2}$\ subspace of $CP^{3}$.

Recall that electromagnetism (either classical or quantum) and gravity start
imposing the sources as independent "objects". But here the solitonic
electron comes with the metric and the distributional closed self-dual
2-form, which contains both its gravittation and electromagnetic field
"dressing" with their sources. Using the generalized function terminology,
we state that the electron is the singular support, and electromagnetism
(and gravity) is the regular support of the soliton configuration, being a
generalized function. The Kerr-Newman manifold has been extensively studied,
but I describe here its electromagnetic field in oblate spheroidal 
\begin{equation}
\begin{array}{l}
x=\sqrt{r^{2}+a^{2}}\cos \varphi \sin \theta \quad ,\quad y=\sqrt{r^{2}+a^{2}%
}\sin \varphi \sin \theta \quad ,\quad z=r\cos \theta \\ 
\\ 
\cos \theta =\frac{z}{r}\quad ,\quad \sin ^{2}\theta =\frac{x^{2}+y^{2}}{%
r^{2}+a^{2}}\quad ,\quad \frac{x^{2}+y^{2}}{r^{2}+a^{2}}+\frac{z^{2}}{r^{2}}%
=1 \\ 
\end{array}
\label{e-15a}
\end{equation}%
and cartesian coordinates, in order to compare its singular part with the
corresponding singular part of the gluonic field of the quark soliton. The
self-dual 2-form is

\begin{equation}
\begin{array}{l}
G^{+}=G-i\ \ast G=\frac{e}{4\pi (r+ia\cos \theta )^{2}}(\ell \wedge
n-m\wedge \overline{m})= \\ 
=\frac{e}{4\pi (r+ia\cos \theta )^{2}}[dt\wedge dr-ia\sin \theta dt\wedge
d\theta +a\sin ^{2}\theta dr\wedge d\varphi -i(r^{2}+a^{2})\sin \theta
d\theta \wedge d\varphi ] \\ 
\end{array}
\label{e-15b}
\end{equation}%
in oblate spheroidal coordinates. In cartesian coordinates its electric $%
\overrightarrow{E}$ and magnetic $\overrightarrow{B}$ fields have the form 
\begin{equation}
\begin{array}{l}
E^{1}=\frac{-exr^{5}}{4\pi (r^{4}+a^{2}z^{2})^{2}}\quad ,\quad E^{2}=\frac{%
-eyr^{5}}{4\pi (r^{4}+a^{2}z^{2})^{2}}\quad ,\quad E^{3}=\frac{%
-ezr^{3}(r^{2}-a^{2})}{4\pi (r^{4}+a^{2}z^{2})^{2}} \\ 
\\ 
B^{1}=\frac{eaxzr^{3}}{4\pi (r^{4}+a^{2}z^{2})^{2}}\quad ,\quad B^{2}=\frac{%
eayzr^{3}}{4\pi (r^{4}+a^{2}z^{2})^{2}}\quad ,\quad B^{3}=\frac{%
ear^{3}(r^{2}+z^{2})}{4\pi (r^{4}+a^{2}z^{2})^{2}} \\ 
\end{array}
\label{e-15c}
\end{equation}%
The singularities occur at the ring ($r,z$)=($0,0$). After a Poincar\'{e}
transformation, this local singularity moves with a constant velocity, as
expected from the solitonic origin of the configuration.

The LCR-structure relations (\ref{l-1}) imply the eikonal relations 
\begin{equation}
\begin{array}{l}
g^{\mu \nu }(\partial _{\mu }z^{\alpha })(\partial _{\nu }z^{\beta })=0\quad
,\quad J^{\mu \nu }(\partial _{\mu }z^{\alpha })(\partial _{\nu }z^{\beta
})=0 \\ 
\\ 
g^{\mu \nu }(\partial _{\mu }z^{\widetilde{\alpha }})(\partial _{\nu }z^{%
\widetilde{\beta }})=0\quad ,\quad J^{\mu \nu }(\partial _{\mu }z^{%
\widetilde{\alpha }})(\partial _{\nu }z^{\widetilde{\beta }})=0 \\ 
\end{array}
\label{e-15d}
\end{equation}%
which indicate wavefront singularities. These are singularities\cite{STR} of
a generalized function determined by the position and the direction of its
Fourier transform ($x;k$). A singular point $x$, in all the directions $k$,
will be called localizing singularity. Typical examples of such
singularities are the delta (Dirac) functions. The singular points $x$ with
precise cones of "bad" directions will be called quantum singularities.
These are essentially singularities implied by the characteristics of wave
equations on their solutions. The electromagnetic (and gravitational)
"dressing potential" singularities are localizing singularities, and we will
treat them as classical solutions determining the "particle". The
characteristics of the differential operators of the free photon and
electron are the quantum modes of photon and electron. The naif way to
consider the general solution of equation (\ref{e-14}) is 
\begin{equation}
\begin{array}{l}
A_{\mu }=A_{\mu }^{C}+A_{\mu }^{Q}\quad ,\quad j^{\mu }=j_{C}^{\mu }+e%
\overline{\psi }\gamma ^{\mu }\psi \\ 
\\ 
\partial _{\mu }F_{C}^{\mu \nu }=j_{C}^{\nu }\quad ,\quad \partial _{\mu
}F_{Q}^{\mu \nu }=e\overline{\psi }\gamma ^{\nu }\psi \\ 
\lbrack \gamma ^{\mu }(i\partial _{\mu }-eA_{\mu }^{C}(x))-m]\psi =e\gamma
^{\mu }A_{\mu }^{Q}\psi \\ 
\end{array}
\label{e-15e}
\end{equation}%
where the first term is the classical solitonic solution with the
localization singularity and the second is the quantum solution with the
wavefront singularity\cite{BOG1975}. The last fermionic equation is a
general self-consistent condition imposed by the current conservation
implied by (\ref{e-15}). Notice that the electromagnetic "dressing"
introduces a repulsive potential of order $e^{2}$. The gravitational
"dressing" may also enter with the classical electron tetrad, but I will
ignore it below.

From the mathematical point of view the classical $A_{\mu }^{C}$ contains
the localizing singular and regular part of the distributional solution and
the quantum part $A_{\mu }^{Q}$ contains the other wavefront singularities
implied by the principal symbol of the pseudo-differential operator (it
represents the wave-particle duality). The quantum part $A_{\mu }^{Q}$ of
the electromagnetic field interacts with the quantum part (wavefront
propagation) of the electron. The classical part $A_{\mu }^{C}$
(electromagnetic dressing of the electron) intervenes through the quantum
electron propagator. The Schwartz proper definition of generalized functions%
\cite{GELF1} lead to rigged Hilbert space\cite{GELF4} and the Hormander
formulation of the wavefront singularities\cite{STR}. Recall that the
positive and negative energy solutions of the free photon with two
polarizations $i=1,2$, 
\begin{equation}
\begin{array}{l}
A_{\mu }^{\pm }=\varepsilon _{\mu }(i,\overrightarrow{k})e^{\pm ikx} \\ 
k^{0}=|\overrightarrow{k}| \\ 
\end{array}
\label{e-15f}
\end{equation}%
are generalized functions with wave front $WF(A_{\mu }^{\pm })=[(t,%
\overrightarrow{x});(\pm |\overrightarrow{k}|,\overrightarrow{k}\neq 
\overrightarrow{0})]$. Hence quantum creation, propagation and annihilation
may be understood as the creation, propagation and annihilation of the
wavefront singularities. The Bogoliubov\cite{BOG1975} reformulation of the
Wightman axioms, using his microcausality relation, clarifies quantum field
theory. The S-matrix is properly defined as a series with coefficients
recursively computed, turning the renormalization problem to an appropriate
definition of the product of the implied distributions\cite{SCH1}. This
formalism is used in the next subsection to derive an effective quantum
electrodynamics, where the local singularity of the electron is incorporated
besides the conventional quantum singularities of the photon and electron.

\subsection{Derivation of quantum electrodynamics}

I will apply the Bogoliubov-Medvedev-Polivanov\cite{BOG1975} axiomatic
formulation of a quantum field theory, viewed as\ a method for the
construction of renormalizable effective quantum field theories. This method
has been extensively described in the Bogoliubov-Shirkov book\cite{BOG1980}.
It approaches the axiomatic formulation of a quantum field theory starting
from the S-matrix and the introduction of a "switching on and off" function $%
c(x)\in \lbrack 0,1]$ and assuming the following expansion of the S-matrix%
\begin{equation}
\begin{array}{l}
S=1+\underset{n\geq 1}{\sum }\frac{1}{n!}\int
S_{n}(x_{1},x_{2}...x_{n})c(x_{1})c(x_{2})...c(x_{n})[dx] \\ 
\end{array}
\label{e-17}
\end{equation}%
where $S_{n}(x_{1},x_{2}...x_{n})$ are generalized functions, which depend
on the complete free field functions (the local Poincar\'{e} representations
of the particles) and not its separate "positive" and "negative" frequency
parts. That is, the S-matrix is an operator valued functional in the Fock
space of free relativistic particles. Apparently this perturbative expansion
needs the existence of a small coupling constant. The imposed axioms are \ 
\begin{equation}
\begin{array}{l}
Poincar\acute{e}\ covariance:\quad
U_{P}S_{n}(x_{1},x_{2}...x_{n})U_{P}^{\dag }=S_{n}(Px_{1},Px_{2}...Px_{n})
\\ 
Unitarity:\quad SS^{\dag }=S^{\dag }S=1 \\ 
Microcausality:\quad \frac{\delta }{\delta c(x)}[\frac{\delta S(c)}{\delta
c(x)}S^{\dag }(c)]=0\quad for\quad x\precsim y \\ 
Correspondance\ principle:\quad S_{1}(x)=iL_{int}[\phi (x)] \\ 
\end{array}
\label{e-18}
\end{equation}%
where $\phi (x)$ denotes the free particle fields and $x\precsim y$ means $%
x^{0}<y^{0}$ or $(x-y)^{2}<0$. A general solution of these conditions is \ 
\begin{equation}
\begin{array}{l}
S=T[\exp (i\mathbf{L}[\phi (x);c(x))] \\ 
\\ 
\mathbf{L}[\phi (x);c(x)]=L_{Int}[\phi (x)]c(x)+\underset{n\geq 1}{\sum }%
\frac{1}{n}\int \Lambda _{n+1}(x,x_{1}...x_{n})c(x)c(x_{1})...c(x_{n})[dx]
\\ 
\end{array}
\label{e-19}
\end{equation}%
where $\Lambda _{n}(x,x_{1}...x_{n})$ are quasilocal quantities (arbitrary
add-ons of generalized functions\cite{GELF1}), which permit the
renormalization process. This order by order construction of a finite
S-matrix (with possibly infinite hamiltonian and lagrangian) provides a well
established algorithm to distinguish renormalizable with non-renormalizable
interaction lagrangians\cite{BOG1980}. The mathematical origin of
renormalization is a non-permitted multiplication of time step functions
with other distributions which appear in the initial form of the action.
Epstein-Glaser showed\cite{SCH1} that the recursive procedure does not
essentially need these non well defined multiplications.

The formalism is based on the well defined rigged Hilbert-Fock space of the
free quantum field representations of the Poincar\'{e} group. The advantage
of the Bogoliubov procedure is that it can be used in the opposite sense.
Knowing the (free) Poincar\'{e} representations, they are identified with
"free particles" with precise mass and spin. Then they are described with
the corresponding free fields, which are used to write down an effective
interaction lagrangian, suggested by the fundamental dynamics. In the
present case, the fundamental dynamics is the PCFT and the particles are the
solitonic solutions and their corresponding potentials which satisfy the
wave equations. The suggested interaction takes the place of the
"correspondence principle" in the Bogoliubov procedure. In the present case
of effective electrodynamics, the suggested interaction is 
\begin{equation}
\begin{array}{l}
L_{EM}=e\overline{\psi }\gamma ^{\mu }\psi A_{\mu } \\ 
\end{array}
\label{e-19a}
\end{equation}%
where $\psi $ is the Dirac field and $A_{\mu }$ is the quantum
electromagnetic field with its propagator implied by (\ref{e-15e}). Notice
that in the derived quantum electrodynamics the electron field is not
exactly the free Bogoliubov field. It continues to be a representation of
the Poincar\'{e} group, but incorporates the electromagnetic dressing of the
electron. The order by order computation introduces counterterms to the
action (with up to first order derivatives). If the number of the forms of
the counterterms is finite, the action is renormalizable and the model is
considered compatible with quantum mechanics, otherwise the whole
construction is rejected as inapplicable. The great value of this
constructive procedure will appear in its application for the construction
of the effective action of the standard model. The perturbative dependence
of the S-matrix on the tempered distributions of the free fields permits the
application of nilpotent Q-charge of Scharf and collaborators\cite{SCH2},
which assures the elimination of negative norm states and the
renormalization of the action.

In the Bogoliubov procedure we do not need all the interactions from the
beginning. The order by order (perturbative) calculation of the S-matrix,
permits the incorporation of all the "needed" additional lagrangian
interactions imposed by the emerging counterterms. The restriction is that
the final implied order-by-order lagrangian must have a finite number of
terms without higher order derivatives, which are the conditions of
renormalizability and compatibility with quantum mechanics. The effective
quantum electrodynamics, derived from the classical photon-electron current
interaction (correspondence principle), does not need additional terms. But
in its extension with gravitational and (some) weak interaction terms,
additional terms and conditions between the masses and the coupling
constants will be needed for the interaction lagrangian to become
self-consistent (renormalizable).

The perturbative approach permits the definition of general dynamical
variables through the generating functional introduced considering the
formal existence of a "classical" current $J(x)$ for every field $\phi (x)$
of the action. The generating functional $Z_{0}(J)$ and the connected
generating functional are

\begin{equation}
\begin{array}{l}
Z_{0}(J)=<0|T[\exp \{i\tint (L_{I}(x)+\phi (x)J(x))d^{4}x\}]|0> \\ 
\\ 
Z_{c}(J)=-i\ln [Z_{0}(J)]%
\end{array}
\label{e-20}
\end{equation}%
Any field $\phi (x)$ defines a generating field $\Phi (x;J)$\ and the
Legendre transformation

\begin{equation}
\begin{array}{l}
\Phi (x;J)=\frac{\delta Z_{c}(J)}{\delta J(x)} \\ 
\\ 
Z_{c}(J)\rightarrow W(\Phi )=Z_{c}(J)-\tint \Phi (x;J)J(x)d^{4}x%
\end{array}
\label{e-21}
\end{equation}%
In the context of the Bogoliubov-Shirkov notation\cite{BOG1980}

\begin{equation}
\begin{array}{l}
\Phi (x;g)=-\frac{\delta H(x;g)}{\delta J(x)}=\frac{-i}{g(x)}(\frac{\delta S%
}{\delta J(x)}\overset{\ast }{S})|_{J=0} \\ 
\\ 
H(x;g):=i(\frac{\delta S(g)}{\delta g(x)}\overset{\ast }{S}(g))%
\end{array}
\label{e-22}
\end{equation}%
where $H(x;g)$ is the "quantum" hamiltonian of the system. The expected
relation of a "dressing" potential of the elementary particles in PCFT and
the above formalism is

\begin{equation}
\begin{array}{l}
A_{1}(x;1)=-\frac{\delta E(J)}{\delta J(x)}|_{J=0}=\frac{-i}{<S>}\overset{%
\ast }{\Phi }_{1}(\frac{\delta S}{\delta J(x)}\overset{\ast }{S})\Phi
_{1}|_{J=0} \\ 
\\ 
\Phi _{1}=(2\pi )^{\frac{3}{2}}a_{\nu }^{+}(\overrightarrow{k})\Phi _{0}%
\end{array}
\label{e-23}
\end{equation}%
where $\Phi _{1}$ is the one-electron state. Notice that the elementary
particle has the same initial and final energies and their creation and
annihilation operators are outside the time ordering. The physical intuition
is that we use the classical current $J(x)$ as a sensor of the potential
generated by a particle. The relativistic field equations are also derived
from the causal perturbative approach and all the experimental results are
properly computed. Hence (\ref{e-23}) is going to provide precise
self-consistency conditions between PCFT and current quantum field theories,
which we will describe below.

The first term of the effective electron potential in conventional quantum
electrodynamics is

\begin{equation}
\begin{array}{l}
A_{1_{\mu }}(x;1)\simeq \frac{-i}{2}\overset{\ast }{\Phi }_{1}\frac{\delta 
\widehat{S}_{2}(J)}{\delta J^{\mu }(x)}\Phi _{1}|_{J^{\mu }=0} \\ 
\\ 
\widehat{S}_{2}(J)=\tint T((L_{I}(x_{1})+A_{\nu }(x_{1})J^{\nu
}(x_{1}))(L_{I}(x_{2})+A_{\nu }(x_{2})J^{\nu }(x_{2}))[dx]%
\end{array}
\label{e-24}
\end{equation}%
which becomes

\begin{equation}
\begin{array}{l}
A_{1}^{\mu }(x)\simeq -e\tint D_{0}^{c}(x-y)\overset{\ast }{\Phi }%
_{1p^{\prime }}:\overline{\psi _{e}}(y)\gamma ^{\mu }\psi _{e}(y):\Phi
_{1p}d^{4}y \\ 
\\ 
\Phi _{1p}=(2\pi )^{\frac{3}{2}}\overset{\ast }{a_{\nu }^{+}}(%
\overrightarrow{p})\Phi _{0}%
\end{array}
\label{e-25}
\end{equation}

The electromagnetic dressing of the electron LCR-manifold in cartesian
coordinates\ 
\begin{equation}
\begin{array}{l}
A=\frac{qr^{3}}{4\pi (r^{4}+a^{2}(x^{3})^{2})}(dx^{0}-\frac{rx^{1}-ax^{2}}{%
r^{2}+a^{2}}dx^{1}-\frac{rx^{2}+ax^{1}}{r^{2}+a^{2}}dx^{2}-\frac{x^{3}}{r}%
dx^{3})= \\ 
\qquad =\frac{qr^{3}}{4\pi (r^{4}+a^{2}(x^{3})^{2})}\ell _{\mu }dx^{\mu } \\ 
\\ 
r^{4}-[(x^{1})^{2}+(x^{2})^{2}+(x^{3})^{2}-a^{2}]r^{2}-a^{2}(x^{3})^{2}=0%
\end{array}
\label{e-26}
\end{equation}%
is proportional to $\ell _{\mu }$ and has the proper asymptotic charge $e$
and magnetic moment $ea$, already computed by Carter without any reference
to quantum electrodynamics. Besides, all its components are locally
integrable functions determining through derivations the "ladder" of the
generalized functions (electric field, magnetic field, etc) . Hence it
strongly suggests to relate this form with the sum of all the orders of the
effective potential of quantum electrodynamics. But (\ref{e-26}) is singular
at the ring with radius $a$, while the perturbative terms (\ref{e-25}) are
singular at the point $\overrightarrow{x}=0$, which emerge after an
expansion of (\ref{e-26}) and the definition of $r$ in powers of $a=\frac{%
\hbar }{2m}$. The emergence of the Planck constant $\hbar $ strongly
indicates\ that (\ref{e-26}) includes the contributions of loop diagrams.

Recall that the Kerr-Newman metric (with the electromagnetic potential)
satisfy the Einstein field equations with

\begin{equation}
\begin{array}{l}
h_{\mu \nu }=2f(x)\ell _{\mu }\ell _{\nu } \\ 
\\ 
g_{\mu \nu }=\eta _{\mu \nu }+\kappa h_{\mu \nu }%
\end{array}
\label{e-26a}
\end{equation}
Therefore causal perturbative approach has to start with gravity too with
initial interaction

\begin{equation}
\begin{array}{l}
L_{I}=\frac{k}{2}h^{\mu \nu }:(\overline{\psi _{e}}\gamma _{\mu }\partial
_{\nu }\psi _{e}-(\partial _{\nu }\overline{\psi _{e}})\gamma _{\mu }\psi
_{e}): \\ 
\end{array}
\label{e-26b}
\end{equation}%
and possibly the first order gravity-gravity interaction, which has already
computed by Scharf and his collaborators\cite{SCH2}, using the nilpotent $Q$
gauge charge method. Their observation that the computed first terms
coincide with the expansion of the Einstein-Hilbert action should be
expected, because the other gravitational scalars contain second order
derivatives, which imply negative norm particle states, removed by their
method.

The causal perturbative approach of quantum field theory provides the
transition amplitudes between the free elementary particles (the stable
asymptotic LCR-manifolds), but it is practically impossible to sum up all
the terms. That is, quantum field theory cannot compute the geometric ring
singularity of the elementary LCR-manifolds, which determines the particles
themselves and the geometry of the background $%
%TCIMACRO{\U{211d} }%
%BeginExpansion
\mathbb{R}
%EndExpansion
\times S^{3}$ universe. That is, the expansion of $A$ and $r$ of (\ref{e-26}%
) hides the ring singularity of the global geometric solution.This
singularity permits us to bypass the Hawking-Penrose singularity theorems
for lorentzian riemannian manifolds, as described in the following
subsection.

\subsection{LCR-ray tracing in the electron LCR-manifold}

The electron mass $M_{e}$, charge $e^{2}$ and spin parameter $a$ have the
values%
\begin{equation}
\begin{array}{l}
M_{e}=4.2\ast 10^{-23} \\ 
e^{2}=\frac{q^{2}}{4\pi \varepsilon _{0}\hbar c}=\frac{1}{137} \\ 
a=\frac{\hbar }{2M_{e}}=2.1\ast 10^{23} \\ 
\\ 
a^{2}>>e^{2}>>M_{e}^{2}%
\end{array}
\label{e-27}
\end{equation}%
in dimensionless units $c=G=\hbar =1$. Hence $a^{2}+e^{2}-M_{e}^{2}>0,$
which implies that its Kerr-Newman metric has a naked essential singularity.
Because of this singularity the Kerr-Newman metric cannot be related with
the electron despite the extraordinary fact of fermionic gyromagnetic ratio $%
g=2$. The purpose of this subsection is to show that the LCR-manifold is
well defined, permitting its identification with the electron.

The static electron is identified with the static axially symmetric
LCR-structure determined with the linear trajectory $\xi ^{a}=(\tau ,0,0,ia)$%
. That is, we have%
\begin{equation}
\begin{array}{l}
X^{mi}=%
\begin{pmatrix}
1 & -z^{\widetilde{1}} \\ 
z^{1} & 1 \\ 
-i(z^{0}-ia) & i(z^{\widetilde{0}}-ia)z^{\widetilde{1}} \\ 
-i(z^{0}+ia)z^{1} & -i(z^{\widetilde{0}}+ia)%
\end{pmatrix}
\\ 
\end{array}
\label{e-28}
\end{equation}%
where ($z^{\alpha };z^{\widetilde{\beta }}$) are now the structure
coordinates. Here I will first derive the "flat" LCR-structure (defined by $%
X^{\dag }E_{U}X=0$) and after I will make a "Kerr-Schild" ansatz adapted to
the LCR-tetrad to finally refind the axially symmetric LCR-structure, which
is identified with the electron. I think this approach will make general
relativists more confident to the final picture of the electron as a
gaussian beam (in the optics terminology) in $U(2)$ spacetime.

This procedure implies first the "flat" LCR-structure coordinates%
\begin{equation}
\begin{array}{l}
z^{0}=t-r+ia\cos \theta \quad ,\quad z^{1}=e^{i\varphi }\tan \frac{\theta }{2%
} \\ 
z^{\widetilde{0}}=t+r-ia\cos \theta \quad ,\quad z^{\widetilde{1}}=\frac{r+ia%
}{r-ia}e^{-i\varphi }\tan \frac{\theta }{2} \\ 
\end{array}
\label{e-29}
\end{equation}%
from which we find the tetrad compatible with the Minkowski metric%
\begin{equation}
\begin{array}{l}
L_{\mu }dx^{\mu }=\Lambda \lbrack dt-dr-a\sin ^{2}\theta d\varphi ] \\ 
N_{\mu }dx^{\mu }=N[dt+\frac{r^{2}+2a^{2}\cos ^{2}\theta -a^{2}}{r^{2}+a^{2}}%
dr-a\sin ^{2}\theta \ d\varphi ] \\ 
M_{\mu }dx^{\mu }=M[-ia\sin \theta \ (dt-dr)+(r^{2}+a^{2}\cos ^{2}\theta
)d\theta + \\ 
\qquad \qquad +i\sin \theta (r^{2}+a^{2})d\varphi ] \\ 
\end{array}
\label{e-30}
\end{equation}%
where the tetrad-Weyl factors are not determined as expected. They are
determined by simply imposing that the tetrad gives the Minkowski metric.
But for that, we have to find first the relation of the cartesian
coordinates with the present convenient "asymmetric" coordinates ($%
t,r,\theta ,\varphi $), which are not the same\cite{CHAND} with the
"symmetric" ones.

The general relation between the projective coordinates and the homogeneous
coordinates of $G_{4,2}$ is found by simply inverting their definition
formula (\ref{l-10}). We finally find%
\begin{equation}
\begin{array}{l}
r^{0}=i\frac{(X^{01}X^{32}-X^{31}X^{02})+(X^{21}X^{12}-X^{11}X^{22})}{%
2(X^{01}X^{12}-X^{11}X^{02})} \\ 
r^{1}=i\frac{(X^{11}X^{32}-X^{31}X^{12})+(X^{21}X^{02}-X^{01}X^{22})}{%
2(X^{01}X^{12}-X^{11}X^{02})} \\ 
r^{2}=\frac{(X^{11}X^{32}-X^{31}X^{12})-(X^{21}X^{02}-X^{01}X^{22})}{%
2(X^{01}X^{12}-X^{11}X^{02})} \\ 
r^{3}=i\frac{(X^{01}X^{32}-X^{31}X^{02})-(X^{21}X^{12}-X^{11}X^{22})}{%
2(X^{01}X^{12}-X^{11}X^{02})}%
\end{array}
\label{e-31}
\end{equation}%
We already know that the imaginary part of $r^{b}=x^{b}+iy^{b}$ determines
the gravitational "dressing", because the algebraic "flatness" condition
implies $y^{b}=0$. The Minkowski coordinates $x^{b}$ are related with the
"asymmetric" ($t,r,\theta ,\varphi $) via the relation%
\begin{equation}
\begin{array}{l}
x^{0}=t \\ 
x^{1}+ix^{2}=(r-ia)\sin \theta e^{i\varphi } \\ 
x^{3}=r\cos \theta \\ 
\\ 
r^{4}-[(x^{1})^{2}+(x^{2})^{2}+(x^{3})^{2}-a^{2}]r^{2}-a^{2}(x^{3})^{2}=0 \\ 
\cos \theta =\frac{x^{3}}{r}\quad ,\quad \sin \theta =\sqrt{\frac{%
(x^{1})^{2}+(x^{2})^{2}}{r^{2}+a^{2}}}%
\end{array}
\label{e-32}
\end{equation}%
with the following diffeomorphic relations%
\begin{equation}
\begin{array}{l}
dx^{0}=dt \\ 
dx^{1}=\sin \theta \cos \varphi dr+\cos \theta (r\cos \varphi +a\sin \varphi
)d\theta -\sin \theta (r\sin \varphi -a\cos \varphi )d\varphi \\ 
dx^{2}=\sin \theta \sin \varphi dr+\cos \theta (r\sin \varphi -a\cos \varphi
)d\theta +\sin \theta (r\cos \varphi +a\sin \varphi )d\varphi \\ 
dx^{3}=\cos \theta dr-r\sin \theta d\theta \\ 
\end{array}
\label{e-33}
\end{equation}%
Their inversion implies%
\begin{equation}
\begin{array}{l}
dt=dx^{0} \\ 
dr=\frac{rx^{1}-ax^{2}}{r^{2}+a^{2}}dx^{1}+\frac{ax^{1}+rx^{2}}{r^{2}+a^{2}}%
dx^{2}+\frac{x^{3}}{r}dx^{3} \\ 
d\theta =\frac{x^{3}(rx^{1}-ax^{2})}{r^{2}\sqrt{%
(r^{2}+a^{2})((x^{1})^{2}+(x^{2})^{2})}}dx^{1}+\frac{x^{3}(ax^{1}+rx^{2})}{%
r^{2}\sqrt{(r^{2}+a^{2})((x^{1})^{2}+(x^{2})^{2})}}dx^{2}-\frac{\sqrt{%
(x^{1})^{2}+(x^{2})^{2}}}{r\sqrt{r^{2}+a^{2}}}dx^{3} \\ 
d\varphi =-\frac{ax^{1}+rx^{2}}{r((x^{1})^{2}+(x^{2})^{2})}dx^{1}+\frac{%
rx^{1}-ax^{2}}{r((x^{1})^{2}+(x^{2})^{2})}dx^{2}%
\end{array}
\label{e-34}
\end{equation}

Hence, we finally find that the conventional tetrad corresponding to the
Minkowski metric is 
\begin{equation}
\begin{array}{l}
L_{\mu }dx^{\mu }=[dt-dr-a\sin ^{2}\theta d\varphi ] \\ 
N_{\mu }dx^{\mu }=\frac{r^{2}+a^{2}}{2(r^{2}+a^{2}\cos ^{2}\theta )}[dt+%
\frac{r^{2}+2a^{2}\cos ^{2}\theta -a^{2}}{r^{2}+a^{2}}dr-a\sin ^{2}\theta \
d\varphi ] \\ 
M_{\mu }dx^{\mu }=\frac{-1}{\sqrt{2}(r+ia\cos \theta )}[-ia\sin \theta \
(dt-dr)+(r^{2}+a^{2}\cos ^{2}\theta )d\theta + \\ 
\qquad \qquad +i\sin \theta (r^{2}+a^{2})d\varphi ] \\ 
\end{array}
\label{e-35}
\end{equation}

The general tetrad is found with the "Kerr-Schild" ansatz adapted to the
LCR-structure formalism%
\begin{equation}
\begin{array}{l}
\ell _{\mu }=L_{\mu }\quad ,\quad m_{\mu }=M_{\mu }\quad ,\quad n_{\mu
}=N_{\mu }+\frac{h(r)}{2(r^{2}+a^{2}\cos ^{2}\theta )}\ L_{\mu } \\ 
\end{array}
\label{e-36}
\end{equation}%
I want to point out that we find the same static LCR-structure looking for
LCR-structures admitting time translation and axisymmetric symmetries.

With the above definition of the coordinates ($t,r,\theta ,\varphi $), the
structure coordinates have the form

\begin{equation}
\begin{array}{l}
z^{0}=t-r+ia\cos \theta \quad ,\quad z^{1}=e^{i\varphi }\tan \frac{\theta }{2%
} \\ 
\\ 
z^{\widetilde{0}}=t+r-ia\cos \theta -2f_{1}\quad ,\quad z^{\widetilde{1}}=%
\frac{r+ia}{r-ia}\ e^{2iaf_{2}}\ e^{-i\varphi }\tan \frac{\theta }{2}%
\end{array}
\label{e-37}
\end{equation}%
where the two new functions are%
\begin{equation}
\begin{array}{l}
f_{1}(r)=\int \frac{h}{r^{2}+a^{2}+h}\ dr\quad ,\quad f_{2}(r)=\int \frac{h}{%
(r^{2}+a^{2}+h)(r^{2}+a^{2})}\ dr \\ 
\end{array}
\label{e-38}
\end{equation}%
The Newman-Penrose spin coefficients are found to be%
\begin{equation}
\begin{tabular}{|l|}
\hline
$\alpha =\frac{ia(1+\sin ^{2}\theta )-r\cos \theta }{2\sqrt{2}\sin \theta \
(r-ia\cos \theta )^{2}}\quad ,\quad \beta =\frac{\cos \theta }{2\sqrt{2}\sin
\theta \ (r+ia\cos \theta )}$ \\ \hline
$\gamma =-\frac{a^{2}+iar\cos \theta +h}{2\rho ^{2}\ (r-ia\cos \theta )}+%
\frac{h^{\prime }}{4\rho ^{2}}\quad ,\quad \varepsilon =0$ \\ \hline
$\mu =-\frac{r^{2}+a^{2}+h}{2\rho ^{2}\ (r-ia\cos \theta )}\quad ,\quad \pi =%
\frac{ia\sin \theta }{\sqrt{2}(r-ia\cos \theta )^{2}}$ \\ \hline
$\rho =-\frac{1}{r-ia\cos \theta }\quad ,\quad \tau =-\frac{ia\sin \theta }{%
\sqrt{2}\rho ^{2}}$ \\ \hline
$\kappa =0\quad ,\quad \sigma =0\quad ,\quad \nu =0\quad ,\quad \lambda =0$
\\ \hline
\end{tabular}
\label{e-39}
\end{equation}%
which will be useful for our computations. Recall that the Kerr-Newman
spacetime has $h(r)=-2Mr+e^{2}$. In this case the integrals are%
\begin{equation}
\begin{array}{l}
f_{1}(r)=\int \frac{-2Mr+e^{2}}{r^{2}+a^{2}-2Mr+e^{2}}\ dr=-M\ln \frac{%
|\Delta |}{r1}+\frac{2M^{2}-e^{2}}{\Theta }\arctan \frac{\Theta }{r-M} \\ 
f_{2}(r)=\int \frac{-2Mr+e^{2}}{(r^{2}+a^{2}-2Mr+e^{2})(r^{2}+a^{2})}\ dr=%
\frac{1}{2ia}\ln [r_{2}\frac{r-ia}{r+ia}(\frac{r-M+i\Theta }{r-M-i\Theta })^{%
\frac{a}{\Theta }}] \\ 
\\ 
\Delta :=r^{2}+a^{2}-2Mr+e^{2}\quad ,\quad \Theta :=\sqrt{a^{2}+e^{2}-M^{2}}%
\end{array}
\label{e-40}
\end{equation}%
where $r1$ and $r2$\ are normalization constants, and the structure
coordinates of the "Kerr-Newman" LCR-manifold are

\begin{equation}
\begin{array}{l}
z^{0}=t-r+ia\cos \theta \quad ,\quad z^{1}=e^{i\varphi }\tan \frac{\theta }{2%
} \\ 
z^{\widetilde{0}}=t+r-ia\cos \theta +2M\ln \frac{|\Delta |}{r_{1}}+\frac{%
2(e^{2}-2M^{2})}{\Theta }\arctan \frac{\Theta }{r-M} \\ 
z^{\widetilde{1}}=r_{2}(\frac{r-M+i\Theta }{r-M-i\Theta })^{\frac{a}{\Theta }%
}\ e^{-i\varphi }\tan \frac{\theta }{2}%
\end{array}
\label{e-41}
\end{equation}%
Notice the singularities in the ambient complex manifold at the two complex
values of $r=M\pm i\Theta $. It is well known to general relativists that
this choice of tetrad-Weyl factors preserve the electromagnetic current and
the energy-momentum and angular momentum currents. Hence, fixing the factors
of the LCR-tetrad (to achieve conservation of the currents) implies a
breaking of the tetrad-Weyl symmetry.

Notice that the electron LCR-structure coordinates (\ref{e-41}) of the
embedding of the LCR-manifold in the ambient complex manifold may be viewed
as a \textbf{anti-}meromorphic deformation

\begin{equation}
\begin{array}{l}
z^{\widetilde{\beta }}=f^{\widetilde{\beta }}(\overline{z^{\alpha }};r) \\ 
\\ 
z^{\widetilde{0}}=\overline{z^{0}}+2(r-f_{1}) \\ 
z^{\widetilde{1}}=r_{2}\overline{z^{1}}(\frac{r-M+i\Theta }{r-M-i\Theta })^{%
\frac{a}{\Theta }}%
\end{array}
\label{e-42}
\end{equation}%
where the deformation parameter is the real variable $r$.

The static axially symmetric LCR-structure (identified with the electron) is
stable, because all its relative invariants 
\begin{equation}
\begin{array}{l}
\Phi _{1}=\frac{\rho -\overline{\rho }}{i}=\frac{-2a\cos \theta }{%
r^{2}+a^{2}\cos ^{2}\theta } \\ 
\Phi _{2}=\frac{\mu -\overline{\mu }}{i}=-\frac{(r^{2}+a^{2}+h)a\cos \theta 
}{(r^{2}+a^{2}\cos ^{2}\theta )^{2}} \\ 
\Phi _{3}=-(\tau +\overline{\pi })=\frac{2iar\sin \theta }{\sqrt{2}(r+ia\cos
\theta )^{2}(r-ia\cos \theta )} \\ 
\end{array}
\label{e-43}
\end{equation}%
do not vanish.

The $L^{\mu }\partial _{\mu }z^{\alpha }=0$ annihilation implies that the
outgoing integral curves (rays) are determined by the surfaces%
\begin{equation}
\begin{array}{l}
s_{1}:=t-r\quad ,\quad s_{2}:=\theta \quad ,\quad s_{3}:=\varphi \\ 
\end{array}
\label{e-44}
\end{equation}%
We use the caustic coordinates ($r,s_{1},s_{2},s_{3}$), which have the
property ($0,s_{1},\frac{\pi }{2},s_{3}$) to be on the caustic. In this
caustic coordinate system the LCR-rays are traced by the relation%
\begin{equation}
\begin{array}{l}
x_{L}^{0}(r)=s_{1}+r \\ 
x_{L}^{1}(r)=(r\cos \varphi +a\sin \varphi )\sin \theta \\ 
x_{L}^{2}(r)=(r\sin \varphi -a\cos \varphi )\sin \theta \\ 
x_{L}^{3}(r)=r\cos \theta \\ 
\\ 
Jacobian=[r^{2}+a^{2}\cos ^{2}\theta ]\sin \theta%
\end{array}
\label{e-45}
\end{equation}%
The source of the LCR-rays are at $r=0$, i.e. the disk%
\begin{equation}
\begin{array}{l}
x_{L}^{0}(0)=s_{1} \\ 
x_{L}^{1}(0)=a\sin \varphi \sin \theta \\ 
x_{L}^{2}(0)=-a\cos \varphi \sin \theta \\ 
x_{L}^{3}(0)=0%
\end{array}
\label{e-46}
\end{equation}

The $N^{\mu }\partial _{\mu }z^{\widetilde{\alpha }}=0$ annihilation implies
that its incoming rays are determined by the surfaces%
\begin{equation}
\begin{array}{l}
s_{1}^{\prime }:=t+r\quad ,\quad s_{2}^{\prime }:=\theta \quad ,\quad
s_{3}^{\prime }:=\varphi +\arctan \frac{2ar}{a^{2}-r^{2}} \\ 
\end{array}
\label{e-47}
\end{equation}%
Then we find the congruence%
\begin{equation}
\begin{array}{l}
x_{N}^{0}(r)=s_{1}^{\prime }-r \\ 
x_{N}^{1}(r)=[r\cos s_{3}^{\prime }-a\sin s_{3}^{\prime }]\sin \theta \\ 
x_{N}^{2}(r)=[r\sin s_{3}^{\prime }+a\cos s_{3}^{\prime }]\sin \theta \\ 
x_{N}^{3}(r)=r\cos \theta \\ 
\\ 
Jacobian=[r^{2}+a^{2}\cos ^{2}\theta ]\sin \theta%
\end{array}
\label{e-48}
\end{equation}%
As expected the velocities $\overset{.}{x}_{L}^{i}(t)$ and $\overset{.}{x}%
_{N}^{i}(t)$ have asymptotically opposite radial directions.

We will now show that the origin of the essential singularity of the Kerr
manifold is the intersection of the two sheets of the static electron
regular quadric (in the unbounded Siegel realization)%
\begin{equation}
\begin{array}{l}
X^{1}X^{2}-X^{0}X^{3}+2aX^{0}X^{1}=0 \\ 
\end{array}
\label{e-49}
\end{equation}%
of $CP^{3}$. In the flatprint case we have 
\begin{equation}
\begin{array}{l}
X^{0}=1\quad ,\quad X^{1}=\lambda \quad ,\quad
X^{2}=-i[(x^{0}-x^{3})-(x^{1}-ix^{2})\lambda ] \\ 
X^{3}=-i[-(x^{1}+ix^{2})+(x^{0}+x^{3})\lambda ] \\ 
\end{array}
\label{e-50}
\end{equation}%
and the Kerr polynomial and its two solutions are 
\begin{equation}
\begin{array}{l}
(x^{1}-ix^{2})\lambda ^{2}+2(x^{3}-ia)\lambda -(x^{1}+ix^{2})=0 \\ 
\lambda _{1,2}=\frac{-(x^{3}-ia)\pm \sqrt{\Delta }}{x-iy}\quad ,\quad \Delta
=(x^{1})^{2}+(x^{2})^{2}+(x^{3})^{2}-a^{2}-2iax^{3} \\ 
\end{array}
\label{e-51}
\end{equation}%
where $\lambda _{1,2}$\ are the two values of $\lambda $ on the two sheets
of the quadric. The intersection curve of these two sheets is 
\begin{equation}
\begin{array}{l}
\Delta =(x^{1})^{2}+(x^{2})^{2}+(x^{3})^{2}-a^{2}-2iax^{3}=0 \\ 
\\ 
x^{3}=0\quad ,\quad (x^{1})^{2}+(x^{2})^{2}=a^{2}%
\end{array}
\label{e-52}
\end{equation}%
which, after the LCR projection to $%
%TCIMACRO{\U{211d} }%
%BeginExpansion
\mathbb{R}
%EndExpansion
^{4}$, becomes the singularity ring of the electron (Kerr-Newman) manifold.
Notice that the quadratic surface is regular and the intersection of the two
branches is implied by the projection. The points of the algebraic
intersection curve (the branch curve) of the (regular) quadric of $CP^{3}$
are regular points like any other point of the quadric.

We have already pointed out that the entire LCR-manifold (universe) is the
spinorial $U(2)$ manifold which needs more than two $%
%TCIMACRO{\U{211d} }%
%BeginExpansion
\mathbb{R}
%EndExpansion
^{4}$ charts. Recall that the one $%
%TCIMACRO{\U{211d} }%
%BeginExpansion
\mathbb{R}
%EndExpansion
^{4}$ chart of the $U(2)\rightarrow 
%TCIMACRO{\U{211d} }%
%BeginExpansion
\mathbb{R}
%EndExpansion
^{4}$ Cayley $2\rightarrow 1$ transformation is (\ref{g-3d}) $x_{+}^{\mu }$
for $s>0$ and the second (non-intersecting chart) is (\ref{g-3e}) $%
x_{-}^{\mu }$ for $s<0$, where the affine parameter is related with%
\begin{equation}
\begin{array}{l}
r=+\left\{ \frac{s^{2}-a^{2}}{2}+\sqrt{[\frac{s^{2}-a^{2}}{2}%
]^{2}+a^{2}(x^{3})^{2}}\right\} ^{\frac{1}{2}}\ for\ s>0 \\ 
r=-\left\{ \frac{s^{2}-a^{2}}{2}+\sqrt{[\frac{s^{2}-a^{2}}{2}%
]^{2}+a^{2}(x^{3})^{2}}\right\} ^{\frac{1}{2}}\ for\ s<0%
\end{array}
\label{e-53}
\end{equation}%
Notice that in the identified region (the disc for both charts) $r=0$ in
both charts. That is, $r=0$ occurs at $x^{3}=0$ and $s^{2}\leq a^{2}$ for
both charts $s\gtrless 0$.

The two LCR-congruences $L^{\mu }=\frac{dx_{L}^{\mu }}{dr}$ and $N^{\mu }=%
\frac{dx_{N}^{\mu }}{dr}$\ of the flatprint electron LCR-manifold can be
easily implied from my calculations of the previous section. The starting
idea is that the structure coordinates $z^{\alpha }(x)$ provide the three
invariants ($s_{1},s_{2},s_{3}$) along the $L$-ray $x_{L}^{\mu }(r)$, and
the structure coordinates $z^{\widetilde{\alpha }}(x)$ provide the
invariants ($s_{1}^{\prime },s_{2}^{\prime },s_{3}^{\prime }$), which\ label
the $N$-ray $x_{N}^{\mu }(r)$. Hence we simply have the same forms, but we
let $r\in (-\infty ,+\infty )$ and at $r=0$ we pass to the second $%
x_{L}^{\prime \mu }(r),x_{N}^{\prime \mu }(r)\in 
%TCIMACRO{\U{211d} }%
%BeginExpansion
\mathbb{R}
%EndExpansion
^{4}$ sheet.

The second way is tracing the rays $w_{L,N}(r;s_{1},s_{2},s_{3})\in U(2)$ in
the complete bounded universe $U(2)$ taking $r\in (-\infty ,+\infty )$ as
the parameter indicating the ray points. From the relation 
\begin{equation}
\begin{array}{l}
Y^{0}=\frac{1}{\sqrt{2}}(X^{0}+X^{2})\quad ,\quad Y^{1}=\frac{1}{\sqrt{2}}%
(X^{1}+X^{3}) \\ 
\\ 
Y^{2}=\frac{1}{\sqrt{2}}(X^{0}-X^{2})\quad ,\quad Y^{3}=\frac{1}{\sqrt{2}}%
(X^{1}-X^{3}) \\ 
\end{array}
\label{e-54}
\end{equation}%
between the bounded $Y^{ni}$ and unbounded $X^{ni}$ homogeneous coordinates
and (\ref{e-28}) we find%
\begin{equation}
\begin{array}{l}
Y^{mi}=\frac{1}{\sqrt{2}}%
\begin{pmatrix}
1-i(z^{0}-ia) & (-1+i(z^{\widetilde{0}}-ia))z^{\widetilde{1}} \\ 
(1-i(z^{0}+ia))z^{1} & 1-i(z^{\widetilde{0}}+ia) \\ 
1+i(z^{0}-ia) & -(1+i(z^{\widetilde{0}}-ia))z^{\widetilde{1}} \\ 
(1+i(z^{0}+ia))z^{1} & 1+i(z^{\widetilde{0}}+ia)%
\end{pmatrix}
\\ 
\end{array}
\label{e-55}
\end{equation}%
Like previously, we use the relations (\ref{e-29}) to find the labels (\ref%
{e-44}) of $L^{\mu }$ rays ($r;s_{1},s_{2},s_{3}$), assuming the $r$
parameter to indicate the points of one ray. The coordinates $z^{\alpha }$
do not depend on $r$, remain invariant along the rays, therefore I keep them
unchanged. Then we express only $z^{\widetilde{\alpha }}$ as functions of
the proper $L^{\mu }$ ray coordinates ($r;s_{1},s_{2},s_{3}$)%
\begin{equation}
\begin{array}{l}
z^{\widetilde{0}}=s_{1}+2r-ia\cos s_{2}\quad ,\quad z^{\widetilde{1}}=\frac{%
r+ia}{r-ia}e^{-is_{3}}\tan \frac{s_{2}}{2} \\ 
\end{array}
\label{e-56}
\end{equation}%
and we find the rays in homogeneous coordinates $Y^{mi}(r;s_{1},s_{2},s_{3})$%
.

In the context of the quadratic $CP^{3}$ hypersurface, along the $L^{\mu }$
integral curves, the one intersection point with the line is preserved
constant and changes the second. For the $N^{\mu }$ integral curves the role
of the intersection points are interchanged. Now using the relation 
\begin{equation}
\begin{array}{l}
w_{11}=\frac{Y^{21}Y^{12}-Y^{11}Y^{22}}{Y^{01}Y^{12}-Y^{11}Y^{02}}\quad
,\quad w_{12}=\frac{Y^{01}Y^{22}-Y^{21}Y^{02}}{Y^{01}Y^{12}-Y^{11}Y^{02}} \\ 
\\ 
w_{21}=\frac{Y^{31}Y^{12}-Y^{11}Y^{32}}{Y^{01}Y^{12}-Y^{11}Y^{02}}\quad
,\quad w_{12}=\frac{Y^{01}Y^{32}-Y^{31}Y^{02}}{Y^{01}Y^{12}-Y^{11}Y^{02}}%
\end{array}
\label{e-57}
\end{equation}%
between the bounded projective $w\in U(2)$ and homogeneous $Y^{ni}$
coordinates, we finally find the rays $w_{L}(r;s_{1},s_{2},s_{3})\in U(2)$
in the complete bounded universe $U(2)$.

The intersection of the two $%
%TCIMACRO{\U{211d} }%
%BeginExpansion
\mathbb{R}
%EndExpansion
^{4}$ charts in $U(2)$ coordinates can be computed by simply making the
Cayley transformation of the cartesian form of the ring singularity. Then we
find that in ($\tau ,\rho ,\sigma ,\chi $) coordinates the ring singularity
(the caustic of the congruence) and its "tube" connecting the two charts is 
\begin{equation}
\begin{array}{l}
\sigma =\frac{\pi }{2}\quad ,\quad R_{0}^{2}\frac{\sin ^{2}\rho }{(\cos \tau
+\cos \rho )^{2}}\leq a^{2} \\ 
-\pi <\rho <\pi \quad ,\quad -\pi <\tau <\pi%
\end{array}
\label{e-58}
\end{equation}%
which apparently contains both rings of the two $%
%TCIMACRO{\U{211d} }%
%BeginExpansion
\mathbb{R}
%EndExpansion
^{4}$ copies.

\section{NEUTRINO AND\ STANDARD\ MODEL}

\setcounter{equation}{0}

The search for the electron-soliton started from the quite general
assumptions to be massive and automorphic relative to time translation and
z-rotation. That is in quantum theory terminology, looking for massive
eigenstates of the hamiltonian and z-component angular momentum. The found
stable LCR-manifold is quite restrictive without any indication for the
existence of other connected massive configuration. On the other hand the
trajectory (\ref{e-10a}) of the Poincar\'{e} group in the set of quadratic
algebraic surfaces of $CP^{3}$ provides two possibilities. The massive
irreducible regular (rank-4) quadratic surface (\ref{e-10a}) with $\det
p\neq 0$, which is identified with the free electron, and the massless
reducible surface $\det p=0$, which is apparently singular. Therefore, I
focus my search for a massless stable LCR-structure, described by this
reducible quadratic surface of $CP^{3}$.

It is computationally easier to first look for a LCR-structure compatible
with a minkowskian class [$\eta _{\mu \nu }$] of metrics (a flatprint in the
terminology of general relativity) and after applying a Kerr-Schild ansatz
to find a curved candidate. So we look for a Kerr polynomial (\ref{e-10a})
with $\det p=0$, which is automorphic relative to the z-rotation (\ref{e-7}%
). No rank-3 quadratic surface of the form (\ref{e-10a}) survives this
condition. For every helicity [$E=\pm p^{3}$] of the neutrino LCR-structure,
I only find the rank-2 union of the following two planes

\begin{equation}
\begin{array}{l}
\lbrack E=-p^{3}]:\quad X^{3}-aX^{1}=0\quad ,\quad X^{0}=0 \\ 
\\ 
\lbrack E=+p^{3}]:\quad X^{1}=0\quad ,\quad X^{2}+bX^{0}=0 \\ 
\end{array}
\label{n-1}
\end{equation}%
in the frame with $p^{1}+ip^{2}=0$.

The union of two planes is singular at their intersection line, if they are
embedded in $CP^{3}$. It is well known in algebraic geometry that this kind
of singularities are resolved with the blowing-up procedure\cite{GRIF}. That
is, this singularity is essentially fictitious implied by the embending of
both hyperplanes in $CP^{3}$. It disappears if they are embedded in larger
projective spaces. The analogous simple example is the union of two lines
embedded in $RP^{2}$, which are singular at their intersection point. But
the union of two non-intersecting lines embedded in $RP^{3}$ are generally
nowhere singular.

The intersection (complex) line of the first two hyperplanes of $CP^{3}$ of (%
\ref{n-1}) is at the infinity of the $X^{0}=1$ affine space, while the
intersection line of the next two hyperplanes of (\ref{n-1}) is at the
infinity of the $X^{1}=1$ affine space. Therefore no neutrino trajectory
(singular line) is seen in the affine space

\begin{equation}
\begin{array}{l}
X^{ni}=%
\begin{pmatrix}
\lambda ^{Ai} \\ 
-ix_{A^{\prime }A}\lambda ^{Ai}%
\end{pmatrix}
\\ 
\end{array}
\label{n-2}
\end{equation}%
of the grassmannian $G_{4,2}$, where the trajectory (\ref{e-10a}) of the
Poincar\'{e} group has been considered. This fact should be interpreted that
the two stationary chiral neutrinos do not have a classical trajectory in
spacetime. These stationary neutrinos have $\Phi _{2}=\Phi _{3}=0$ vanishing
relative invariants. Besides, unlike the electron LCR-structure, the
corresponding 2-form (\ref{e-13}) does not admit electromagnetic sources,
hence they are chargeless.

It is very instructive to consider the following linear Kerr polynomials

\begin{equation}
\begin{array}{l}
X^{31}-aX^{11}=0\quad ,\quad X^{22}+bX^{02}=0 \\ 
\end{array}
\label{n-3}
\end{equation}%
for the left and right columns of the homogeneous coordinates. It is not
stationary, but it has a singular trajectory

\begin{equation}
\begin{array}{l}
-(x^{1}+ix^{2})\lambda ^{01}+(x^{0}+x^{3}-ia)\lambda ^{11}=0 \\ 
(x^{0}-x^{3}+ib)\lambda ^{02}-(x^{1}-ix^{2})\lambda ^{12}=0 \\ 
\\ 
\lambda ^{01}:\lambda ^{11}\sim \lambda ^{02}:\lambda ^{12}\Longrightarrow
\\ 
(b+a)x^{3}+(b-a)x^{0}=0\quad ,\quad (x^{1})^{2}+(x^{2})^{2}=ab[\frac{%
(2x^{0})^{2}}{(a+b)^{2}}+1]%
\end{array}
\label{n-4}
\end{equation}%
For $ab>0$\ it is not realistic, because the singularity ring "explodes".
But if $b=0$ it has a massless line trajectory. For $a\neq 0\neq b$ the
LCR-structure conditions are

\begin{equation}
\begin{array}{l}
\overline{X^{mi}}E_{mn}X^{nj}=0 \\ 
\\ 
\frac{z^{0}-\overline{z^{0}}}{2i}+az^{1}\overline{z^{1}}=0\quad ,\quad z^{%
\widetilde{1}}\overline{z^{0}}+(z^{\widetilde{0}}+ia)\overline{z^{1}}=0\quad
,\quad \frac{z^{\widetilde{0}}-\overline{z^{\widetilde{0}}}}{2i}-bz^{%
\widetilde{1}}\overline{z^{\widetilde{1}}}=0 \\ 
\end{array}
\label{n-6}
\end{equation}%
This is like a twisted natural $U(2)$ LCR-structure\cite{RAG2013b}, because
its third relative invariant vanishes, $\Phi _{3}=0$. The LCR-tetrad can be
directly found.

For $b=0$,\ the structure coordinates are

\begin{equation}
\begin{array}{l}
z^{0}\equiv i\frac{X^{21}}{X^{01}}\quad ,\quad z^{1}\equiv \frac{X^{11}}{%
X^{01}}\quad ,\quad z^{\widetilde{0}}\equiv i\frac{X^{32}}{X^{12}}\quad
,\quad z^{\widetilde{1}}\equiv -\frac{X^{02}}{X^{12}} \\ 
\\ 
z^{0}=x^{0}-x^{3}-\frac{(x^{1})^{2}+(x^{2})^{2}}{x^{0}+x^{3}-ia}\quad ,\quad
z^{1}=\frac{x^{1}+ix^{2}}{x^{0}+x^{3}-ia} \\ 
z^{\widetilde{0}}=x^{0}+x^{3}-\frac{(x^{1})^{2}+(x^{2})^{2}}{x^{0}-x^{3}}%
\quad ,\quad z^{\widetilde{1}}=\frac{x^{1}-ix^{2}}{x^{0}-x^{3}} \\ 
\\ 
u=x^{0}-x^{3}-\frac{(x^{1})^{2}+(x^{2})^{2}}{(x^{0}+x^{3})^{2}+a^{2}}%
(x^{0}+x^{3}) \\ 
v=x^{0}+x^{3}-\frac{(x^{1})^{2}+(x^{2})^{2}}{x^{0}-x^{3}}%
\end{array}
\label{n-5}
\end{equation}%
The convenient structure coordinates are

\begin{equation}
\begin{array}{l}
z^{\prime 0}=-\frac{1}{z^{0}}=u^{\prime }-iaz^{\prime 1}\overline{z^{\prime
1}}\quad ,\quad z^{\prime 1}=-\frac{z^{1}}{z^{0}}=\zeta ^{\prime } \\ 
z^{\prime \widetilde{0}}=z^{\widetilde{0}}=v^{\prime }\quad ,\quad z^{\prime 
\widetilde{1}}=\frac{z^{\widetilde{1}}}{z^{\widetilde{0}}+ia}=\overline{%
z^{\prime 1}} \\ 
\\ 
z^{\prime \widetilde{1}}=\frac{z^{\widetilde{1}}}{z^{\widetilde{0}}+ia}=-%
\frac{\overline{z^{1}}}{\overline{z^{0}}}=\overline{z^{\prime 1}}%
\end{array}
\label{n-7}
\end{equation}%
In these coordinates the following tetrad can be easily computed

\begin{equation}
\begin{array}{l}
L=du^{\prime }+ia\overline{\zeta ^{\prime }}d\zeta ^{\prime }-ia\zeta
^{\prime }d\overline{\zeta ^{\prime }}\quad ,\quad N=dv^{\prime }\quad
,\quad M=d\zeta ^{\prime } \\ 
dL=-2iaM\wedge \overline{M}\quad ,\quad dN=0\quad ,\quad dM=0 \\ 
\\ 
L^{\mu }\partial _{\mu }=\partial _{v^{\prime }}\quad ,\quad N^{\mu
}\partial _{\mu }=\partial _{u^{\prime }}\quad ,\quad M^{\mu }\partial _{\mu
}=-ia\zeta ^{\prime }\partial _{u^{\prime }}-\partial _{\overline{\zeta
^{\prime }}}%
\end{array}
\label{n-8}
\end{equation}%
Note that this flat LCR-structure has vanishing $\Phi _{2}=\Phi _{3}=0$
relative invariants.

The Kerr-Schild ansatz may be applied either on $L$ or on $N$. I will
consider the later case, in order to show an interesting effect of gravity.
Let

\begin{equation}
\begin{array}{l}
\ell =L\quad ,\quad m=M\quad ,\quad n=N+fL \\ 
\end{array}
\label{n-9}
\end{equation}%
The LCR-structure condition fixes the form of $f(x)$ because

\begin{equation}
\begin{array}{l}
dn=df\wedge L+fdL=Z_{2}\wedge n+i\Phi _{2}m\wedge \overline{m} \\ 
\\ 
M^{\mu }\partial _{\mu }f=0\quad \Longrightarrow \quad \partial _{u^{\prime
}}f=0,\quad \partial _{\zeta ^{\prime }}f=0%
\end{array}
\label{n-10}
\end{equation}%
Hence $f=f(v^{\prime })$ depends only on $v^{\prime }$. Notice that the
structure relations now take the form

\begin{equation}
\begin{array}{l}
d\ell =-2iam\wedge \overline{m}\quad ,\quad dm=0 \\ 
dn=df\wedge L+fdL=Z_{2}\wedge n+i\Phi _{2}m\wedge \overline{m} \\ 
\end{array}
\label{n-11}
\end{equation}%
with non-vanishing relative invariants $\Phi _{1}\neq 0\neq \Phi _{2}=-2iaf$%
. That is gravity may generate a right chirality and the neutrino flatprint
LCR-structure may not be a smooth deformation of the curved one, as it
happens for the electron LCR-structure.

The next question we have to answer is whether the neutrino admits an
electromagnetic potential. By analogy to the electron LCR-structure, the
neutrino (\ref{n-9}) "electromagnetic field" should be defined by the
self-dual 2-form

\begin{equation}
\begin{array}{l}
F^{+}=Ce^{2iav^{\prime }}(\ell \wedge n-m\wedge \overline{m}%
)=Ce^{2iav^{\prime }}(L\wedge N-M\wedge \overline{M}) \\ 
\end{array}
\label{n-12}
\end{equation}%
where $C$ is an arbitrary complex constant. It is closed and exact, because
a straightforward application of Stokes' theorem on the $t$ and $r$ constant
sphere implies no sources. One can see it by simply observing that it is an
exact form, because the imaginary term $ia$ does not permit any singularity.
Hence, I conclude that the "charge" of the neutrino (\ref{n-3}) vanishes.

We have already pointed out that the apparent trajectories of the elementary
leptonic particles strongly indicate that they are ruled surfaces (l-12c) of 
$CP^{3}$. The radiating electron is naturally described by their Newman
complex trajectory. The relation (l-12e) proves that a ruled surface is
characterized by a massive and a massless trajectory. Hence, the emergence
of massive-massless pairs is a characteristic property of the ruled
surfaces. In this context the three leptonic generations (electonic, muonic
and tauonic) could correspond to quadric (already studied above), cubic and
quartic curves $Z^{n}(\tau )$ of the ruled surface.

\subsection{The electroweak-U(2) gauge fields}

In conventional field theory the interactions have to be imposed as
connections. In the computation of the electron and its neutrino
distributional LCR-structures, these fields appear as gravitational and
electromagnetic dressing distributions with precise compact singular
support. The purpose of the present subsection is to provide the algorithmic
derivation of the weak connection\cite{RAG2021}. This generalizes the
surprising identification of the electron electromagnetic dressing $A_{\mu
}(x)$ with the vector $\ell _{\mu }$ of the LCR-tetrad (and the induced
gravitational dressing).

The general solution of a realizable LCR-structure is a special totally real
submanifold of $%
%TCIMACRO{\U{2102} }%
%BeginExpansion
\mathbb{C}
%EndExpansion
^{4}$, determined by the conditions (\ref{I-4}). Its characteristic local
coframe of the surface contains the normal bundle $d\rho _{ij}$ and the
tangent 1-forms \ 
\begin{equation}
\begin{array}{l}
\ell =i(\partial -\overline{\partial })\rho _{11}\quad ,\quad n=i(\partial -%
\overline{\partial })\rho _{22}\quad ,\quad m=i(\partial -\overline{\partial 
})\overline{\rho _{12}} \\ 
\\ 
\begin{pmatrix}
\ell & \overline{m} \\ 
m & n%
\end{pmatrix}%
=i(\partial -\overline{\partial })%
\begin{pmatrix}
\rho _{11} & \rho _{12} \\ 
\overline{\rho _{12}} & \rho _{22}%
\end{pmatrix}%
\end{array}
\label{n-13}
\end{equation}%
arranged to a 2$\times $2 hermitian matrix, and considered as a function on
the algebra of the Lie group $U(2)$. Hence, it may be considered as an
electroweak Cartan connection.

The "natural U(2)" LCR-structure is \ 
\begin{equation}
\begin{array}{l}
e=-iw^{-1}dw=:%
\begin{pmatrix}
\ell & \overline{m} \\ 
m & n%
\end{pmatrix}%
\quad ,\quad de-ie\wedge e=0 \\ 
\\ 
d\ell =im\wedge \overline{m}\quad ,\quad dn=-im\wedge \overline{m}\quad
,\quad dm=i(\ell -n)\wedge m%
\end{array}
\label{n-14}
\end{equation}%
This form strongly suggests to osculate the LCR-structure with the $U(2)$\
group. The first step of that is to cast a LCR-tetrad into the hermitian
matrix \ 
\begin{equation}
\begin{array}{l}
e^{\prime }:=%
\begin{pmatrix}
\ell ^{\prime } & \overline{m^{\prime }} \\ 
m^{\prime } & n^{\prime }%
\end{pmatrix}%
=i(\partial -\overline{\partial })%
\begin{pmatrix}
\rho _{11} & \rho _{12} \\ 
\overline{\rho _{12}} & \rho _{22}%
\end{pmatrix}
\\ 
\end{array}
\label{n-15}
\end{equation}%
Hence the electroweak gauge fields and the corresponding curvature are \ 
\begin{equation}
\begin{array}{l}
B=B_{I\mu }dx^{\mu }t_{I}=%
\begin{pmatrix}
\ell ^{\prime } & \overline{m^{\prime }} \\ 
m^{\prime } & n^{\prime }%
\end{pmatrix}%
\quad ,\quad \lbrack t_{I},t_{J}]=iC_{IJK}t_{K} \\ 
\\ 
F=dB-iB\wedge B\ \longrightarrow DF:=\ dF+iB\wedge F-iF\wedge B=0 \\ 
\end{array}
\label{n-16}
\end{equation}%
where $t_{J}$ are generators of $U(2)$. Apparently a gauge transformation
breaks the tetrad-Weyl symmetry, because the implied tetrad is an
LCR-tetrad. Therefore we chose the LCR-tetrad $e^{\prime }$ such that $\Phi
_{1}^{\prime }=1=-\Phi _{2}^{\prime }$.\ That is, we partly fix the
tetrad-Weyl symmetry for non-trivial LCR-structures with $\Phi _{1}\neq
0\neq \Phi _{2}$. Recall the general tetrad-Weyl transformation \ 
\begin{equation}
\begin{array}{l}
\ell ^{\prime }=\Lambda \ell \quad ,\quad n^{\prime }=Nn\quad ,\quad
m^{\prime }=Mm \\ 
\\ 
Z_{1}^{\prime }=Z_{1}+d(\ln \Lambda )\quad ,\quad \Phi _{1}^{\prime }=\frac{%
\Lambda }{M\overline{M}}\Phi _{1} \\ 
Z_{2}^{\prime }=Z_{2}+d(\ln N)\quad ,\quad \Phi _{2}^{\prime }=\frac{N}{M%
\overline{M}}\Phi _{2} \\ 
Z_{3}^{\prime }=Z_{3}+d(\ln M)\quad ,\quad \Phi _{3}^{\prime }=\frac{M}{%
\Lambda N}\Phi _{1}%
\end{array}
\label{n-17}
\end{equation}%
In the case of the following generators \ 
\begin{equation}
\begin{array}{l}
t_{0}=I\quad ,\quad t_{k}=\frac{\sigma _{k}}{2}\ \rightarrow \
C_{ijk}=\epsilon _{ijk} \\ 
\end{array}
\label{n-18}
\end{equation}%
we have\ 
\begin{equation}
\begin{array}{l}
B_{0\mu }+\frac{1}{2}B_{3\mu }=\ell _{\mu }^{\prime }\quad ,\quad B_{0\mu }-%
\frac{1}{2}B_{3\mu }=n_{\mu }^{\prime }\quad ,\quad \frac{1}{2}(B_{1\mu
}+iB_{2\mu })=m_{\mu }^{\prime } \\ 
\\ 
F_{0\mu \nu }=\partial _{\mu }B_{0\nu }-\partial _{\nu }B_{0\mu } \\ 
F_{i\mu \nu }=\partial _{\mu }B_{i\nu }-\partial _{\nu }B_{i\mu }-\epsilon
_{ijk}B_{j\mu }B_{k\nu }%
\end{array}
\label{n-19}
\end{equation}%
Notice the direct relation of the gravitational tetrad with the electroweak
potentials of the standard model.

In the case of the electron LCR-tetrad 
\begin{equation}
\begin{array}{l}
\Phi _{1}=\frac{-2a\cos \theta }{r^{2}+a^{2}\cos ^{2}\theta } \\ 
\Phi _{2}=-\frac{(r^{2}+a^{2}+h)a\cos \theta }{(r^{2}+a^{2}\cos ^{2}\theta
)^{2}} \\ 
\Phi _{3}=\frac{2iar\sin \theta }{\sqrt{2}(r+ia\cos \theta )^{2}(r-ia\cos
\theta )} \\ 
\end{array}
\label{n-20}
\end{equation}%
we make first the tetrad-Weyl transformation to reach the condition $\Phi
_{1}^{\prime }=1=-\Phi _{2}^{\prime }$. We find \ \ 
\begin{equation}
\begin{array}{l}
N=-\frac{r^{2}+a^{2}\cos ^{2}\theta }{r^{2}+a^{2}+h}\Lambda \\ 
M\overline{M}=-\frac{2a\cos \theta }{r^{2}+a^{2}\cos ^{2}\theta }\Lambda \\ 
\end{array}
\label{n-21}
\end{equation}%
The electromagnetic dressing is found with $\Lambda =\frac{qr}{%
r^{2}+a^{2}\cos ^{2}\theta }$. Then the connection $B$ is found with\ \ 
\begin{equation}
\begin{array}{l}
\Lambda =\frac{qr}{r^{2}+a^{2}\cos ^{2}\theta } \\ 
N=-\frac{qr}{4\pi (r^{2}+a^{2}+h)} \\ 
M\overline{M}=-\frac{qra\cos \theta }{2\pi (r^{2}+a^{2}\cos ^{2}\theta )^{2}}
\\ 
\end{array}
\label{n-22}
\end{equation}%
up to $M$ phase tetrad-Weyl transformation. We see that the $U(2)$ gauge
field is directly related to the LCR-tetrad and the Higgs field is related
with the $\Phi _{i}$\ factors, which determine the relative invariants of
the LCR-structure.

We will now prove that the above definition of theelectroweak connection
permits the emergence of distributional singularities in the embedded
LCR-manifold. The starting point is to write the surface $\rho _{ij}=0$,
using the regular coordinates 
\begin{equation}
\begin{array}{l}
\func{Im}z^{0}=\phi _{11}(\overline{z^{1}},z^{1},\func{Re}z^{0})\ ,\ \func{Im%
}z^{\widetilde{0}}=\phi _{22}(\overline{z^{\widetilde{1}}},z^{\widetilde{1}},%
\func{Re}z^{\widetilde{0}})\ ,\ z^{\widetilde{1}}-\overline{z^{1}}=\phi
_{12}(\overline{z^{\beta }},z^{\widetilde{0}}) \\ 
\\ 
\phi _{11}(p)=\phi _{22}(p)=\phi _{12}(p)=0\quad ,\quad d\phi _{11}(p)=d\phi
_{22}(p)=d\phi _{12}(p)=0%
\end{array}
\label{n-23}
\end{equation}%
in a neighborhood of a point $p$. But the LCR-structure is a special totally
real CR-structure, which at a real analytic neighborhood admits\cite{BAOU} a
general analytic transformation $r^{b}=f^{b}(z^{c})$, which makes it trivial 
\begin{equation}
\begin{array}{l}
\frac{r^{a}-\overline{r^{a}}}{2i}=0\  \\ 
\end{array}
\label{n-24}
\end{equation}%
This last analytic transformation is not generally an LCR-transformation.
Hence, it breaks the LCR-structure, but there is no reason to worry for that
now, because we are going to look for a connection, which has already broken
the LCR-symmetry. The essential point here is the neighborhood of $p$ in the
ambient complex manifold, where the analytic transformation can be extended.
The case of the distributional electron (and neutrino) indicates that the
analytic transformation cannot be extended around their location. Besides
the entire region of the LCR-manifold can be described by a distribution
with a representative (locally integrable function), which at the regular
point $p$ appears as a regular potential with each source at the location of
the electron. The location of the electron is not a real analytic region of
the LCR-manifold, because it does not admit analytic extensions in both
sides of the real surface $\rho _{ij}=0$ in the ambient complex manifold.
Recall that it is the Sato's definition of generalized functions.

\subsection{Derivation of standard model action}

The successful application of the Bogoliubov recursive procedure\cite%
{BOG1980} to build up an effective quantum electrodynamics and its
extraordinary experimental verification, suggest us to extend it including
the massless neutrino soliton as a left-hand field $\frac{1-\gamma _{5}}{2}%
\psi _{\nu }$, and all the permitted charged and neutral currents. No
neutrino electromagnetic interaction should be introduced or permit it to
appear through the Bogoliubov recursive procedure. It has already been shown%
\cite{SCH2} that assuming the existence of all the standard model particles
(for every generation separately) the implied standard model lagrangian is a
consequence of the renormalizability condition. The appearance of the Poincar%
\'{e} group and the Schwartz distributions imposes the use of the basis of
the rigged Hilbert space of the temered distributions. A free field is an
operator valued distribution in the appropriate Lorentz group representation
corresponding to every elementary particle and all its dressings. Let us now
enumerate the fields and the interactions we will consider in the beginning
(correspondence principle) of the Bogoliubov procedure, indicating their
existence in the context of PCFT:

1) The massive Dirac electron field $\psi _{e}(x)$, which satisfies the free
Dirac equation and hence it implies a free massive Dirac propagator in the
time ordering term (\ref{e-19}). This fermionic solitonic LCR-manifold (with 
$g=2$ gyromagnetic ratio) has been extensively studied.

2) The left-hand part of the massless Dirac neutrino field $\frac{1-\gamma
_{5}}{2}\psi _{\nu }$, which satisfies the free Dirac equation and hence it
implies a free massless Dirac propagator in the time ordering term. All the
considered currents will contain only the left-hand part of the neutrino
field. This is the massless developable surface which corresponds to the
massive electron ruled surface.

3) The massless electromagnetic field $A_{\mu }(x)$, which implies the
corresponding massless propagator. It is the potential of the real part of
the closed self-dual 2-form (\ref{e-13}).

4) The electromagnetic interaction between the electron and the photon as
indicated by the electromagnetic dressing of the electron LCR-structure.

5) The $U(2)$ gauge fields (\ref{n-19}) viewed as an extension of
electromagnetism. These fields are properly coupled with the charged and
neutral currents of the electron neutrino pair. The most convenient method
seems to be the Scharf\cite{SCH2} Q-charge operator with the $U(2)$ breaking
will be implied by the difference of the electron and neutrino masses.

6) The derivation of the Cartan $U(2)$ connection strongly suggests that the
Higgs scalar field is related to the relative invariants $\Phi _{i}$ of the
LCR-structure. The real $\Phi _{1}$ and $\Phi _{2}$ are fixed with (\ref%
{n-22}), and the phase of the complex $\Phi _{3}$ can be absorbed by $m_{\mu
}dx^{\mu }$. Hence the radial of $\Phi _{3}$ should be identified with the
scalar Higgs field of the standard model.

Including all these assumptions in the initial action through the
"correspondence principle", the Bogoliubov procedure implies a closed
lagrangian form only if the well known relations between the coupling
constants and the masses of particles are valid\cite{SCH2}. This means that
the "internal symmetry" $U(2)$ breaking mechanism, is a consequence of the
initial mass difference between the electron and the neutrino and the
renormalizability condition of the Bogoliubov procedure, viewed as an
effective action generating mechanism! There is no initial internal group in
PCFT. The ad hock assumption of a fundamental $U(2)$ internal symmetry
misled the scientific research to grand unified theories and their
supersymmetric extensions, which have not been observed.

Let us now turn to the origin\cite{RAG1999} of the three elementary particle
generations (families). The three particle generations are a consequence of
the gravity potentials of these solitons which emerge through the Einstein
metric $g_{\mu \nu }$ (\ref{g-1}). It is well known that the Einstein metric 
$g_{\mu \nu }=\eta _{ab}e_{\mu }^{a}e_{\nu }^{b}$, where $e_{\mu }^{a}$\ are
the four Cartan moving frames. They are defined\cite{CHAND} up to a local $%
SO(1,3)$ transformation $e_{\mu }^{\prime a}=S_{b}^{a}e_{\mu }^{b}$ which
generates and relates the Cartan connection with the ordinary metric $g_{\mu
\nu }$ connection. Newman and Penrose\cite{P-R} have noticed that assuming a
null tetrad, the Cartan formalism acquires very useful properties easily
applied to the radiation problems. In this formalism the LCR-structure
coincides with the existence of two geodetic and shear-free null
congruences, which have the simple form $\kappa =\sigma =\lambda =\nu =0$.
Besides, the use of the spinor dyad ($o^{A},\imath ^{B}$) through the
relations (\ref{g-10}) imply the spinorial formulation of general
relativity. I have already pointed out that a metric does not always admit
two geodetic and shear-free congruences. In this case of metrics, using an
arbitrary non geodetic and shear-free null tetrad, the spinor form of the
conformal tensor $\Psi _{ABCD}$ can always be defined, and it admits two
spinors ($\lambda ^{A1},\lambda ^{B2}$), which satisfy the relations (\ref%
{g-17}). In the linearized gravity approximation they become the spinors of
the first two rows of the homogeneous coordinates of $G_{4,2}$. Hence
locally, a non-conformally flat metric compatible with an LCR-structure has
at most four geodetic and shear-free null congruences, i.e. at most four
branches (sheets). Every two of them determine a LCR-structure. From the
Petrov classification\cite{P-R}, we have the types of spacetime with four
(type I), three (type II), two double (type D) and a triple (type III)
principal null directions. Apparently the electron and the neutrino solitons
correspond to type D spacetimes. The fact that I have not found\cite%
{RAG2008b} static LCR-structures for cubic and quartic Kerr polynomials,
suggests us to correspond the decaying muon and tau generations to the two
Petrov types II and I respectively. The soliton stability is assured by the
different degrees of the quadric (electron and its neutrino), cubic (muon
and its neutrino) and quartic (tau and its neutrino) algebraic surfaces. The
internal stability of each flavor is assured by the different relative
invariants.

\section{THE UP AND DOWN\ QUARKS}

\setcounter{equation}{0}

Concerning the electromagnetic and weak interactions, the hadronic sector of
the elementary particles is (about) a copy of the leptonic sector. Quarks
simply have the additional strong interaction, which should provide a
confining mechanism. The standard model does not explain the general
copy-picture, while the artificial add-on of the $SU(3)$ gauge group gives
some answers to some phenomena, but it fails to imply (in the continuum)
confinement and chirality breaking, which are the characteristic properties
of strong interactions.

PCFT is mathematically a vector bundle (with a gauge field) over a
lorentzian CR-manifold. The gluon field is identified with the gauge field
of the action and the LCR-structure describes (contains) gravity,
electromagnetic and weak interactions as outlined in the previous section,
where we have assumed that the found distributional solitons have vanishing
gluon field configuration. In this section I will explicitly find stable
gluonic configurations for the electron and the neutrino LCR-manifolds,
which I will identify with down and up quarks. That is, the origin of the
observed general copy-picture between the leptons and quarks is simply their
common LCR-structure (which contains gravitational, electromagnetic and weak
interactions), they differ to the non-vanishing "gluonic" dressing of the
quarks.

Variation of the actions (\ref{p-1}) relative to the gauge field implies the
field equations 
\begin{equation}
\begin{array}{l}
I_{R}\quad \rightarrow \quad \frac{1}{\sqrt{-g}}(D_{\mu })_{ij}(\sqrt{-g}%
(\Gamma ^{\mu \nu \rho \sigma }-\overline{\Gamma ^{\mu \nu \rho \sigma }}%
)F_{j\rho \sigma })=0 \\ 
I_{I}\quad \rightarrow \quad \frac{1}{\sqrt{-g}}(D_{\mu })_{ij}(\sqrt{-g}%
(\Gamma ^{\mu \nu \rho \sigma }+\overline{\Gamma ^{\mu \nu \rho \sigma }}%
)F_{j\rho \sigma })=0 \\ 
\\ 
\Gamma ^{\mu \nu \rho \sigma }=\frac{1}{2}[(\ell ^{\mu }m^{\nu }-\ell ^{\nu
}m^{\mu })(n^{\rho }\overline{m}^{\sigma }-n^{\sigma }\overline{m}^{\rho
})+(n^{\mu }\overline{m}^{\nu }-n^{\nu }\overline{m}^{\mu })(\ell ^{\rho
}m^{\sigma }-\ell ^{\sigma }m^{\rho })] \\ 
(D_{\mu })_{ij}=\delta _{ij}\partial _{\mu }-\gamma f_{ikj}A_{k\mu }%
\end{array}
\label{q-1}
\end{equation}%
Recall that the derivation of quantum electrodynamics (as an affective field
theory) was triggered by the existence of a source in the closed self-dual
antisymmetric tensor of the massive static soliton. But the above (\ref{q-1}%
) both field equations are exact. We cannot replace (ad hoc) the zero of the
second part of the equation with a source, because the symmetries of the
action will be destroyed, and subsequently the renormalizability of the
action will be destroyed too. The solution\cite{RAG2018b} to this
obstruction comes after a close look at the form of the field equations \ref%
{q-1}. Notice that they are the sum or difference of two complex conjugate
terms. This does not permit us to apply the complexification (necessary for
the application of the Frobenius theorem) and use the convenient form that
the LCR-structure tetrad takes in the ambient complex manifold through the
structure coordinates $z^{a}(x)$.

Therefore I find convenient to give the PDEs (\ref{q-1}) the following
equivalent forms 
\begin{equation}
\begin{array}{l}
I_{R}\quad \rightarrow \quad \frac{1}{\sqrt{-g}}(D_{\mu })_{ij}\{\sqrt{-g}%
[(\ell ^{\mu }m^{\nu }-\ell ^{\nu }m^{\mu })(n^{\rho }\overline{m}^{\sigma
}F_{j\rho \sigma })+ \\ 
\qquad \qquad +(n^{\mu }\overline{m}^{\nu }-n^{\nu }\overline{m}^{\mu
})(\ell ^{\rho }m^{\sigma }F_{j\rho \sigma })]\}=-k_{i}^{\nu } \\ 
\\ 
I_{I}\quad \rightarrow \quad \frac{1}{\sqrt{-g}}(D_{\mu })_{ij}\{\sqrt{-g}%
[(\ell ^{\mu }m^{\nu }-\ell ^{\nu }m^{\mu })(n^{\rho }\overline{m}^{\sigma
}F_{j\rho \sigma })+ \\ 
\qquad \qquad +(n^{\mu }\overline{m}^{\nu }-n^{\nu }\overline{m}^{\mu
})(\ell ^{\rho }m^{\sigma }F_{j\rho \sigma })]\}=-ik_{i}^{\nu }%
\end{array}
\label{q-3}
\end{equation}%
where $k_{i}^{\nu }(x)$ is a \textbf{real} vector field. The PDEs look like
the equations of a gauge field with a color-electric and color-magnetic
source respectively. Notice the natural emergence of the sources. I will
solve these partial differential equations in the static (electron)
LCR-structure (\ref{l-13}). This is possible, because the LCR-structure
defining equations completely decouple from the gauge field equations. The
LCR-structure is first fixed (via the Lagrange multipliers) and after we
proceed to the solution of the field equations, which involve the gauge
field. This property of PCFT is essentially behind the physical observation
of the lepton-quark correspondence! That is a quark has the same
LCR-structure with the corresponding lepton. But the quark has in addition a
stable non-vanishing distributional gauge field configuration (from which it
gets its color), while the lepton has vanishing gauge field.

Recall that a distribution has two parts. The singular part and the regular
part. A classical solution of the gauge field with a singular compact source
will be interpreted as a colored soliton (the quark) with its gluon
potential ("dressing") being the regular part of the generalized function%
\cite{GELF1}. If we apply again with the gauge covariant derivative $(D_{\nu
})_{ij}$ and use the commutation relation

\begin{equation}
\begin{array}{l}
\lbrack (D_{\mu }),(D_{\nu })]_{ik}=-\gamma f_{ijk}F_{j\mu \nu } \\ 
\end{array}
\label{q-4}
\end{equation}%
we find that the current must be gauge covariantly conserved $(D_{\nu
})_{ij}k_{j}^{\nu }=0$ for a classical solution to exist. We will look for
fundamental distributional solutions which have compact singular sources,
which may be interpreted as localized "particles". I will work out the
derivation of a (null) distributional solution for the first PDE (action $%
I_{R}$), where such a solution can exist, and I will simply indicate why the
second PDE (action $I_{I}$) does not admit a corresponding color-magnetic
solution.

In the case of gravity and electromagnetism we found distributional
(fundamental) solutions, where the singular part is compact and located at
the ring singularity. It is identified with the electron, while its
gravitational and electromagnetic fields are the regular part of the
distribution (the gravitational and electromagnetic dressings) located
outside the singular support of the source (the electron). Now we will apply
the same point of view for the computation of the quark and its dressing
gluonic field strength. Outside the compact singular support of color
sources, the current $k_{j}^{\nu }=0$ vanishes. In this region we can make
the complexification of the real coordinate variable $x$ of the (real)
LCR-manifold and after we can make an holomorphic transformation to the
LCR-structure coordinates ($z^{\alpha }(x),z^{\widetilde{\alpha }}(x)$), and
use their following powerful properties

\begin{equation}
\begin{array}{l}
dz^{\alpha }=f_{0}^{\alpha }\ \ell _{\mu }dx^{\mu }+f_{1}^{\alpha }\ m_{\mu
}dx^{\mu }\quad ,\quad dz^{\widetilde{\alpha }}=f_{\widetilde{0}}^{%
\widetilde{\alpha }}\ n_{\mu }dx^{\mu }+f_{\widetilde{1}}^{\widetilde{\alpha 
}}\ \widetilde{m}_{\mu }dx^{\mu } \\ 
\\ 
\ell _{\mu }dx^{\mu }=\ell _{\alpha }dz^{\alpha }\quad ,\quad m_{\mu
}dx^{\mu }=m_{\alpha }dz^{\alpha }\quad ,\quad n_{\mu }dx^{\mu }=n_{%
\widetilde{\alpha }}dz^{\widetilde{\alpha }}\quad ,\quad \overline{m}_{\mu
}dx^{\mu }=\widetilde{m}_{\widetilde{\alpha }}dz^{\widetilde{\alpha }} \\ 
\ell ^{\mu }\partial _{\mu }=\ell ^{\widetilde{\alpha }}\partial _{%
\widetilde{\alpha }}\quad ,\quad m^{\mu }\partial _{\mu }=m^{\widetilde{%
\alpha }}\partial _{\widetilde{\alpha }}\quad ,\quad n^{\mu }\partial _{\mu
}=n^{\alpha }\partial _{\alpha }\quad ,\quad \overline{m}^{\mu }\partial
_{\mu }=\widetilde{m}^{\alpha }\partial _{\alpha } \\ 
\end{array}
\label{q-5}
\end{equation}%
In these complex coordinates, the metric takes the off-diagonal form (\ref%
{i-3}) and

\begin{equation}
\begin{array}{l}
\sqrt{-g}dx^{0}\wedge dx^{1}\wedge dx^{2}\wedge dx^{3}=-i\ell \wedge m\wedge
n\wedge \widetilde{m}=-i\widehat{g}dz^{0}\wedge dz^{1}\wedge dz^{\widehat{0}%
}\wedge dz^{\widehat{1}} \\ 
\\ 
g_{ab}=%
\begin{pmatrix}
0 & \widehat{g}_{\alpha \widetilde{\beta }} \\ 
\widehat{g}_{\beta \widetilde{\alpha }} & 0%
\end{pmatrix}%
\;\;\;\;,\;\;\;g^{ab}=%
\begin{pmatrix}
0 & \widehat{g}^{\alpha \widetilde{\beta }} \\ 
\widehat{g}^{\beta \widetilde{\alpha }} & 0%
\end{pmatrix}
\\ 
\ \widehat{g}_{\alpha \widetilde{\beta }}=\ell _{\alpha }n_{\widetilde{\beta 
}}-m_{\alpha }\overline{m}_{\widetilde{\beta }}\;\;\;\;,\;\;\;\widehat{g}%
^{\alpha \widetilde{\beta }}=n^{\alpha }\ell ^{\widetilde{\beta }}-\overline{%
m}^{\alpha }m^{\widetilde{\beta }}\;\;\;,\;\;\;\widehat{g}\equiv \det 
\widehat{g}_{\alpha \widetilde{\beta }} \\ 
(\ell _{0}m_{1}-m_{0}\ell _{1})(n_{\widetilde{0}}\overline{m}_{\widetilde{1}%
}-\overline{m}_{\widetilde{0}}n_{\widetilde{1}})=-\widehat{g}%
\;\;\;,\;\;\;(\ell ^{\widetilde{0}}m^{\widetilde{1}}-m^{\widetilde{0}}\ell ^{%
\widetilde{1}})(n^{0}\overline{m}^{1}-\overline{m}^{0}n^{1})=-\frac{1}{%
\widehat{g}}%
\end{array}
\label{q-6}
\end{equation}%
Hence after the complexification we have to replace $\sqrt{-g}\rightarrow -i%
\widehat{g}$. Notice, that now we deal with a complex metric
(pseudo-metric), and we must not take complex conjugations before returning
back to real $x$. Then (\ref{q-3}) takes the form 
\begin{equation}
\begin{array}{l}
For\ b=0\quad ,\quad \partial _{1}F_{i\widetilde{0}\widetilde{1}}-\gamma
f_{ikj}A_{k1}F_{j\widetilde{0}\widetilde{1}}=(D_{1})_{ij}F_{j\widetilde{0}%
\widetilde{1}}=-\widehat{g}k_{i}^{0} \\ 
For\ b=1\quad ,\quad \partial _{0}F_{i\widetilde{0}\widetilde{1}}-\gamma
f_{ikj}A_{k0}F_{j\widetilde{0}\widetilde{1}}=(D_{0})_{ij}F_{j\widetilde{0}%
\widetilde{1}}=\widehat{g}k_{i}^{1} \\ 
For\ b=\widetilde{0}\quad ,\quad \partial _{\widetilde{1}}F_{i01}-\gamma
f_{ikj}A_{k\widetilde{1}}F_{j01}=(D_{\widetilde{1}})_{ij}F_{j01}=-\widehat{g}%
k_{i}^{\widetilde{0}} \\ 
For\ b=\widetilde{1}\quad ,\quad \partial _{\widetilde{0}}F_{i01}-\gamma
f_{ikj}A_{k\widetilde{0}}F_{j01}=(D_{\widetilde{0}})_{ij}F_{j01}=\widehat{g}%
k_{i}^{\widetilde{1}} \\ 
\end{array}
\label{q-7}
\end{equation}%
written separately for every structure coordinate in order to help a
non-familiar reader to understand the subsequent mathematical operations.
The integrability conditions imply 
\begin{equation}
\begin{array}{l}
\lbrack (D_{0}),(D_{1})]_{ik}F_{k\widetilde{0}\widetilde{1}}=-\gamma
f_{ijk}F_{j01}F_{k\widetilde{0}\widetilde{1}}=-(D_{\alpha })_{ij}(\widehat{g}%
k_{j}^{\alpha }) \\ 
\lbrack (D_{\widetilde{0}}),(D_{\widetilde{1}})]_{ik}F_{k01}=-\gamma
f_{ijk}F_{j\widetilde{0}\widetilde{1}}F_{k01}=-(D_{\widetilde{\alpha }%
})_{ij}(\widehat{g}k_{j}^{\widetilde{\alpha }}) \\ 
\end{array}
\label{q-8}
\end{equation}%
They vanish outside the compact singular gluonic source.

As expected, the written in LCR-structure coordinates equations do not
contain the complexified "metric" $g_{\alpha \widetilde{\beta }}$,\ and
contain only the self-dual left-hand component $F_{j01}$\ and right-hand
component $F_{j\widetilde{0}\widetilde{1}}$ of the gauge field strength,
because the present gauge field action has been constructed to be metric
independent.

It is evident that if 
\begin{equation}
\begin{array}{l}
f_{ijk}F_{j01}F_{k\widetilde{0}\widetilde{1}}=-\frac{1}{\widehat{g}}%
f_{ijk}(n^{\mu }\widetilde{m}^{\nu }F_{j\mu \nu })(\ell ^{\mu }m^{\nu
}F_{k\mu \nu })\neq 0 \\ 
\end{array}
\label{q-9}
\end{equation}%
does not vanish outside the sources, the differential equations (\ref{q-7})
do not accept (fundamental) solutions with compact sources. Hence my
conclusion is that, outside the singular compact part (the quark) of the
generalized function, we may have solutions only if $F_{i01}$\ or $F_{j%
\widetilde{0}\widetilde{1}}$ vanish for non vanishing $f_{ijk}$. That is, we
may have the following solutions\ 
\begin{equation}
\begin{array}{l}
A_{\alpha }=\frac{1}{\gamma }(\partial _{\alpha }U)U^{-1}\quad ,\quad (\ell
^{\mu }m^{\nu }F_{k\mu \nu })=(\ell ^{\widetilde{0}}m^{\widetilde{1}}-\ell ^{%
\widetilde{1}}m^{\widetilde{0}})F_{k\widetilde{0}\widetilde{1}}\neq 0 \\ 
\\ 
(n^{\mu }\widetilde{m}^{\nu }F_{k\mu \nu })=(n^{0}\widetilde{m}^{1}-n^{1}%
\widetilde{m}^{0})F_{k01}\neq 0\quad ,\quad A_{\widetilde{\alpha }}=\frac{1}{%
\gamma }(\partial _{\widetilde{\alpha }}U^{\prime })U^{\prime -1} \\ 
\end{array}
\label{q-10}
\end{equation}%
where $U$ and $U^{\prime }$ are arbitrary elements of the gauge group in a
prescribed gauge group representation.

Hence, the two gauge field equations become abelian 
\begin{equation}
\begin{array}{l}
\partial _{\widetilde{\alpha }}F_{01}-\gamma \lbrack A_{\widetilde{\alpha }%
},F_{01}]=0\quad \Rightarrow \quad \partial _{\widetilde{\alpha }%
}F_{01}^{\prime }=0\quad ,\quad F_{01}=U^{\prime }F_{01}^{\prime }U^{\prime
-1} \\ 
\\ 
\partial _{\alpha }F_{\widetilde{0}\widetilde{1}}-\gamma \lbrack A_{\alpha
},F_{\widetilde{0}\widetilde{1}}]=0\quad \Rightarrow \quad \partial _{\alpha
}F_{\widetilde{0}\widetilde{1}}^{\prime }=0\quad ,\quad F_{\widetilde{0}%
\widetilde{1}}=UF_{\widetilde{0}\widetilde{1}}^{\prime }U^{-1} \\ 
\end{array}
\label{q-13}
\end{equation}%
Now returning back in the real LCR-manifold these two partial differential
equations apparently coincide with the (abelian) equations 
\begin{equation}
\begin{array}{l}
\frac{1}{\sqrt{-g}}\partial _{\mu }\{\sqrt{-g}(\ell ^{\mu }m^{\nu }-\ell
^{\nu }m^{\mu })(n^{\rho }\overline{m}^{\sigma }F_{j\rho \sigma
})\}=-k_{j}^{\nu }\quad ,\quad \ell ^{\mu }m^{\nu }F_{j_{\mu \nu }}=0 \\ 
\\ 
n^{\mu }\overline{m}^{\nu }F_{j_{\mu \nu }}=0\quad ,\quad \frac{1}{\sqrt{-g}}%
\partial _{\mu }\{\sqrt{-g}(n^{\mu }\overline{m}^{\nu }-n^{\nu }\overline{m}%
^{\mu })(\ell ^{\rho }m^{\sigma }F_{j\rho \sigma })\}=-k_{i}^{\nu } \\ 
\end{array}
\label{q-13a}
\end{equation}%
Notice that the essential non-vanishing term in both solutions is null,
therefore we will look for completely null solutions, i.e. $(\ell ^{\rho
}n^{\sigma }-m^{\rho }\overline{m}^{\sigma })F_{j\rho \sigma }=0$. Hence we
will look for null abelian solutions which satisfy the equations 
\begin{equation}
\begin{array}{l}
d\{\ell \wedge m(n^{\rho }\overline{m}^{\sigma }F_{j\rho \sigma })\}=i\ast
k_{j}\quad ,\quad \ell ^{\mu }m^{\nu }F_{j_{\mu \nu }}=0\quad ,\quad (\ell
^{\rho }n^{\sigma }-m^{\rho }\overline{m}^{\sigma })F_{j\rho \sigma }=0 \\ 
\\ 
n^{\mu }\overline{m}^{\nu }F_{j_{\mu \nu }}=0\quad ,\quad d\{(n\wedge 
\overline{m}(\ell ^{\rho }m^{\sigma }F_{j\rho \sigma })\}=i\ast k_{i}^{\nu
}\quad ,\quad (\ell ^{\rho }n^{\sigma }-m^{\rho }\overline{m}^{\sigma
})F_{j\rho \sigma }=0 \\ 
\end{array}
\label{q-13b}
\end{equation}

The LCR-structure coordinates (\ref{q-5}) determine the two characteristic
2-forms of the static quadratic surface of $CP^{3}$. 
\begin{equation}
\begin{array}{l}
dz^{0}\wedge dz^{1}=(f_{0}^{0}f_{1}^{1}-f_{1}^{0}f_{0}^{1})\ell \wedge
m\quad ,\quad dz^{\widetilde{0}}\wedge dz^{\widetilde{1}}=(f_{\widetilde{0}%
}^{\widetilde{0}}f_{\widetilde{1}}^{\widetilde{1}}-f_{\widetilde{1}}^{%
\widetilde{0}}f_{\widetilde{0}}^{\widetilde{1}})n\wedge \overline{m} \\ 
\end{array}
\label{q-14}
\end{equation}%
These two surfaces are one-side extensions of the corresponding searched
solutions on $%
%TCIMACRO{\U{211d} }%
%BeginExpansion
\mathbb{R}
%EndExpansion
^{4}$ boundary of the classical domain. Hence they can be solved.

The non-vanishing closed 2-forms (with sources) are found to be

\begin{equation}
\begin{array}{l}
d\{\frac{C_{j}^{\prime }}{\sin \theta (r-ia\cos \theta )}\ell \wedge
m\}=i\ast k_{j}^{\prime } \\ 
\\ 
d(\frac{C_{j}^{\prime \prime }(r-ia\cos \theta )}{(r^{2}+a^{2})\sin \theta }%
n\wedge \overline{m})=i\ast k_{j}^{\prime \prime } \\ 
\end{array}
\label{q-15}
\end{equation}%
where $C_{j}^{\prime }$ and $C_{j}^{\prime \prime }$ are arbitrary complex
constants, which are fixed using Stokes' theorem. In the oblate spheroidal
coordinates the solutions have the explicit forms

\begin{equation}
\begin{array}{l}
\frac{C_{j}^{\prime }}{\sin \theta (r-ia\cos \theta )}\ell \wedge m=\frac{%
C_{j}^{\prime }}{\sqrt{2}}[\frac{ia}{(r^{2}+a^{2})}dt\wedge dr-\frac{1}{\sin
\theta }dt\wedge d\theta -idt\wedge d\varphi + \\ 
\qquad +\frac{\rho ^{2}}{(r^{2}+a^{2})\sin \theta }dr\wedge d\theta
+idr\wedge d\varphi -a\sin \theta d\theta \wedge d\varphi ] \\ 
\\ 
\frac{C_{j}^{\prime \prime }(r-ia\cos \theta )}{(r^{2}+a^{2})\sin \theta }%
n\wedge \overline{m}=\frac{C_{j}^{\prime \prime }}{2\sqrt{2}}[\frac{ia}{%
(r^{2}+a^{2})}dt\wedge dr-\frac{1}{\sin \theta }dt\wedge d\theta +idt\wedge
d\varphi - \\ 
\qquad -\frac{\rho ^{2}}{(r^{2}+a^{2})\sin \theta }dr\wedge d\theta
+idr\wedge d\varphi -a\sin \theta d\theta \wedge d\varphi ] \\ 
\end{array}
\label{q-15a}
\end{equation}%
After a straightforward calculation I find

\begin{equation}
\begin{array}{l}
\tint\limits_{t,r=const}\frac{C_{j}^{\prime }}{\sin \theta (r-ia\cos \theta )%
}\ell \wedge m=-2\sqrt{2}\pi C_{j}^{\prime }a\equiv \gamma _{j}^{\prime } \\ 
\\ 
\tint\limits_{t,r=const}\frac{C_{j}^{\prime \prime }(r-ia\cos \theta )}{%
(r^{2}-2Mr+a^{2}+q^{2})\sin \theta }n\wedge \overline{m}=-\sqrt{2}\pi
C_{j}^{\prime \prime }a\equiv \gamma _{j}^{\prime \prime }%
\end{array}
\label{q-16}
\end{equation}%
which implies that the constants must be real for the sources to be real and
the original field equations to be satisfied. Notice that they are
proportional to the coefficient $a$ (the spin of the soliton) implying that
the scalar LCR-structures (\ref{g-3}) do not define colored configurations
with sources. The physical meaning of this remark is that PCFT does not
permit glueballs.

We finally find the solutions

\begin{equation}
\begin{array}{l}
F_{j}^{\prime }=\frac{-\gamma _{j}^{\prime }}{\pi a\sqrt{2}}[\frac{a}{%
r^{2}+a^{2}}dt\wedge dr-d(t-r)\wedge d\varphi ]= \\ 
\qquad =d[\frac{-\gamma _{j}}{\pi a\sqrt{2}}(t-r)(\frac{a}{r^{2}+a^{2}}%
dr-d\varphi )] \\ 
\\ 
F_{j}^{\prime \prime }=\frac{-\gamma _{j}^{\prime \prime }\sqrt{2}}{\pi a}[%
\frac{a}{r^{2}+a^{2}}dt\wedge dr+d(t+r)\wedge d\varphi ]= \\ 
\qquad =d[\frac{-\gamma _{j}^{\prime \prime }\sqrt{2}}{\pi a}(t+r)(\frac{a}{%
r^{2}+a^{2}}dr+d\varphi )] \\ 
\end{array}
\label{q-17}
\end{equation}%
with the corresponding potentials been apparent. Notice that the parameter $%
a $ (the radius of the singularity ring) appears in the denominator. This
seems to be the origin of the confining potential asymptotic solution (\ref%
{i-5)} of the gluonic equations in contradiction to the electromagnetic
solution.

The second PDE of (\ref{q-3}), which is implied by the action $I_{I}$, may
be written as 
\begin{equation}
\begin{array}{l}
I_{I}\quad \rightarrow \quad \frac{1}{\sqrt{-g}}(D_{\mu })_{ij}\{\sqrt{-g}%
[(\ell ^{\mu }m^{\nu }-\ell ^{\nu }m^{\mu })(n^{\rho }\overline{m}^{\sigma
}\ast F_{j\rho \sigma })+ \\ 
\qquad \qquad +(n^{\mu }\overline{m}^{\nu }-n^{\nu }\overline{m}^{\mu
})(\ell ^{\rho }m^{\sigma }\ast F_{j\rho \sigma })]\}=-ik_{i}^{\nu }%
\end{array}
\label{q-17a}
\end{equation}%
because $\ell ^{\lbrack \rho }m^{\sigma ]}$ and $n^{[\rho }\overline{m}%
^{\sigma ]}$are self-dual. This has exactly the form of the first PDE, with
the gauge field tensor replaced by its dual. Hence the solutions of the
second PDE will be $-\ast F_{j}^{\prime }$ and $-\ast F_{j}^{\prime \prime }$%
, which is impossible, because they have sources, i.e. $d\ast F_{j}^{\prime
}\neq 0\neq d\ast F_{j}^{\prime \prime }$.

The left $F_{i01}$ and right $F_{j\widetilde{0}\widetilde{1}}$ solutions may
coexist in the same region if they do not vanish for $i$ and $j$\ in the
abelian subalgebra. In the physically interesting case of the $su(3)$ Lie
algebra can happen if $i$ and $j$\ take the values $3$ and $8$. But in this
case the final classical solution will not be null. It is a non-null
solution with precise gluonic charges. The final form of the non-null
gluonic solution is%
\begin{equation}
\begin{array}{l}
A_{j}^{(g)}=\frac{-\gamma _{j}}{4\pi a}(\tan ^{-1}\frac{r}{a}dt+rd\varphi )=
\\ 
\qquad =\frac{-\gamma _{j}}{4\pi a}(\tan ^{-1}\frac{r}{a}dx^{0}-\frac{%
ax^{1}+rx^{2}}{(x^{1})^{2}+(x^{2})^{2}}dx^{1}+\frac{rx^{1}-ax^{2}}{%
(x^{1})^{2}+(x^{2})^{2}}dx^{2}) \\ 
\\ 
A^{(e)}=\frac{qr^{3}}{4\pi (r^{4}+a^{2}(x^{3})^{2})}(dx^{0}-\frac{%
rx^{1}-ax^{2}}{r^{2}+a^{2}}dx^{1}-\frac{rx^{2}+ax^{1}}{r^{2}+a^{2}}dx^{2}-%
\frac{x^{3}}{r}dx^{3})%
\end{array}
\label{17ab}
\end{equation}%
where the last formula is the electromagnetic dressing. The gluonic monopole
potential is singular at $a=0$ of the spin parameter and and its magnetic
part is linear in $r$. These characteristics do not appear in the
electromagnetic potential and may provide confinement.

The second quark of the massless LCR-structure can be found using the same
procedure. Let us consider the first ($[E=-p^{3}]$) LCR-structure of (\ref%
{n-1}) with the corresponding structure coordinates and tetrad

\begin{equation}
\begin{array}{l}
\lbrack E=-p^{3}]:\quad X^{3}-aX^{1}=0\quad ,\quad X^{0}=0 \\ 
\\ 
z^{0}=x^{0}-ia-x^{3}-\frac{(x^{1})^{2}+(x^{2})^{2}}{x^{0}+x^{3}-ia}\quad
,\quad z^{1}=\frac{x^{1}+ix^{2}}{x^{0}+x^{3}-ia} \\ 
z^{\widetilde{0}}=x^{0}+x^{3}-\frac{(x^{1})^{2}+(x^{2})^{2}}{x^{0}-x^{3}}%
\quad ,\quad z^{\widetilde{1}}=\frac{x^{1}-ix^{2}}{x^{0}-x^{3}}%
\end{array}
\label{q-17b}
\end{equation}%
The two solutions with sources are expected to have the forms

\begin{equation}
\begin{array}{l}
F_{j}^{\prime }=f_{j}(z^{0},z^{1})dz^{0}\wedge dz^{1}\quad \rightarrow \quad
dF_{j}^{\prime }=-\ast k_{j}^{\prime } \\ 
\\ 
F_{j}^{\prime \prime }=f_{j}(z^{\widetilde{0}},z^{\widetilde{1}})dz^{%
\widetilde{0}}\wedge dz^{\widetilde{1}}\quad \rightarrow \quad
dF_{j}^{\prime \prime }=-\ast k_{j}^{\prime \prime }%
\end{array}
\label{q-17c}
\end{equation}%
But the naive Stokes' theorem does not apply. This problem has to be treated
in the "unphysical" grassmannian chart (\ref{g-3c}). Apparently, we may
bypass this difficulty by assuming a mass term and repeat the preceding
calculations, in order to experimentally check PCFT.

In analogy to electrodynamics we may introducing the quark field as the
source. The implied self-consistent equations become 
\begin{equation}
\begin{array}{l}
(D_{\mu })_{ij}H_{j}^{\mu \nu }=\gamma \overline{q}\gamma ^{\nu }\tau
_{i}q\quad ,\quad (D_{\mu })_{ij}\ast H_{j}^{\mu \nu }=0\quad ,\quad \gamma
^{\mu }(i\partial _{\mu }-\gamma A_{i\mu }\tau _{i})q-mq=0 \\ 
\\ 
H_{j}^{\mu \nu }\equiv \frac{1}{2}[(\ell ^{\mu }m^{\nu }-\ell ^{\nu }m^{\mu
})(n^{\rho }\overline{m}^{\sigma }F_{j\rho \sigma })+(\ell ^{\mu }\overline{m%
}^{\nu }-\ell ^{\nu }\overline{m}^{\mu })(n^{\rho }m^{\sigma }F_{j\rho
\sigma })+ \\ 
\qquad +(n^{\mu }\overline{m}^{\nu }-n^{\nu }\overline{m}^{\mu })(\ell
^{\rho }m^{\sigma }F_{j\rho \sigma })+(n^{\mu }\overline{m}^{\nu }-n^{\nu }%
\overline{m}^{\mu })(\ell ^{\rho }m^{\sigma }F_{j\rho \sigma })] \\ 
\ast H_{j}^{\mu \nu }\equiv \frac{i}{2}[(\ell ^{\mu }m^{\nu }-\ell ^{\nu
}m^{\mu })(n^{\rho }\overline{m}^{\sigma }F_{j\rho \sigma })-(\ell ^{\mu }%
\overline{m}^{\nu }-\ell ^{\nu }\overline{m}^{\mu })(n^{\rho }m^{\sigma
}F_{j\rho \sigma })+ \\ 
\qquad +(n^{\mu }\overline{m}^{\nu }-n^{\nu }\overline{m}^{\mu })(\ell
^{\rho }m^{\sigma }F_{j\rho \sigma })-(n^{\mu }\overline{m}^{\nu }-n^{\nu }%
\overline{m}^{\mu })(\ell ^{\rho }m^{\sigma }F_{j\rho \sigma })]%
\end{array}
\label{q-18}
\end{equation}%
It is not clear to me what this set of PDEs represent.

\subsection{A quark confining mechanism}

In order to understand the implied confinement, we have to understand the
mathematical framework of PCFT. Therefore I think it will be helpful to the
reader, if I briefly recapitulate it, using now the Sato's hyperfunction
point of view\cite{MOR} for the distributions (generalized functions).

The LCR-manifold is a special totally real 4-dimensional submanifold of a
complex 4-dimensional manifold satisfying the relations%
\begin{equation}
\begin{array}{l}
\rho =%
\begin{pmatrix}
\rho _{11}(\overline{z^{\alpha }},z^{\alpha }) & \rho _{12}(\overline{%
z^{\alpha }},z^{\widetilde{\alpha }}) \\ 
\overline{\rho _{12}(\overline{z^{\alpha }},z^{\widetilde{\alpha }})} & \rho
_{22}(\overline{z^{\widetilde{\alpha }}},z^{\widetilde{\alpha }})%
\end{pmatrix}%
=0 \\ 
\end{array}
\label{q-19}
\end{equation}%
in a neighborhood of $z^{b}(p)$. The CR-submanifolds are usually considered
as boundaries of domains of holomorphy. The LCR-manifold may be considered
as the boundary of the domain%
\begin{equation}
\begin{array}{l}
\rho =%
\begin{pmatrix}
\rho _{11}(\overline{z^{\alpha }},z^{\alpha }) & \rho _{12}(\overline{%
z^{\alpha }},z^{\widetilde{\alpha }}) \\ 
\overline{\rho _{12}(\overline{z^{\alpha }},z^{\widetilde{\alpha }})} & \rho
_{22}(\overline{z^{\widetilde{\alpha }}},z^{\widetilde{\alpha }})%
\end{pmatrix}%
\succ 0 \\ 
\\ 
\det \rho >0\quad ,\quad trace(\rho )>0 \\ 
\end{array}
\label{q-20}
\end{equation}%
where the symbol $\succ $\ means that the matrix $\rho $\ is positive
definite, which is equivalent to the last two conditions. In the zero
gravity approximation, the present ambient complex manifold is the $SU(2,2)$
classical (Cartan) and its characteristic (Shilov) boundary is $S^{1}\times
S^{3}$, which is a double cover of $%
%TCIMACRO{\U{211d} }%
%BeginExpansion
\mathbb{R}
%EndExpansion
^{4}$, as boundaries of the corresponding Siegel domains.

If the ambient complex manifold is a projective variety, precise special
dependence of the defining functions from the structure coordinates ($%
z^{\alpha },z^{\widetilde{\beta }}$) suggest us to consider it to be the
lines of $CP^{3}$ (points of the grassmannian manifold $G_{4,2}$), which
intersect two sheets (branches) of a hypersurface $K(Z^{m})=0$. The
structure coordinates $z^{\alpha }$ determine the one intersection point at
the one branch and $z^{\widetilde{\alpha }}$ determine the other
intersection point at another branch. The two branches intersect at a branch
curve of $CP^{3}$, which corresponds to the branch points of the Riemann
surfaces (algebraic curves) of $CP^{2}$. Recall that in $CP^{2}$,
analyticity is restored by using a branch cut that joins two branch points.
In the present case of $CP^{3}$, the cut is done at a surface, which has the
branch curve as boundary. The LCR-structure essentially projects this
picture down to the LCR-manifold (the boundary of the domain). Then the
holomorphic functions on the domains of holomorphy become generalized
functions (Sato's hyperfunctions) on the LCR-manifold (the real spacetime)
with real analytic forms in some regions and distributional sources into
others\cite{GRAF}. Below I will perform these calculations in the case of
the static solitonic LCR-manifold (with zero gravity) and the colored
solutions of the gauge field. These calculations may be elementary for the
mathematicians, but we (particle physicists) are not familiar with these
techniques.

The precise subset of quadratic polynomials, which is closed relative to the
Poincar\'{e} transformations have the form%
\begin{equation}
\begin{array}{l}
A_{mn}Z^{m}Z^{n}=0 \\ 
\\ 
A_{mn}=%
\begin{pmatrix}
\omega & P \\ 
P^{\top } & 0%
\end{pmatrix}%
\quad ,\quad P=%
\begin{pmatrix}
-(p^{1}-ip^{2}) & -p^{0}+p^{3} \\ 
p^{0}+p^{3} & (p^{1}+ip^{2})%
\end{pmatrix}%
=-p\epsilon \\ 
p=%
\begin{pmatrix}
p^{0}-p^{3} & -(p^{1}-ip^{2}) \\ 
-(p^{1}+ip^{2}) & p^{0}+p^{3}%
\end{pmatrix}%
\quad ,\quad \epsilon =%
\begin{pmatrix}
0 & 1 \\ 
-1 & 0%
\end{pmatrix}%
\end{array}
\label{q-21}
\end{equation}%
where $p^{\mu }$ is the (real) 4-momentum and $\omega $ the $2\times 2$
spin-matrix of the solitonic LCR-structure. The projection (from an external
point) of the quadric to a $CP^{2}\subset CP^{3}$ is a double cover (it has
two sheets). Let us consider the following projection of the quadric 
\begin{equation}
\begin{array}{l}
Z=X+\tau Y\quad ,\quad X=%
\begin{pmatrix}
1 \\ 
-1 \\ 
0 \\ 
0%
\end{pmatrix}%
\quad ,\quad Y=%
\begin{pmatrix}
Y^{1} \\ 
Y^{1} \\ 
Y^{2} \\ 
Y^{3}%
\end{pmatrix}
\\ 
\\ 
X^{m}X^{n}A_{mn}+2\tau X^{m}Y^{n}A_{mn}+\tau ^{2}Y^{m}Y^{n}A_{mn}=0 \\ 
\end{array}
\label{q-22}
\end{equation}%
The two roots of $\tau $ determine the two sheets of the quadric. In our
case, it is more convenient to consider the two intersection points between
the two branches and the line determined by $r\in G_{4,2}$ as the two roots
of the (projective) equation%
\begin{equation}
\begin{array}{l}
Z^{m}=%
\begin{pmatrix}
\lambda \\ 
-ir\lambda%
\end{pmatrix}
\\ 
A_{mn}Z^{m}Z^{n}=\lambda ^{T}(\omega -iPr-ir^{\top }P^{\top })\lambda =0 \\ 
\end{array}
\label{q-23}
\end{equation}%
The branch curve is given by the double root, i.e. it is%
\begin{equation}
\begin{array}{l}
\det (\omega -iPr-ir^{\top }P^{\top })=0 \\ 
\end{array}
\label{q-24}
\end{equation}%
It intersects the zero gravity LCR-submanifold of $G_{4,2}$ , when $r=x$ is
hermitian matrix. Recall that in zero gravity the domain becomes the $%
SU(2,2) $ symmetric classical domain and its boundary (in the chiral Siegel
realization) is now the real $%
%TCIMACRO{\U{211d} }%
%BeginExpansion
\mathbb{R}
%EndExpansion
^{4}$ submanifold.

We already have found that in one from the two branches the gauge field
strength $F_{\mu \nu }$ must vanish. Otherwise the gauge field does not
admit null solution with localized distributional source. Let us start with
the first left-hand solution of (\ref{q-17}), which at the one side of the
boundary it is 
\begin{equation}
\begin{array}{l}
F_{j01}(z^{\alpha })dz^{0}\wedge dz^{1}=\frac{-\gamma _{j}^{\prime }}{2\sqrt{%
2}\pi a\sin \theta (r-ia\cos \theta )}\ell \wedge m \\ 
d\{\frac{-\gamma _{j}^{\prime }}{2\sqrt{2}\pi a\sin \theta (r-ia\cos \theta )%
}\ell \wedge m\}=i\ast k_{j}^{\prime } \\ 
\end{array}
\label{q-25}
\end{equation}%
and at the other side it vanishes.

It is more convenient to use cartesian coordinates (\ref{e-15a}), where the
differential forms are 
\begin{equation}
\begin{array}{l}
dx=\frac{r\cos \varphi \sin \theta dr}{\sqrt{r^{2}+a^{2}}}+\sqrt{r^{2}+a^{2}}%
\cos \varphi \cos \theta d\theta -\sqrt{r^{2}+a^{2}}\sin \varphi \sin \theta
d\varphi \\ 
dy=\frac{r\sin \varphi \sin \theta dr}{\sqrt{r^{2}+a^{2}}}+\sqrt{r^{2}+a^{2}}%
\sin \varphi \cos \theta d\theta +\sqrt{r^{2}+a^{2}}\cos \varphi \sin \theta
d\varphi \\ 
dz=\cos \theta dr-r\sin \theta d\theta \\ 
\end{array}
\label{q-26}
\end{equation}%
which are inverted to 
\begin{equation}
\begin{array}{l}
d\varphi =\frac{1}{\sqrt{r^{2}+a^{2}}\sin \theta }(\cos \varphi dy-\sin
\varphi dx) \\ 
d\theta =\frac{\sqrt{r^{2}+a^{2}}\cos \theta \cos \varphi }{\rho ^{2}}dx+%
\frac{\sqrt{r^{2}+a^{2}}\cos \theta \sin \varphi }{\rho ^{2}}dy-\frac{r\sin
\theta }{\rho ^{2}}dz \\ 
dr=\frac{r\sqrt{r^{2}+a^{2}}\sin \theta \cos \varphi }{\rho ^{2}}dx+\frac{r%
\sqrt{r^{2}+a^{2}}\sin \theta \sin \varphi }{\rho ^{2}}dy+\frac{%
(r^{2}+a^{2})\cos \theta }{\rho ^{2}}dz \\ 
\end{array}
\label{q-27}
\end{equation}%
Then the flatprint of the LCR-tetrad takes the form%
\begin{equation}
\begin{array}{l}
\ell _{\mu }dx^{\mu }=dt+\frac{ay-rx}{r^{2}+a^{2}}dx-\frac{ry+ax}{r^{2}+a^{2}%
}dy-\frac{z}{r}dz \\ 
n_{\mu }dx^{\mu }=\frac{r^{2}(r^{2}+a^{2})}{2(r^{4}+a^{2}z^{2})}[dt+\frac{%
ay+rx}{r^{2}+a^{2}}dx+\frac{ry-ax}{r^{2}+a^{2}}dy+\frac{z}{r}dz] \\ 
\frac{1}{\overline{\eta }\sin \theta }m_{\mu }dx^{\mu }=\frac{r^{2}}{%
(r^{4}+a^{2}z^{2})\sqrt{2}}[iadt+\frac{r^{2}+a^{2}}{x^{2}+y^{2}}(-\frac{xz}{r%
}+iy)dx \\ 
\qquad -\frac{r^{2}+a^{2}}{x^{2}+y^{2}}(\frac{yz}{r}+ix)dy+rdz] \\ 
\end{array}
\label{q-28}
\end{equation}%
from which I will compute the self-dual 2-form

\begin{equation}
\begin{array}{l}
G_{j}-i\ast G_{j}\equiv \frac{-\gamma _{j}}{2\sqrt{2}\pi a\sin \theta
(r-ia\cos \theta )}\ell \wedge m=-(E_{j}^{1}+iB_{j}^{1})dt\wedge dx- \\ 
\qquad -(E_{j}^{2}+iB_{j}^{2})dt\wedge dy-(E_{j}^{3}+iB_{j}^{3})dt\wedge
dz-i(E_{j}^{3}+iB_{j}^{3})dx\wedge dy+ \\ 
\qquad +i(E_{j}^{2}+iB_{j}^{2})dx\wedge dz-i(E_{j}^{1}+iB_{j}^{1})dy\wedge dz
\\ 
\end{array}
\label{q-29}
\end{equation}%
Hence it defines an effective real 2-form $G_{j}$, where $\overrightarrow{%
E_{j}}$ is its color electric component and $\overrightarrow{B_{j}}$ is its
color magnetic component. The color electric and magnetic fields of the
first solution are 
\begin{equation}
\begin{array}{l}
\overrightarrow{E_{j}^{\prime }}=\frac{\gamma _{j}^{\prime }r^{2}}{%
(r^{4}+a^{2}z^{2})\pi a\sqrt{2}}%
\begin{pmatrix}
\frac{a(ay-rx)}{r^{2}+a^{2}}-\frac{y(r^{2}+a^{2})}{x^{2}+y^{2}} \\ 
\frac{x(r^{2}+a^{2})}{x^{2}+y^{2}}-\frac{a(ax+ry)}{r^{2}+a^{2}} \\ 
-\frac{az}{r}%
\end{pmatrix}
\\ 
\end{array}
\label{q-30}
\end{equation}%
and 
\begin{equation}
\begin{array}{l}
\overrightarrow{B_{j}^{\prime }}=\frac{\gamma _{j}^{\prime }r}{%
(r^{4}+a^{2}z^{2})\pi a\sqrt{2}}%
\begin{pmatrix}
\frac{(r^{2}+a^{2})xz}{(x^{2}+y^{2})} \\ 
\frac{(r^{2}+a^{2})yz}{(x^{2}+y^{2})} \\ 
-1%
\end{pmatrix}
\\ 
\end{array}
\label{q-31}
\end{equation}%
respectively.

Recall that the variable $r$ is an oblate spheroidal coordinate and
satisfies the relation 
\begin{equation}
\begin{array}{l}
\frac{x^{2}+y^{2}}{r^{2}+a^{2}}+\frac{z^{2}}{r^{2}}=1 \\ 
\end{array}
\label{q-32}
\end{equation}%
Now we are ready to reveal the singularities of the solution, which occur at 
\begin{equation}
\begin{array}{l}
(r,z)=(0,0)\quad \rightarrow \quad x^{2}+y^{2}=a^{2} \\ 
and \\ 
x^{2}+y^{2}=0\quad \rightarrow \quad z=\pm r \\ 
\end{array}
\label{q-33}
\end{equation}%
This is the ring circle and the z-axis. An additional singularity
(discontinuity) exists on the branch cut ($x^{2}+y^{2}<0$), which is not
seen by the field outside the branch cut. At the one side of the branch cut
the color gauge field does not vanish, while at the other side it does
vanish, because crossing the disk branch cut we pass to the other branch
with vanishing gauge field. Recall that a point of the grassmannian space
corresponds to two points of $CP^{3}$, which belong to different branches,
branched at the branch cut. In brief the quark is located at the ring
singularity and the positive (or negative) z-axis, where it has
non-vanishing gluon field. Compare the singularities of the present gluonic
field with the corresponding singularities of the electromagnetic field (\ref%
{e-15c}). The gluonic field has the additional singularity at the z-axis
which obstructs its free existence. Hence, unlike the electron, the quark
cannot be free.

In the case of a quark and antiquark system, located at the two end rings of
a finite tube and having non-vanishing gluon field inside the tube and
vanishing outside the tube, form a "fat" Nielsen-Olesen string\cite{FELS},
which implies confinement. Recall that the non-vanishing component of the
gluonic field inside the string will be either the left or the right one,
because there is no solution with both being non-zero. Hence the scalar
meson bound state (pion) will have a precise chirality and it will be a
pseudoscalar.

The second (right-hand) solution of (\ref{q-17}), which at the one side of
the boundary it is

\begin{equation}
\begin{array}{l}
F_{j\widetilde{0}\widetilde{1}}(z^{\widetilde{\alpha }})dz^{\widetilde{0}%
}\wedge dz^{\widetilde{1}}=\frac{-\gamma _{j}^{\prime \prime }(r-ia\cos
\theta )}{\sqrt{2}\pi a(r^{2}+a^{2})\sin \theta }n\wedge \overline{m} \\ 
G_{j}^{\prime \prime }-i\ast G_{j}^{\prime \prime }\equiv \frac{-\gamma
_{j}(r-ia\cos \theta )}{\sqrt{2}\pi a(r^{2}+a^{2})\sin \theta }n\wedge 
\overline{m} \\ 
\end{array}
\label{q-34}
\end{equation}%
where $\gamma _{j}^{\prime \prime }$ is the effective color-electric charge.

The color electric and magnetic fields of the second solution are 
\begin{equation}
\begin{array}{l}
\overrightarrow{E_{j}^{\prime \prime }}=\frac{\gamma _{j}^{\prime \prime
}r^{2}\sqrt{2}}{(r^{4}+a^{2}z^{2})\pi a}%
\begin{pmatrix}
\frac{y(r^{2}+a^{2})}{x^{2}+y^{2}}-\frac{a(ay+rx)}{r^{2}+a^{2}} \\ 
\frac{a(ax-ry)}{r^{2}+a^{2}}-\frac{x(r^{2}+a^{2})}{x^{2}+y^{2}} \\ 
-\frac{az}{r}%
\end{pmatrix}
\\ 
\end{array}
\label{q-35}
\end{equation}%
and 
\begin{equation}
\begin{array}{l}
\overrightarrow{B_{j}^{\prime \prime }}=\frac{\gamma _{j}^{\prime \prime }r%
\sqrt{2}}{(r^{4}+a^{2}z^{2})\pi a}%
\begin{pmatrix}
\frac{(r^{2}+a^{2})xz}{(x^{2}+y^{2})} \\ 
\frac{(r^{2}+a^{2})yz}{(x^{2}+y^{2})} \\ 
-1%
\end{pmatrix}
\\ 
\end{array}
\label{q-36}
\end{equation}%
respectively. This non-vanishing right-hand solution is slightly different
than the left-hand one, but it has the same singularities.

Let us now see that it does not seem possible to incorporate this gluonic
solution to the standard model action, using the Bogoliubov recursive
procedure. One can check that the color electric and magnetic vectors
satisfy the following null relations 
\begin{equation}
\begin{array}{l}
(\overrightarrow{E_{j}^{\prime }})^{2}-(\overrightarrow{B_{j}^{\prime }}%
)^{2}=0\quad ,\quad \overrightarrow{E_{j}^{\prime }}\cdot \overrightarrow{%
B_{j}^{\prime }}=0\quad ,\quad No\ j\ summation \\ 
\\ 
(\overrightarrow{E_{j}^{\prime \prime }})^{2}-(\overrightarrow{B_{j}^{\prime
\prime }})^{2}=0\quad ,\quad \overrightarrow{E_{j}^{\prime \prime }}\cdot 
\overrightarrow{B_{j}^{\prime \prime }}=0\quad ,\quad No\ j\ summation%
\end{array}
\label{q-37}
\end{equation}%
and the effective field equations 
\begin{equation}
\begin{array}{l}
dG_{j}^{\prime }=0\quad ,\quad d\ast G_{j}^{\prime }=i\ast k_{j}^{\prime }
\\ 
\\ 
\overrightarrow{\nabla }\cdot \overrightarrow{B_{j}^{\prime }}=0\quad ,\quad
\partial _{t}\overrightarrow{B_{j}^{\prime }}+\overrightarrow{\nabla }\times 
\overrightarrow{E_{j}^{\prime }}=0 \\ 
\overrightarrow{\nabla }\cdot \overrightarrow{E_{j}^{\prime }}=k_{j}^{\prime
0}\quad ,\quad \partial _{t}\overrightarrow{E_{j}^{\prime }}-\overrightarrow{%
\nabla }\times \overrightarrow{B_{j}^{\prime }}=-\overrightarrow{%
k_{j}^{\prime }}%
\end{array}
\label{q-38}
\end{equation}%
where the Minkowski metric is assumed. The non-vanishing right-hand solution
satisfies the same equations.

\section{A SU(3) CONNECTION\ FROM\ THE\ LCR-STRUTURE}

\setcounter{equation}{0}

Recall that Einstein was looking for a geometric structure, which could
replace the lorentzian metric, and produce all the interactions. His higher
dimensional Kaluza-Klein model did not succeed in describing
electomagnetism. The same fate had the metric with torsion suggested by
Cartan. We already showed that the LCR-structure implies the metric
structure and the electroweak $U(2)$ connection, which are manifestations of
the LCR-tetrad. In the previous sections, we found distributional solutions
of the gauge field related to the static LCR-structure. If we also succeed
to derive the $SU(3)$ connection from the LCR-structure, Einstein could be
justified. Everything (gravity, electromagnetism, weak and strong
interactions and the fermionic particles) are manifestations of the pure
geometric LCR-structure, without any additional gauge field. The suggestion
of this section is that a $SU(3)$ Cartan connection exists, which could
provide the distributional solutions found above.

A realizable LCR-structure is based on hypersurfaces of $CP^{3}$, which are
covariant relative to $SL(4,%
%TCIMACRO{\U{2102} }%
%BeginExpansion
\mathbb{C}
%EndExpansion
)$ transformation. That is $SL(4,%
%TCIMACRO{\U{2102} }%
%BeginExpansion
\mathbb{C}
%EndExpansion
)$ preserves LCR-structure. Following the Griffiths-Harris\cite{GrHa2}
moving frame approach, we will look for a possible emergence of the gluonic
connection.

Let a frame \{$A_{0},\ A_{1},\ A_{2},\ A_{3}$\} of $CP^{3}$ determined by
the corresponding four vectors of $%
%TCIMACRO{\U{2102} }%
%BeginExpansion
\mathbb{C}
%EndExpansion
^{4}$, where $A_{0}$ determines the point where the $CP^{3}$ defining lines
of $%
%TCIMACRO{\U{2102} }%
%BeginExpansion
\mathbb{C}
%EndExpansion
^{4}$ pass through. Assuming it to be a unitary basis, the Cartan moving
frame relations are%
\begin{equation}
\begin{array}{l}
dA_{m}=\omega _{mn}A_{n}\quad ,\quad d\omega _{mn}=\omega _{ml}\wedge \omega
_{ln}\quad ,\quad \omega _{mn}=\overline{\omega _{nm}} \\ 
l,m,n=0,1,2,3 \\ 
\end{array}
\label{c-1}
\end{equation}%
$\omega _{ml}$ are 1-forms, which take values in the Lie algebra of $SU(4)$. 
$CP^{3}$ is determined by the annihilation of the 1-forms $\omega _{0i}$,
(which is a basis of the cotangent space $T^{\ast }(CP^{3})$) because of the
Frobenius relation%
\begin{equation}
\begin{array}{l}
d\omega _{0i}=\omega _{00}\wedge \omega _{0i}+\omega _{0j}\wedge \omega _{ji}
\\ 
\end{array}
\label{c-2}
\end{equation}%
Hence the projection%
\begin{equation}
\begin{array}{l}
A_{0}:\ U(4)\rightarrow CP^{3} \\ 
\end{array}
\label{c-3}
\end{equation}%
gives a principal $U(1)\times U(3)$ fibration with corresponding vector
bundles, the line bundle $L_{A_{0}}=\overrightarrow{OA_{0}}$ and the
universal quotient bundle $Q_{A_{0}}=%
%TCIMACRO{\U{2102} }%
%BeginExpansion
\mathbb{C}
%EndExpansion
^{4}/L_{A_{0}}$. In our case, the Kerr function (\ref{l-9}) generates a
Darboux unitary frame 
\begin{equation}
\begin{array}{l}
\widehat{Z}_{0}(\tau ,s),\ \widehat{Z}_{1}(\tau ,s),\ \widehat{Z}_{2}(\tau
,s),\ \widehat{Z}_{3}(\tau ,s) \\ 
d\widehat{Z}_{0}=\widehat{\theta }_{00}\widehat{Z}_{0}+\widehat{\theta }_{0i}%
\widehat{Z}_{i} \\ 
d\widehat{\theta }_{0i}-iA_{ij}\wedge \widehat{\theta }_{0j}=0\quad ,\quad 
\overline{A_{ij}}=A_{ji} \\ 
A=(A_{ij})=\tsum\limits_{I=1}^{8}A_{I\beta }dz^{\beta }(t_{I})_{ij}\quad
,\quad \lbrack t_{I},t_{J}]=if_{IJK}t_{K} \\ 
\\ 
F=\partial A-iA\wedge A\ \longrightarrow DF:=\ \partial F+iA\wedge
F-iF\wedge A=0%
\end{array}
\label{c-4}
\end{equation}%
where the general form of antihermitian connection has been replaced with
the usual hermitian gauge field $A=(A_{ij})$, $t_{J}$ are the generators of $%
SU(3)$, and $\partial (A_{I\beta }dz^{\beta })=\frac{\partial A_{I\beta }}{%
\partial z^{\alpha }}dz^{\alpha }\wedge dz^{\alpha }$. The explicit form of
the curvature is\ 
\begin{equation}
\begin{array}{l}
F_{I\alpha \beta }=\partial _{\alpha }A_{I\beta }-\partial _{\beta
}A_{I\alpha }-f_{IJK}A_{J\alpha }A_{K\beta } \\ 
\\ 
F_{I01}=\partial _{0}A_{I1}-\partial _{1}A_{I0}-f_{IJK}A_{J0}A_{K1}%
\end{array}
\label{c-5}
\end{equation}%
Notice that the Bianchi identity is identically satisfied, because of the
(complex) dimension-2 of the analytic hypersurface. In this context the
leptonic particles correspond to $F_{I\alpha \beta }=0$ and the quarks
(hadrons) to $F_{I\alpha \beta }\neq 0$.

Recall that a complex point ($z^{\alpha },z^{\widetilde{\beta }}$) in the
4-dimensional ambient Kaehler manifold is determined by the two complex
points $z^{\alpha }$ and $z^{\widetilde{\beta }}$ of the hypersurface of $%
CP^{3}$. Hence the above holomorphic connection adapted to the analytic
surface implies the following section\ 
\begin{equation}
\begin{array}{l}
A_{J}=A_{J\alpha }(z^{\beta })dz^{\alpha }+A_{J\widetilde{\alpha }}(z^{%
\widetilde{\beta }})dz^{\widetilde{\alpha }} \\ 
\\ 
F_{I\alpha \beta }(z^{\beta })=\partial _{\alpha }A_{I\beta }-\partial
_{\beta }A_{I\alpha }-f_{IJK}A_{J\alpha }A_{K\beta } \\ 
F_{I\widetilde{\alpha }\widetilde{\beta }}(z^{\widetilde{\beta }})=\partial
_{\widetilde{\alpha }}A_{I\widetilde{\beta }}-\partial _{\widetilde{\beta }%
}A_{I\widetilde{\alpha }}-f_{IJK}A_{J\widetilde{\alpha }}A_{K\widetilde{%
\beta }} \\ 
G_{I\alpha \widetilde{\beta }}=0%
\end{array}
\label{c-6}
\end{equation}%
which is reduced down to the (real) 4-dimensional LCR-manifold\ 
\begin{equation}
\begin{array}{l}
A_{J}=A_{J\alpha }(z^{\beta }(x))\frac{\partial z^{\alpha }}{\partial x^{\mu
}}dx^{\mu }+A_{J\widetilde{\alpha }}(z^{\widetilde{\beta }}(x))\frac{dz^{%
\widetilde{\alpha }}}{\partial x^{\mu }}dx^{\mu } \\ 
\end{array}
\label{c-7}
\end{equation}

Achieving such a framework could provide Einstein's objective to show that
all the interactions observed in nature have a geometric origin. Besides the
color group is fixed to the observed in nature $SU(3)$ group.

\section{PERSPECTIVES}

\setcounter{equation}{0}

The recent experimental results of the LHC experiments at CERN show that
supersymmetric particles do not exist and subsequently, quantum string
theory does not describe nature. Hence, the 4-dimensional PCFT remains the
only known model, compatible with quantum theory, which provides the general
experimentally observed framework, without needing supersymmetry to
introduce fermions and internal symmetries.

Paraphrasing Euclid (of Alexandria) we may say that "there is no royal road
to .... the theory of everything too". We already realized the background
algebraic geometric structure in string theory based on the algebraic
curves. PCFT has essentially the same mathematical basis as the Polyakov
action, it simply transfers the mathematics to the algebraic surfaces of $%
CP^{3}$ and their intersection with the lines, which constitute the
grassmannian space $G_{4,2}$. The lorentzian CR-structure projection to a
real 4-dimensional submanifold complicates further the mathematical
problems, but it clarifies the general physical picture. No moving strings
and hidden dimensions are needed. The particles and their potentials
(dressings) are distributional solitons. The singular part of the
distribution (the source) is the fermionic particle and the regular part is
its "potential". The observed spacetime is just the "superposition" of these
elementary solitons. Mathematically they are real 4-dimentional manifolds
(boundaries of domains of holomorphy) (\ref{g-8}). The Minkowski spacetime
is the Shilov boundary of this $SU(2,2)$ classical domain in its unbounded
realization and precisely its universal cover $%
%TCIMACRO{\U{211d} }%
%BeginExpansion
\mathbb{R}
%EndExpansion
\times S^{3}$.

The mathematical structure is simple and beautiful, because the action is
fully geometric and apparently renormalizable by simple conventional power
counting. It is just the vector bundle on a 4-dimensional lorentzian
CR-structure. The riemannian metric of general relativity is replaced by the
LCR-structure as the fundamental quantity in PCFT. Around the notion of the
LCR-structure we have to build up again the appropriate computational
techniques to derive the observed phenomena. Up to now, I used the
conventional solitonic techniques, which seem to be quite limited. Only the
static (electron) and stationary (neutrino) solitons, and the corresponding
quarks have been revealed. The three particle generations and the
correspondence between leptons and quarks have been easily derived, but I do
not actually see the way to compute "numbers". The computation of the masses
and coupling constants of the effective standard model should be the
ultimate goal of PCFT. They should emerge from the intimate relation between
the electroweak gauge fields and the geodetic and shear free null tangent
vectors of Einstein's gravity. The apparent difference between the
conventional quantum chromodynamics with the present derived gluonic
interaction could be used to provide precise experimental tests of PCFT in
HL-LHC experiments.

The Bogoliubov axiomatic formulation of quantum field theory is essentially
based on the rigged Hilbert spaces (the Gelfand triplet) of the
representations of the Poincar\'{e} group. The Epstein-Glaser observation
makes the Bogoliubov mathematical formulation intimately related with the
Schwartz distributions. In the context of PCFT, the elementary particles are
distributional solitons. They are essentially the wavefront singularities of
the LCR-manifolds (the leptons) and the compatible gauge field (the quarks).
Notice that the "free" particles emerge with their corresponding
gravitational, electromagnetic and gluonic (for the quarks) dressings. In
conventional QFT, it is clear that the generalized functions is the adequate
framework to formulate quantum field theory. PCFT seems to invert the
reasoning. That is, the particle-wave duality (quantum mechanics) may also
be a consequence of the distributional solitonic nature of the structures of
PCFT. I think that we should start investigating\cite{RAG2021} this
provoking possibility too.


\begin{thebibliography}{99}
\bibitem{BAOU} M. S. Baouendi, P. Ebenfelt and L. Rothschild, "Real
submanifolds in complex space and their mappings", Princeton University
Press, Princeton, (1999).

\bibitem{BOG1975} N. N. Bogoliubov, A.A. Logunov and I.T. Todorov,
"Introduction to Axiomatic Quantum Field Theory", W.A. Benjamin Publishing
Company, Inc. USA (1975).

\bibitem{BOG1980} N. N. Bogoliubov and D. V. Shirkov, "Introduction to the
Theory of Quantized Fields", John Wiley and sons, Inc. USA, (1980).

\bibitem{CARTAN} E. Cartan, Ann. Math. Pure Appl. (4) \underline{11} (1932),
17.

\bibitem{CARTER} B. Carter, Phys. Rev. \underline{174} (1968), 1559.

\bibitem{CHAND} S. Chandrasekhar, \textquotedblleft The Mathematical Theory
of Black Holes\textquotedblright , Clarendon, Oxford, (1983).

\bibitem{FELS} Felsager B., \textquotedblleft Geometry, Particles and
Fields\textquotedblright , Odense Univ. Press, (1981).

\bibitem{FLAHE1974} E. J. Jr Flaherty, Phys. Lett. \underline{A46}, (1974)
391.

\bibitem{FLAHE1976} E. J. Jr Flaherty, \textquotedblleft Hermitian and K\"{a}%
hlerian geometry in Relativity\textquotedblright , Lecture Notes in Physics 
\underline{46}, Springer, Berlin, (1976).

\bibitem{GELF1} I. M. Gel'fand and G. E. Shilov, \textquotedblleft
Generalized Functions", vol. 1, Academic Press Inc., New York, (1964).

\bibitem{GELF4} I. M. Gel'fand and N. Ya. Vilenkin, \textquotedblleft
Generalized Functions", vol. 4, Academic Press Inc., New York, (1964).

\bibitem{GRAF} U. Graf, \textquotedblleft Introduction to Hyperfunctions and
Their Integral Transforms\textquotedblright , The Birkhauser/Springer Basel
AG, (2010).

\bibitem{GRIF} P. Griffiths and J. Harris, "Principles of Algebraic
Geometry", John Willey and sons, Inc. New York, (1978).

\bibitem{GrHa2} P. Griffiths and J. Harris, "Algebraic Geometry and Local
Differential Geometry", Ann. scient. Ec. Norm. Sup. vol. 12 (1979), 355.

\bibitem{MOR} M. Morimoto, \textquotedblleft An Introduction to Sato's
Hyperfunctions\textquotedblright , Translation of AMS, (1993).

\bibitem{MTW} C. N. Misner, K. S. Thorn and J. A. Wheeler, "GRAVITATION", W.
H. Freeman and Co, (1973).

\bibitem{NEWM1973} E. T. Newman, J. Math. Phys. \underline{14}, (1973), 102.

\bibitem{NEWM2016} E. T. Newman, "Assymptotically flat space-time and its
hidden recesses: An enigma from GR", arXiv:[gr-qc]/1602.07218v1.

\bibitem{P-R} Penrose R. and Rindler W., \textquotedblleft Spinors and
space-time\textquotedblright , vol. I and II, Cambridge Univ. Press,
Cambridge, (1984).

\bibitem{POL} J. Polchinski, "STRING\ THEORY", vol. I, Cambridge Univ.
Press, Cambridge, (2005).

\bibitem{PIAT} I. I. Pyatetskii-Shapiro, \textquotedblleft Automorphic
functions and the geometry of classical domains\textquotedblright , Gordon
and Breach, New York, (1969).

\bibitem{RAG1990} C. N. Ragiadakos, "A Four Dimensional Extended Conformal
Model", Phys. Lett. \underline{B251}, (1990), 94.

\bibitem{RAG1991} C. N. Ragiadakos, "Solitons in a Four Dimensional
Generally Covariant Conformal Model" Phys. Lett. \underline{B269}, (1991),
325.

\bibitem{RAG1992} C. N. Ragiadakos, "Quantization of a Four Dimensional
Generally Covariant Conformal Model", J. Math. Phys. \underline{33}, (1992),
122.

\bibitem{RAG1999} C. N. Ragiadakos, "Geometrodynamic solitons", Int. J.
Math. Phys. \underline{A14}, (1999), 2607.

\bibitem{RAG2008a} C. N. Ragiadakos (2008), \textquotedblleft
Renormalizability of a modified generally covariant Yang-Mills
action\textquotedblright , arXiv:hep-th/0802.3966v2.

\bibitem{RAG2008b} C. N. Ragiadakos (2008), \textquotedblleft A modified Y-M
action with three families of fermionic solitons and perturbative
confinement\textquotedblright , arXiv:hep-th/0804.3183v1.

\bibitem{RAG2013b} C. N. Ragiadakos (2013), "Lorentzian CR stuctures",
arXiv:hep-th/1310.7252.

\bibitem{RAG2017} C. N. Ragiadakos (2017), "Pseudo-conformal Field Theory",
arXiv:hep-th/1704.00321.

\bibitem{RAG2018a} C. N. Ragiadakos (2018), "Standard Model Derivation from
a 4-d Pseudo-conformal Field Theory", arXiv:hep-th/1805.11966.

\bibitem{RAG2018b} C. N. Ragiadakos (2018), "Hadronic Sector in 4-d
Pseudo-conformal Field Theory", arXiv:hep-th/1811.04428.

\bibitem{RAG2021} C. N. Ragiadakos, Research eBook of "Pseudo-Conformal
Field Theory" in my personal page www.pcft.gr.

\bibitem{SCH1} G. Scharf, \textquotedblleft Finite Quantum Electrodynamics:
The causal approach", Springer-Verlag, Berlin, (1995).

\bibitem{SCH2} G. Scharf, \textquotedblleft Quantum Gauge Theorie: A true
ghost story", John Wiley \& Sons, Inc. USA, (2001).

\bibitem{STR} R. E. Strichartz, \textquotedblleft A Guide to Distribution
Theory and Fourier Transforms", CRC Press Inc., Florida, (1994).
\end{thebibliography}
\end{document}